\documentclass[oneside,11pt]{article}

\usepackage[utf8]{inputenc}
\usepackage[T1]{fontenc}

\usepackage{amsthm}
\usepackage{amsmath}
\usepackage{amsfonts}
\usepackage{amssymb}

\usepackage{graphicx} 
\usepackage{float}
\usepackage{booktabs}
\usepackage{multirow}
\usepackage{rotating}
\usepackage{array}
\usepackage{xcolor}
\usepackage{subfig}
\usepackage[ruled]{algorithm2e}

\newcolumntype{R}[1]{>{\raggedleft\let\newline\\\arraybackslash\hspace{0pt}}m{#1}}

\usepackage{breakcites}

\usepackage{enumitem}

\usepackage[top=1.5cm,bottom=2cm,right=2.5cm,left=2.5cm]{geometry}
\usepackage{titling}

\usepackage{natbib}

\usepackage{ifplatform}

\usepackage{textcomp}

\ifwindows
\fi

\ifwindows
\usepackage{bbm}
\else
\iflinux
\usepackage{bbm}
\else
\usepackage{bbm}
\fi
\fi

\newcommand{\lp}{\left(}
\newcommand{\rp}{\right)}
\newcommand{\lc}{\left[}
\newcommand{\rc}{\right]}

\newcommand{\R}{\mathbb{R}}

\newcommand{\bx}{\mathbf{x}}

\newcommand{\bX}{\mathbf{X}}

\newcommand{\bmu}{\boldsymbol\mu}
\newcommand{\bu}{\mathbf{u}}
\newcommand{\bv}{\mathbf{v}}

\newcommand{\zero}{\mathbf{0}}
\newcommand{\bxi}{\boldsymbol\xi}

\newcommand{\bbeta}{\boldsymbol\eta}
\newcommand{\bzeta}{\boldsymbol\zeta}
\newcommand{\bga}{\boldsymbol\gamma}

\newcommand{\btheta}{\boldsymbol\theta}

\newcommand{\bsigma}{\boldsymbol\sigma}

\newcommand{\Hcal}{\mathcal{H}}

\newcommand{\bB}{\mathbf{B}}

\newcommand{\bI}{\mathbf{I}}

\newcommand{\inlaw}{\stackrel{d}{\rightsquigarrow}}

\newcommand{\lrp}[1]{\left(#1\right)}
\newcommand{\lrc}[1]{\left[#1\right]}
\newcommand{\lrb}[1]{\left\{#1\right\}}

\newcommand{\E}[1]{\mathbb{E}\lp #1\rp}

\newcommand{\Es}[2]{\mathbb{E}_{#2}\lc #1\rc}

\newcommand{\abs}[1]{\left| #1\right|}

\newcommand{\Om}[1]{\Omega_{#1}}

\DeclareFontFamily{OT1}{pzc}{}
\DeclareFontShape{OT1}{pzc}{m}{it}{<-> s * [1.10] pzcmi7t}{}
\DeclareMathAlphabet{\mathpzc}{OT1}{pzc}{m}{it}

\usepackage[pdftex,final,bookmarksnumbered,bookmarksopen=false,breaklinks,colorlinks]{hyperref}
\hypersetup{
	final,
	bookmarksnumbered=true,
	bookmarksopen=true,
	bookmarksopenlevel=0,
	unicode=false,          
	pdftoolbar=true,        
	pdfmenubar=true,        
	pdffitwindow=false,     
	pdftitle={On a projection-based class of uniformity tests on the hypersphere},    
	pdfdisplaydoctitle=true, 
	pdftoolbar=true,
	pdfmenubar=true,
	pdflang={English},
	pdfauthor={Eduardo Garcia-Portugues, Paula Navarro-Esteban, Juan A. Cuesta-Albertos},     
	pdfsubject={arXiv paper},   
	pdfcreator={Eduardo Garcia-Portugues},   
	pdfproducer={Eduardo Garcia-Portugues}, 
	pdfkeywords={Circular data}{Directional data}{Hypersphere}{Sobolev tests}{Uniformity}, 
	pdfnewwindow=true,      
	breaklinks=true,
	hidelinks,       
	linkcolor=black,          
	citecolor=black,        
	filecolor=magenta,      
	urlcolor=cyan           
}

\newtheorem{definition}{Definition}
\newtheorem{theorem}{Theorem}
\newtheorem{corollary}{Corollary}
\newtheorem{remark}{Remark}
\newtheorem{proposition}{Proposition}
\newtheorem{lemma}{Lemma}
\newtheorem{algo}{Algorithm}

\renewenvironment{proof}[1]{\textit{Proof#1.}}{\qed\\} 

\allowdisplaybreaks

\newif\ifmain
\maintrue
\ifmain

\else

\fi
\newif\ifsupplement
\supplementtrue
\newif\iffigstabs
\figstabstrue

\begin{document}

\ifmain

\title{On a projection-based class of uniformity tests on the hypersphere}
\setlength{\droptitle}{-1cm}
\predate{}%
\postdate{}%

\date{}

\author{Eduardo Garc\'ia-Portugu\'es$^{1,2,4}$, Paula Navarro-Esteban$^{3}$, and Juan A. Cuesta-Albertos$^{3}$}

\footnotetext[1]{Department of Statistics, Carlos III University of Madrid (Spain).}
\footnotetext[2]{UC3M-Santander Big Data Institute, Carlos III University of Madrid (Spain).}
\footnotetext[3]{Department of Mathematics, Statistics and Computer Science, University of Cantabria (Spain).}
\footnotetext[4]{Corresponding author. e-mail: \href{mailto:edgarcia@est-econ.uc3m.es}{edgarcia@est-econ.uc3m.es}.}

\maketitle


\begin{abstract}
	We propose a projection-based class of uniformity tests on the hypersphere using statistics that integrate, along all possible directions, the weighted quadratic discrepancy between the empirical cumulative distribution function of the projected data and the projected uniform distribution. Simple expressions for several test statistics are obtained for the circle and sphere, and relatively tractable forms for higher dimensions. Despite its different origin, the proposed class is shown to be related with the well-studied Sobolev class of uniformity tests. Our new class proves itself advantageous by allowing to derive new tests for hyperspherical data that neatly extend the circular tests by Watson, Ajne, and Rothman, and by introducing the first instance of an Anderson--Darling-like test for such data. The asymptotic distributions and the local optimality against certain alternatives of the new tests are obtained. A simulation study evaluates the theoretical findings and evidences that, for certain scenarios, the new tests are competitive against previous proposals. The new tests are employed in three astronomical applications.
\end{abstract}
\begin{flushleft}
	\small\textbf{Keywords:} Circular data; Directional data; Hypersphere; Sobolev tests; Uniformity.
\end{flushleft}

\section{Introduction}
\label{sec:intro}

Testing the uniformity of a sample $\bX_1,\ldots,\bX_n$ supported on the unit hypersphere $\Om{q}:=\{\bx \in \R^{q+1}: \bx' \bx=1\}$ of $\R^{q+1}$, with $q\geq1$, is one of the first steps when analysing multivariate data for which only the directions (and not the magnitudes) are of interest -- the so-called \textit{directional data}. Directional data arise in many applied disciplines, such as protein bioinformatics, environmental science, and biology; see \cite{Ley2018} for an overview of recent case studies. Due to the peculiarity of the support, a rigorous analysis of directional data requires from the consideration of adapted statistical methods, with \cite{Mardia2000} and \cite{Ley2017} being the current reference monographs on \textit{directional statistics}, and \cite{Pewsey2020} reviewing the advances on the field for the last two decades. \\

Since the second half of the 20th century, a sizeable number of tests for assessing uniformity on $\Om{q}$ have been proposed. These contributions range notably in generality (arbitrary dimension vs. specific-dimension tests; consistency against all kind of deviations vs. consistency only against certain alternatives) and underlying methodology (parametric vs. nonparametric tests); see \cite{Garcia-Portugues:review} for an updated review. In addition to its self-importance, uniformity tests on $\Om{q}$ are important auxiliary tools for, among others, the following statistical problems: (\textit{i}) testing for spherically-symmetric distributions on $\R^{q+1}$ (see, e.g., \cite{Cai2013}); (\textit{ii}) goodness-of-fit tests on $\Om{1}$ via the probability integral transform \cite[Section 6.4.2]{Mardia2000}; (\textit{iii}) goodness-of-fit tests on $\Om{q}$, $q\geq1$, via an almost-canonical transformation \citep[Proposition 1]{Jupp2020}; (\textit{iv}) testing for rotational symmetry on $\Om{q+1}$ (see, e.g., \nopagebreak[4]\cite{Garcia-Portugues:optimal}).\\

Testing uniformity on $\Om{q}$ is formalized as the testing of 
\begin{align*}
\Hcal_0: {\rm P}=\nu_q\quad\text{vs.}\quad\Hcal_1:{\rm P}\neq \nu_q,
\end{align*}
from a sample $\mathbf{X}_1,\ldots,\mathbf{X}_n$ of independent and identically distributed (iid) observations of $\bX$, where ${\rm P}$ is the distribution of $\bX$ (henceforth $\bX\sim{\rm P}$) and $\nu_q$ is the uniform distribution on $\Om{q}$. The probability density function (pdf) of $\nu_q$ with respect to the Lebesgue measure on $\Om{q}$, denoted by $\omega_q$, is $1/\omega_q(\Om{q})$. In an abuse of notation, we also refer by $\omega_q$ to the surface area of $\Om{q}$, $\omega_q(\Om{q})=2\pi^{\frac{q+1}{2}}\big/\Gamma\big(\tfrac{q+1}{2}\big)$, when there is no possible\nolinebreak[4] ambiguity.\\

The rest of this paper is organized as follows. Section \ref{sec:test} introduces the new class of uniformity tests and explores important particular cases. Section \ref{sec:Sobolev} analyses the relation of the new class with the Sobolev class and gives its asymptotics. Simulation studies are performed in Section \ref{sec:simus}, while astronomical data applications using the new tests are carried out in Section \ref{sec:realdata}. Closing remarks are given in Section \ref{sec:discussion}. The proofs are relegated to the Supplementary Material, which also contains further simulation results.

\section{Projected-ecdf statistics: A new approach to testing uniformity}
\label{sec:test}

\subsection{Genesis}
\label{sec:genesis}

Our proposal is inspired by the projection-based test of \cite{Cuesta-Albertos2009}. Their test is based on a characterization, with probability one, of the distribution of $\mathbf{X}$ by means of the one-dimensional distribution of $\bga'\bX$, where $\bga \sim \nu_q$ is a randomly-sampled direction. As a consequence of this characterization, testing $\Hcal_0$ is (almost surely) equivalent to testing $\Hcal^{\bga}_0: \bga'\bX\sim F_{q}$, where $F_{q}$ (see \eqref{eq:Fq} below) is the common cumulative distribution function (cdf) of the sample of random projections $\bga'\bX_1,\ldots,\bga'\bX_n$ under $\Hcal_0$, and $\bga \sim  \nu_q$ is independent of the sample. Denoting by $F_{n,\bga}$ to the empirical cumulative distribution function (ecdf) of the projections $\bga'\bX_1,\ldots,\bga'\bX_n$, the test rejects $\Hcal_0^{\bga}$, and consequently $\Hcal_0$, for large values of the Kolmogorov--Smirnov statistic
\begin{align}
D_{n,\bga}:=\sup_{x \in [-1,1]} |F_{n,\bga}(x)- F_{q}(x)|.\label{eq:Dn}
\end{align}

The test outcome clearly depends on the random direction $\bga$. To alleviate this, \cite{Cuesta-Albertos2009} considered $k$ random projections on the iid sample $\bga_1,\ldots,\bga_k$ and used as \textit{aggregated} test statistic $C_{n,\bga_1,\ldots,\bga_k}:=\min\{p_{\bga_1},\ldots,p_{\bga_k}\}$, where $p_{\bga_j}$ stands for the $p$-value of the test performed with the random direction $\bga_j$. The test rejects for low values of $C_{n,\bga_1,\ldots,\bga_k}$. The distribution of $C_{n,\bga_1,\ldots,\bga_k}$ under $\Hcal_0$ is unknown, but the authors approximate it by Monte Carlo (conditionally on a selection of $\bga_1,\ldots,\bga_k$).\\

Within the same approach of \cite{Cuesta-Albertos2009}, an alternative to \eqref{eq:Dn} is the well-known weighted quadratic norm by \cite{Anderson1952}:
\begin{align}
Q_{n,q,\bga}^w:=n\int_{-1}^1 \lrp{F_{n,\bga}(x)- F_{q}(x)}^2w(F_q(x))\,\mathrm{d}F_q(x),\label{eq:Qn}
\end{align}
where the specifications $w\equiv1$ and $w(u)=1/(u(1-u))$ for the weight function $w:[0,1]\to\mathbb{R}$ yield the Cram\'er--von Mises and Anderson--Darling statistics, respectively. In addition to the flexibility provided by $w$ in \eqref{eq:Qn} in comparison with \eqref{eq:Dn}, the folklore in goodness-of-fit (see, e.g., \citet[Section 5]{Stephens1974} or \citet[page 110]{DAgostino1986}) typically advises that, in practice,\nopagebreak[4] quadratic norms tend to provide higher powers than sup-norms.\\

Our class of test statistics is based on $Q_{n,q,\bga}^w$. Yet, rather than drawing several random directions and aggregating afterwards the outcomes of the tests, our statistic itself gathers information from \emph{all} the directions on $\Om{q}$: it is defined as the expectation of $Q_{n,q,\bga}^w$ with respect to $\bga\sim \nu_q$. Therefore, the proposed test rejects $\Hcal_0$ for large values\nolinebreak[4] of
\begin{align}
P_{n,q}^w:=\Es{Q_{n,q,\bga}^w}{\bga}=n
\int_{\Om{q}}\lrc{\int_{-1}^1 \lrp{F_{n,\bga}(x)- F_{q}(x)}^2w(F_q(x))\,\mathrm{d}F_q(x)}\nu_q(\mathrm{d}\bga).\label{eq:Pn}
\end{align}
The choice of $\nu_q$ as the distribution for $\bga$ is canonical: it is the only sample-independent distribution that guarantees the invariance of \eqref{eq:Pn} against any rotation of the sample, this being the fundamental property that any uniformity test on $\Om{q}$ must have. \\

Integrating the test statistic along all unit-norm directions was firstly considered, within the regression context, by \cite{Escanciano2006} in his \textit{Projected Cram\'er--von Mises} goodness-of-fit test. His test has been exported to the functional data context by \cite{Garcia-Portugues:flm} and, relevant to the consideration of \eqref{eq:Pn}, this test was compared in \cite{Cuesta-Albertos:gofflm} against a test based on the ``randomly-project-then-test-then-aggregate'' paradigm of \cite{Cuesta-Albertos2009}. Empirical results evidenced that integrating along all the directions within the statistic, as \eqref{eq:Pn} does, tends to provide superior power in practice, in spite of a much slower computation. \\

An obvious generalization of $P_{n,q}^w$ follows by substituting $w$ with the integration with respect to a (positive) $\sigma$-finite Borel measure $W$ on $[0,1]$, giving the statistic
\begin{align}\label{eq:Pn.W}
P_{n,q}^W :=n\Es{\int_{-1}^1 \lrp{F_{n,\bga}(x)- F_{q}(x)}^2 \mathrm{d}W(F_q(x))}{\bga}.
\end{align}

Notice that if $W$ is a probability measure, we integrate with respect to a probability whose cdf is $x \mapsto W\{[0,F_q(x)]\}$. This generalisation, for instance, allows for weighting schemes that are concentrated on denumerable or finite sets. We will consider the generalized formulation \eqref{eq:Pn.W} henceforth in the paper, since it unifies many tests statistics (see Sections \ref{sec:Watson}--\ref{sec:AD}) under the projection-based view. For the sake of notation simplicity, we will refer indistinctly by $W$ to a measure or to its cdf when there is no possible ambiguity. Furthermore, we will focus only on symmetric measures with respect to $1/2$, as our first result shows that, due to the construction of $P_{n,q}^W$, this restriction does not imply a loss in generality. 

\begin{proposition} \label{prop:sym}
	Let $W$ and $\tilde{W}$ be $\sigma$-finite Borel measures on $[0,1]$ such that $\tilde{W}\{(0,t)\}=\tfrac{1}{2}(W\{(0,t)\}+W\{(1-t,1)\})$, $t\in[0,1]$. Then, $P_{n,q}^W=P_{n,q}^{\tilde{W}}$.	
\end{proposition}

\subsection{Projected uniform distribution}
\label{sec:proj}

The distribution of the projection $\Pi_q:=\bga'\bX$, $\bX\sim \nu_q$ is thoroughly employed along the paper. We provide some useful properties next. Using the tangent-normal decomposition (see, e.g., \citet[page 161]{Mardia2000}), it readily follows that the distribution of $\Pi_q$ does not depend on $\bga$ and its pdf is
\begin{align*}
\frac{\omega_{q-1}}{\omega_q}(1-t^2)^{q/2-1}=\mathrm{B}\left(\tfrac{1}{2},\tfrac{q}{2}\right)^{-1}(1-t^2)^{q/2-1}, \quad t\in[-1,1],
\end{align*}
where $\mathrm{B}(a,b):=\Gamma(a)\Gamma(b)/\Gamma(a+b)$. Therefore, $\Pi_q^2\sim\mathrm{Beta}\left(\tfrac{1}{2},\tfrac{q}{2}\right)$ and the cdf of $\Pi_q$ is 
\begin{align}
F_q(x):=\mathrm{B}\left(\tfrac{1}{2},\tfrac{q}{2}\right)^{-1}\int_{-1}^x(1-t^2)^{q/2-1}\,\mathrm{d}t
=\frac{1}{2}\left\{1+\mathrm{sign}(x)\mathrm{I}_{x^2}\left(\tfrac{1}{2},\tfrac{q}{2}\right)\right\},\label{eq:Fq}
\end{align}
where $\mathrm{I}_x(a,b):=\mathrm{B}(a,b)^{-1}\int_{0}^xt^{a-1}(1-t)^{b-1}\,\mathrm{d}t$, $a,b>0$, is the regularized incomplete beta function. We also have that, for $x\in[-1,1]$, $F_{1}(x)=1-\frac{1}{\pi}\cos^{-1}(x)$ and $F_{2}(x)=\frac{1}{2}(x+1)$. Also, due to the recurrence properties of $x\mapsto I_x(a,b)$, for $q\geq3$ we have that
\begin{align}
F_{q}(x)=&\;F_{q-2}(x) + \frac{x(1-x^2)^{q/2-1}}{(q-2)\mathrm{B}\big(\tfrac{1}{2},\tfrac{q-2}{2}\big)}.\label{eq:Fqrec}
\end{align}

\subsection{\texorpdfstring{$U$-statistic form of $P_{n,q}^W$}{U-statistic form of PnqW}}
\label{sec:Ustat}

Simple computations show that
\begin{align}
P_{n,q}^W
=&\;n\int_{-1}^{1}\left\{\Es{F_{n,\bga}(x)^2}{\bga}-F_q(x)^2\right\}\,\mathrm{d}W(F_q(x))\label{eq:pcvm1}\\
=&\;\int_{-1}^{1}\bigg\{\frac{1}{n}\sum_{i\neq j}A_{ij}(x)+F_q(x)(1-nF_q(x))\bigg\}\,\mathrm{d}W(F_q(x)),\label{eq:pcvm2}
\end{align}
where \eqref{eq:pcvm1} follows from $\Es{F_{n,\bga}(x)}{\bga}=F_q(x)$ and \eqref{eq:pcvm2} from $\Es{F_{n,\bga}(x)^2}{\bga}=n^{-1}F_q(x)+n^{-2}\sum_{i\neq j}A_{ij}(x)$, where
\begin{align}
A_{ij}(x):=\int_{\Om{q}}1_{\{\bga'\bX_i\leq x,\bga'\bX_j\leq x\}}\, \mathrm{d}\nu_q(\bga).\label{eq:aij}
\end{align}
If $W$ is not a finite measure, splitting \eqref{eq:pcvm2} into two addends is not possible, as neither would be finite. This is precisely the case for the Anderson--Darling weight.\\

The term in \eqref{eq:aij} is the driver of the $P_{n,q}^W$ statistic. Geometrically, it is the proportion of the hypersphere $\Om{q}$ covered by the intersection of two hyperspherical caps centred at $\bX_i$ and $\bX_j$, respectively, and with solid angle $\theta_x:=\pi-\cos^{-1}(x)$. Evaluating $A_{ij}(x)$, for arbitrary $x\in[-1,1]$, $q\geq1$, and $i\neq j$, is not straightforward. Formulae for the area of the intersection of two hyperspherical caps are available in \citet[page 4]{Lee2014}, though those are impractical as they require univariate integrals on $F_q$ and involve 10 possible cases out of 25 (precisely, cases 1, 2, 4, 5, 6, 8, 14, 15, 23, 25) for evaluating \eqref{eq:aij}, depending on the values of $\theta_x$ and $\theta_{ij}:=\cos^{-1}(\bX_i'\bX_j)$. \\

We simplify the computation of $A_{ij}(x)$, henceforth denoted by $A(\theta_{ij},x)$ due to its dependence on $\theta_{ij}$, with the next result.

\begin{proposition}\label{prop:Aij}
	Let $x\geq 0$. Then, $A(\theta,-x)=A(\theta,x)+1-2F_q(x)$ and
	\begin{align*} 
	A(\theta,x)&=\begin{cases}
	2F_1(x)-1+ \frac{1}{\pi}\lrp{\cos^{-1}(x)-\tfrac{\theta}{2}}_+,&q=1,\\
	2\int_{-1}^{x} F_{q-1}\lrp{\frac{t\tan\lrp{\theta/2}}{(1-t^2)^{1/2}}}\,\mathrm{d}F_q(t),&q\geq2,
	\end{cases}
	\end{align*}
	where $(\cdot)_+$ is the positive part function. Alternatively, for $q\geq2$, the expression
	\begin{align}\label{eq:Aij_piecewise} 
	A(\theta,x)&=\begin{cases}
	\frac{1}{2}-\frac{\theta}{2\pi}+2\int_{0}^{x}F_{q-1}\lrp{\tfrac{t\tan\lrp{\theta/2}}{(1-t^2)^{1/2}}}\,\mathrm{d}F_q(t), & 0\leq \theta< 2\cos^{-1}(x),\\
	2F_q(x)-1, & 2\cos^{-1}(x)\leq\theta\leq\pi
	\end{cases}
	\end{align}
	does not implicitly truncate the argument of $F_{q-1}$ to the interval $[-1,1]$. 
\end{proposition}	

Form \eqref{eq:pcvm2} is not computationally pleasant. For that reason, we provide next alternative expressions for $P_{n,q}^W$ that expose its $U$-statistic nature. We do so for the fairly natural and general case in which $W$ is a positive \textit{finite} measure on $[0,1]$, standardized to a probability measure on $[0,1]$ without loss of generality. We firstly write our statistic, from \eqref{eq:pcvm2}, as
\begin{align}
P_{n,q}^W
%
%
=&\;\frac{1}{n}\sum_{i\neq j}\psi^W_q(\theta_{ij})+\int_{-1}^{1}F_q(x)(1-nF_q(x))\,\mathrm{d}W(F_q(x)),\label{eq:pcvm_aux}\\
\psi^W_q(\theta):=&\;\int_{-1}^{1}A(\theta,x)\,\mathrm{d}W(F_q(x)).\label{def:psi_theta}
\end{align}
Notice that, since $A(0,x)=F_q(x)$, an equivalent expression to \eqref{eq:pcvm_aux} is 
\begin{align}
P_{n,q}^W=&\;\frac{1}{n}\sum_{i,j=1}^n\tilde{\psi}^W_q(\theta_{ij}),\quad \tilde{\psi}^W_q(\theta):=\int_{-1}^{1}\lrp{A(\theta,x)-F_q(x)^2}\,\mathrm{d}W(F_q(x)).\label{eq:psi_tilde_theta}
\end{align}

\begin{proposition} \label{prop:psi_q:w}
	If $W$ is a cdf on $[0,1]$, then, for $\theta\in[0,\pi]$ and $q\geq 2$, 	
	\begin{align*}
	\psi^W_1(\theta)=&\;\frac{1}{2}-\frac{\theta}{2\pi}+2\int_{0}^{\theta/(2\pi)}W(u)\,\mathrm{d}u,\\
	\psi^W_q(\theta)=&-\frac{1}{2}+\frac{\theta}{2\pi}+2\int_{0}^{1/2}W(u)\,\mathrm{d}u+4\int_0^{\cos(\theta/2)}W(F_q(t))\lrp{1-F_{q-1}\lrp{\frac{t\tan\lrp{\theta/2}}{(1-t^2)^{1/2}}}}\mathrm{d}F_q(t).
	\end{align*}
\end{proposition}	

The above ideas open the way for generating new uniformity tests based on particular choices of $W$. For example, the consideration of $W_{a,b}(x):=\mathrm{I}_x(a,b)$ generates a fairly general two-parameter family of uniformity tests on $\Om{q}$, although with somehow challenging forms for $\psi^W_q$. Among the many $W$-specific instances of $P_{n,q}^W$ that could be constructed, we focus on investigating in-depth some weights delivering new tests that relate and extend previous proposals of uniformity tests.

\subsection{Extending the Watson test}
\label{sec:Watson}

One of the simplest measures that can be considered in Proposition \ref{prop:psi_q:w} is given by the cdf $W(x)=x$, $x\in[0,1]$, which is the one associated to the Cram\'er--von Mises (CvM) weight. As seen next, $P_{n,q}^W$ yields the celebrated \cite{Watson1961} test of circular uniformity when $q=1$ and connects it with the chordal-based test on $\Om{2}$ by \cite{Bakshaev2010}. Therefore, since the Watson test is regarded as the ``circular CvM'', the $P_{n,q}^W$-based test can be seen as a generalization of the former to $\Om{q}$, $q\geq2$.

\begin{proposition}[An extension of the Watson test] \label{coro:psi_q:CvM}
	Consider the CvM cdf $W(x)= x$, $x\in[0,1]$. Then,
	\begin{align*}
	P_{n,q}^\mathrm{CvM}=&\;\frac{2}{n}\sum_{i<j} \psi^\mathrm{CvM}_q(\theta_{ij})+\frac{3-2n}{6},
	\end{align*}
	where, for $\theta\in[0,\pi]$, 
	\begin{align*}
	\psi^\mathrm{CvM}_q(\theta) = \begin{cases}
	\frac{1}{2}+\frac{\theta}{2\pi}\left(\frac{\theta}{2\pi}-1\right), & q=1,\\
	\frac{1}{2}-\frac{1}{4}\sin\lrp{\frac{\theta}{2}}, & q=2,\\
	\psi_1^\mathrm{CvM}(\theta)+\frac{1}{4\pi^2}\lrp{(\pi-\theta)\tan\lrp{\frac{\theta}{2}}-2\sin^2\lrp{\frac{\theta}{2}}}, & q=3,
	\end{cases}
	\end{align*}
	and, if $q\geq 2$,
	\begin{align*}
	\psi^\mathrm{CvM}_q(\theta)=-\frac{3}{4}+\frac{\theta}{2\pi}+2F_q^2\lrp{\cos\lrp{\tfrac{\theta}{2}}}-4\int_0^{\cos\lrp{\theta/2}}F_q(t)F_{q-1}\lrp{\frac{t\tan\lrp{\theta/2}}{(1-t^2)^{1/2}}}\mathrm{d}F_q(t).
	\end{align*}
\end{proposition}

\begin{remark}\label{rem:1}
	The test based on $P_{n,1}^\mathrm{CvM}$ is equivalent to the \cite{Watson1961} test. Precisely, $P_{n,1}^\mathrm{CvM}=\frac{1}{2}U_n^2$, where $U_n^2$ is the Watson statistic defined as
	\begin{align}
	U_n^2:=&\,n\int_0^{2\pi}\lrb{G_n(\theta)-G(\theta)-\int_0^{2\pi}\lrp{G_n(\varphi)-G(\varphi)}\,\mathrm{d}G(\varphi)}^2\,\mathrm{d}G(\theta),\label{eq:Un2}
	\end{align}
	where $G_n(\theta):=(1/n)\sum_{i=1}^n1_{\lrb{\Theta_i\leq\theta}}$ is the ecdf of the circular sample $\Theta_1,\ldots,\Theta_n$ in $[0,2\pi)$, $G(\theta):=\theta/(2\pi)$ is the uniform cdf on $[0,2\pi)$, and the origin is implicitly assumed to be $0$. The statistic $U_n^2$ has several neat connections with other circular and linear uniformity tests; particularly it can be regarded as the rotation-invariant version of the CvM statistic that selects the origin in such a way that the discrepancy of the sample with respect to $\Hcal_0$ is minimized (see, e.g., \cite{Garcia-Portugues:review}). The relation of $P_{n,1}^\mathrm{CvM}$ and $U_n^2$ stems from Proposition \ref{coro:psi_q:CvM} and the following alternative form for $U_n^2$ (see, e.g., \citet[page 111]{Mardia2000}):
	\begin{align*}
	U_n^2= \frac{1}{n}\sum_{i,j=1}^n h\lrp{\theta_{ij}}=\frac{2}{n}\sum_{i<j}h\lrp{\theta_{ij}}+\frac{1}{12},\quad h(\theta)=\frac{1}{2}\lrp{\frac{\theta^2}{4 \pi^2}- \frac{\theta}{2 \pi}+\frac{1}{6}},
	\end{align*}
	where $\theta_{ij}=\cos^{-1}(\cos(\Theta_i-\Theta_j))\in[0,\pi]$ is the shortest angle distance between $\Theta_i$ and $\Theta_j$. The connection has two main implications: (\textit{i}) it introduces an interpretation of $P_{n,q}^\mathrm{CvM}$ as a natural extension of the well-known $U_n^2$ to an arbitrary dimension $q$; (\textit{ii}) it adds yet another interesting connection between $U_n^2$ and the CvM test.
\end{remark}

\begin{remark}\label{rem:3}
	\cite{Bakshaev2010} proposed the chordal-based test statistic
	\begin{align*}
	N_{n,q}:=n\E{\|\bX_1-\bX_2\|}-\frac{1}{n}\sum_{i,j=1}^n \|\bX_i-\bX_j\|
	\end{align*}
	for testing uniformity on $\Om{q}$. Observing that $\|\bX_i-\bX_j\|=\sqrt{2-2\cos(\theta_{ij})}=2\sin\lrp{\theta_{ij}/2}$ and that $\E{\|\bX_i-\bX_j\|}=\sqrt{2}\int_{-1}^1\sqrt{1-t}\,\mathrm{d}F_q(t)$ under $\Hcal_0$, when $q=2$ the statistic can be written as
	\begin{align*}
	N_{n,2}=\frac{4n}{3}-\frac{4}{n}\sum_{i<j} \sin\lrp{\tfrac{\theta_{ij}}{2}}.
	\end{align*}
	Therefore, it is evident that $P_{n,2}^\mathrm{CvM}=\frac{1}{8}N_{n,2}$.
\end{remark}

\begin{remark}\label{rem:3b}
	We highlight a perhaps intriguing behaviour of $P_{n,q}^\mathrm{CvM}$: it simultaneously yields as particular cases the Watson test for $q=1$ and the \cite{Bakshaev2010} test for $q=2$, despite the $N_{n,1}$-based test being actually \emph{different} from the Watson test. This phenomenon is explained by the dimension-dependence of $\psi_q^W$, a distinctive feature of $P_{n,q}^W$, naturally arising from its construction, that sharply contrasts with dimension-independent kernels of other uniformity test statistics, such as $N_{n,q}$. Therefore, $P_{n,q}^W$ allows the extension of circular uniformity tests in a dimension-dependent manner.
\end{remark}

\subsection{Extending the Rothman test}
\label{sec:Rothman}

We turn now to the consideration of a discrete measure $W$ that yields as particular cases the circular uniformity tests by \cite{Ajne1968} and \cite{Rothman1972} (extension of the former), and that delivers extensions of them to $\Om{q}$, $q\geq2$. \\

Within the circular setting presented in Remark \ref{rem:1}, the test statistic by \cite{Rothman1972} compares the number of expected and observed data points in arcs of $\Om{1}$ of length $2\pi t$ in a rotationally-invariant \nolinebreak[4]way:
\begin{align}
R_{n,t}:=\frac{1}{2\pi n}\int_0^{2\pi}\left(N(\alpha,t)-nt\right)^2\,\mathrm{d}\alpha,\label{eq:Roth}
\end{align}
where $N(\alpha,t):=\#\left\{\Theta_1,\ldots,\Theta_n:\cos^{-1}(\cos(\Theta_i-(\alpha+t\pi)))<t\pi, i=1,\ldots,n\right\}$ represents the number of observations in the arc $[\alpha,\alpha+2\pi t)$, for $\alpha\in[0,2\pi)$ and $t\in(0,1)$. The \cite{Ajne1968} test statistic arises as a particular case of $R_{n,t}$ with $t=1/2$:
\begin{align}
A_n:=\frac{1}{2\pi n}\int_0^{2\pi}\left(N\left(\alpha,\tfrac{1}{2}\right)-\frac{n}{2}\right)^2\,\mathrm{d}\alpha=\frac{n}{4}-\frac{1}{n\pi}\sum_{i<j}\theta_{ij}.\label{eq:Ajne}
\end{align}

The following result provides a computationally amenable form for \eqref{eq:Roth}, in the spirit of \eqref{eq:Ajne}, that is required for its comparison with \eqref{eq:pcvm_aux}.

\begin{proposition}[Computation of the Rothman test] \label{prop:comp:Roth}
	Let $t_m:=\min(t,1-t)$ for $t\in(0,1)$. The statistic \eqref{eq:Roth} can be expressed as
	\begin{align}
	R_{n,t}=t(1-t)+\frac{2}{n}\sum_{i< j}h_t\lrp{\theta_{ij}},\quad
	h_t(\theta):=\lrp{t_m-\tfrac{\theta}{2\pi}}_+-t_m^2.\label{eq:Rothcomp}
	\end{align}
\end{proposition}

\begin{proposition}[An extension of the Rothman test] \label{prop:psi_q:Roth}
	Consider the cdf 
	\begin{align*}
	W_t(x):=\frac{1}{2}\lrp{1_{\{ t_m\leq x\}}+1_{\{1-t_m\leq x\}}},
	\end{align*}
	where $t_m$ is defined in Proposition \ref{prop:comp:Roth}. Then,
	\begin{align*}
	P_{n,q}^{\mathrm{R}_t}=&\;\frac{2}{n}\sum_{i<j} \psi_{q}^{\mathrm{R}_t}(\theta_{ij})+\frac{1-n}{2}+nt(1-t),
	\end{align*}
	where, for $\theta \in [0,\pi]$ and $q=1$,
	\begin{align*}
	\psi_1^{\mathrm{R}_t}(\theta)=h_t(\theta)+\frac{1}{2}-t(1-t),
	\end{align*}
	if $q=2,3$,
	\begin{align*}
	\psi_{2}^{\mathrm{R}_t}(\theta)=&\;\begin{cases} 
	-t_m+\frac{1}{2}-\frac{1-2t_m}{\pi}\cos^{-1}\lrp{\frac{(1/2-t_m)\tan(\theta/2)}{\lrp{t_m(1-t_m)}^{1/2}}}&\\
	\quad+\frac{1}{\pi}\tan^{-1}\lrp{\frac{\lrp{\cos^2(\theta/2)-(1-2t_m)^2}^{1/2}}{\sin(\theta/2)}},& \theta<\theta_{t_m},\\
	1/2-t_m,& \theta\geq \theta_{t_m},
	\end{cases}\\
	\psi_{3}^{\mathrm{R}_t}(\theta)=&\;\begin{cases} 
	\frac{1}{2}+t_m-\frac{\theta+\theta_{t_m}}{2\pi}+\frac{1}{\pi}\left\{\frac{\sin\lrp{\theta_{t_m}}}{2}+ \tan\lrp{\frac{\theta}{2}}\cos^2\big(\tfrac{\theta_{t_m}}{2}\big)\right\},& \theta<\theta_{t_m},\\
	1/2-t_m,& \theta\geq \theta_{t_m},
	\end{cases}
	\end{align*}
	where $\theta_{t_m}:=2\cos^{-1}\lrp{F_q^{-1}(1-t_m)}$, and, if $q\geq 2$,
	\begin{align*}
	\psi_{q}^{\mathrm{R}_t}(\theta)=\begin{cases} 
	t_m-\frac{\theta}{2\pi}+2\int_0^{\cos(\theta_{t_m}/2)}F_{q-1}\lrp{\frac{u\tan\lrp{\theta/2}}{(1-u^2)^{1/2}}}\mathrm{d}F_q(u),& \theta<\theta_{t_m},\\
	1/2-t_m,& \theta\geq \theta_{t_m}.
	\end{cases}
	\end{align*}
\end{proposition}

\begin{remark}
	$P_{n,1}^{\mathrm{R}_t}$ equals the Rothman statistic, $P_{n,1}^{\mathrm{R}_t}=R_{n,t}$. Therefore, $P_{n,q}^{\mathrm{R}_t}$ can be regarded as an extension of the Rothman statistic to $\Om{q}$, $q\geq 2$. This extension is coherent with \cite{Prentice1978}'s extension of \cite{Ajne1968}'s $A_n$ to $\Om{q}$ for $q\geq2$, given by the test statistic $A_{n,q}:=n/4-1/(n\pi)\sum_{i<j}\theta_{ij}$. Indeed, considering $t=1/2$, Lemma \ref{lemma:aux} in the Supplementary Material shows that the kernel of $P_{n,q}^{\mathrm{R}_{1/2}}$ is $\psi_q(\theta)=1/2-\theta/(2\pi)$, $q\geq 1$, and therefore the tests based on $P_{n,q}^{\mathrm{R}_{1/2}}$ and $A_{n,q}$ are equivalent.
\end{remark}

\begin{remark}\label{rem:6}
	$P_{n,q}^{\mathrm{R}_t}$ only depends on $t_m$, hence $P_{n,q}^{\mathrm{R}_t}=P_{n,q}^{\mathrm{R}_{1-t}}$, $t\in(0,1)$. Also, since $F_{n,\bga}(F^{-1}(1-t_m))=F_{n,\bga}(-F^{-1}(t_m))=1-F_{n,-\bga}(F^{-1}(t_m)^-)$, $P_{n,q}^{\mathrm{R}_{t}}$ can be expressed as 
	\begin{align*}
	P_{n,q}^{\mathrm{R}_{t}}%
	=&\;\frac{1}{n}\int_{\Om{q}}\lrp{N_q\lrp{\bga,\cos^{-1}(F_q^{-1}(1-t_m))}-nt_m}^2\,\nu_q(\mathrm{d}\bga),
	\end{align*}
	where $N_q(\bga,\theta):=\#\{\bX_1,\ldots,\bX_n:\bX_i\in C_q(\bga,\theta)\}$ and $C_q(\bga,\theta)$ is the hyperspherical cap centred at $\bga$ and with solid angle $\theta\in[0,\pi]$. Therefore, geometrically $P_{n,q}^{\mathrm{R}_{t}}$ is comparing the number of observed and expected points in $C_q\lrp{\bga,\cos^{-1}(F_q^{-1}(1-t_m))}$ under $\Hcal_0$, for all $\bga\in\Om{q}$, hence its connection with \eqref{eq:Roth}.
\end{remark}

\subsection{An Anderson--Darling-like test}
\label{sec:AD}

We introduce now a new test based on the Anderson--Darling weight. Contrary to the Kolmogorov--Smirnov and CvM tests, this is the first adaptation of the celebrated \cite{Anderson1954} test to deal with directional data, despite being one of the three most well-known ecdf-based goodness-of-fit tests.

\begin{proposition}[An Anderson--Darling test] \label{prop:psi_q:AD}
	If $w(u)=1/(u(1-u))$. Then,
	\begin{align*}
	P_{n,q}^{\mathrm{AD}}=\frac{2}{n}\sum_{i<j} \psi_q^{\mathrm{AD}}(\theta_{ij})+n,
	\end{align*}
	where, for $\theta\in(0,\pi]$,
	\begin{align*}
	\psi_q^{\mathrm{AD}}(\theta) = \begin{cases}
	-2\log\lrp{2\pi}+\frac{1}{\pi}\lrb{\theta\log(\theta)+(2\pi-\theta)\log(2\pi-\theta)}, & q=1,
	\\
	-\log(4)+\frac{2}{\pi}\int_0^{\cos\lrp{\theta/2}}\log\lrp{\frac{1+t}{1-t}} \cos^{-1}\lrp{\frac{t\tan\lrp{\theta/2}}{(1-t^2)^{1/2}}}\mathrm{d}t, & q=2,
	\\
	-2\log\lrp{2\pi}+\frac{1}{\pi}\big\{s(\theta)\log(s(\theta))+(2\pi-s(\theta))\log(2\pi-s(\theta))\big\}\\
	\quad-\frac{4}{\pi}\tan\lrp{\tfrac{\theta}{2}}\int_0^{\cos(\theta/2)}t\log\lrp{\frac{\pi}{\cos^{-1}(t)-t(1-t^2)^{1/2}}-1}\,\mathrm{d}t, & q=3,
	\end{cases}
	\end{align*}
	with $s(\theta):=\theta-\sin(\theta)$, and, if $q\geq 2$,
	\begin{align}
	\psi_q^{\mathrm{AD}}(\theta)=&-\log(4)\nonumber\\
	&+4\int_{0}^{\cos\lrp{\theta/2}}\log\lrp{\frac{F_{q}(t)}{1-F_{q}(t)}}\lrp{1-F_{q-1}\lrp{\frac{t\tan\lrp{\theta/2}}{(1-t^2)^{1/2}}}}\mathrm{d}F_{q}(t).\label{eq:psi:q2}
	\end{align}
\end{proposition}

\begin{remark}
	Because $\lim_{\theta\to0^+}\psi_q^{\mathrm{AD}}(\theta)=0$, by continuity, $\psi_q^{\mathrm{AD}}(0):=0$.
\end{remark}

\begin{remark}
	$P_{n,q}^\mathrm{AD}$ does \emph{not} emerge by taking $\mathrm{d}W(G(\cdot))$ instead of $\mathrm{d}G(\cdot)$ in \eqref{eq:Un2}.
\end{remark}

We conclude this section by summarising in Table \ref{tab:newtests} the new test statistics and by showing in Figure \ref{fig:psiq} the kernel functions of $P_{n,q}^\mathrm{CvM}$, $P_{n,q}^{\mathrm{R}_t}$, and $P_{n,q}^\mathrm{AD}$. 

\begin{table}[h]
	\centering
	\small
	\begin{tabular}{lll}
		\toprule
		Test & $\Om{1}$- or $\Om{2}$-specific test & $\Omega_q$-test, $q\geq 1$ \\
		\midrule
		Watson & $U_n^2$ for $\Om{1}$ \citep{Watson1961} & $P_{n,q}^\mathrm{CvM}$ \\
		Ajne & $A_n$  for $\Om{1}$ \citep{Ajne1968} & $P_{n,q}^{\mathrm{R}_{1/2}}$ \, \footnotemark \\
		Rothman & $R_{n,t}$ for $\Om{1}$ \citep{Rothman1972} & $P_{n,q}^{\mathrm{R}_t}$ \\
		Chordal-based & $N_{n,2}$ for $\Om{2}$ \citep{Bakshaev2010} & $P_{n,q}^{\mathrm{CvM}}$\\
		Anderson--Darling & --- & $P_{n,q}^{\mathrm{AD}}$ \\
		\bottomrule
	\end{tabular}
	\caption{\label{tab:newtests}\small Extensions and connections given by the new family of projected tests.}
\end{table}

\footnotetext{$P_{n,q}^{\mathrm{R}_{1/2}}$ coincides with \cite{Prentice1978}'s extension of $A_n$ to $\Om{q}$.}

\begin{figure}[!h]
	\centering
	\includegraphics[width=0.33\textwidth]{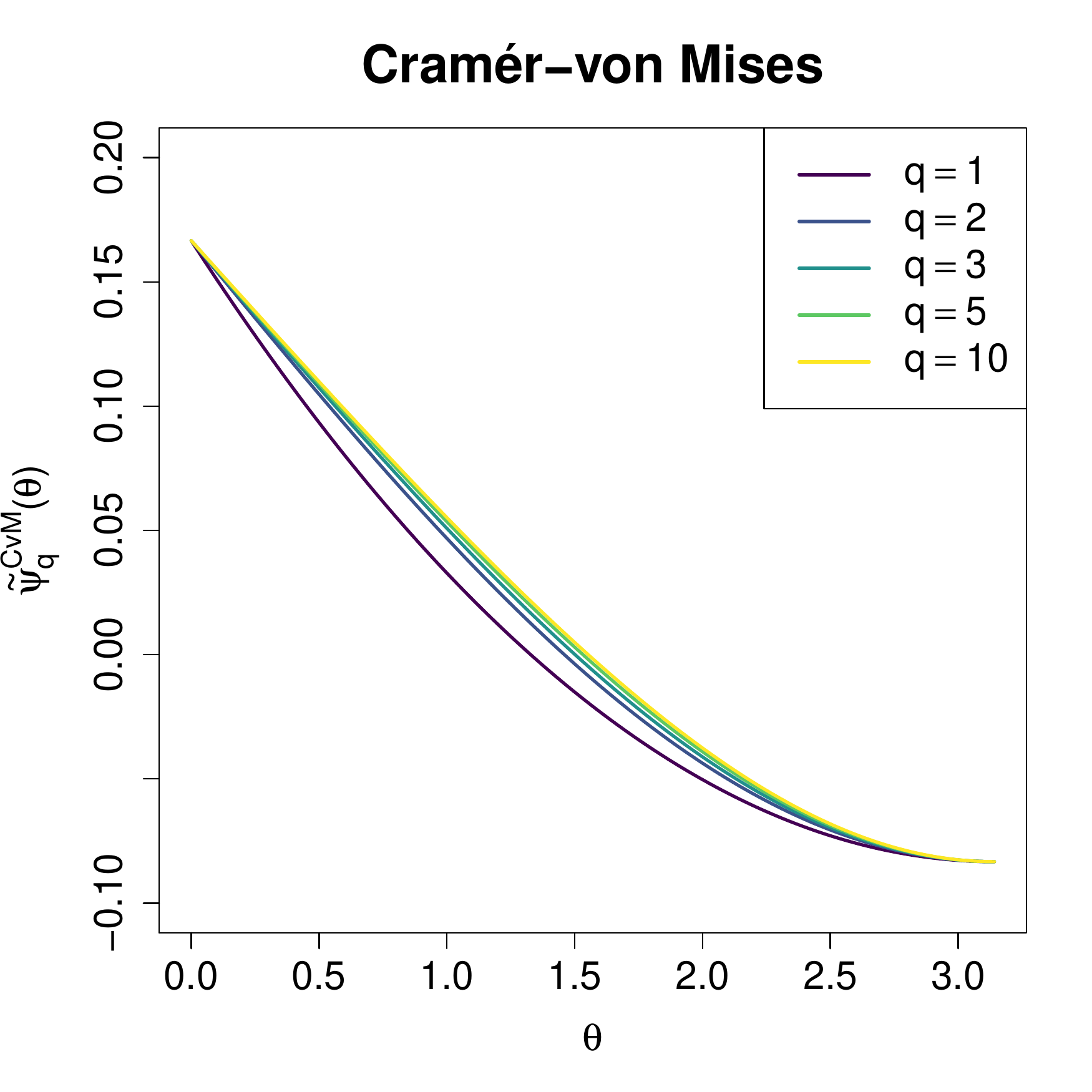}\includegraphics[width=0.33\textwidth]{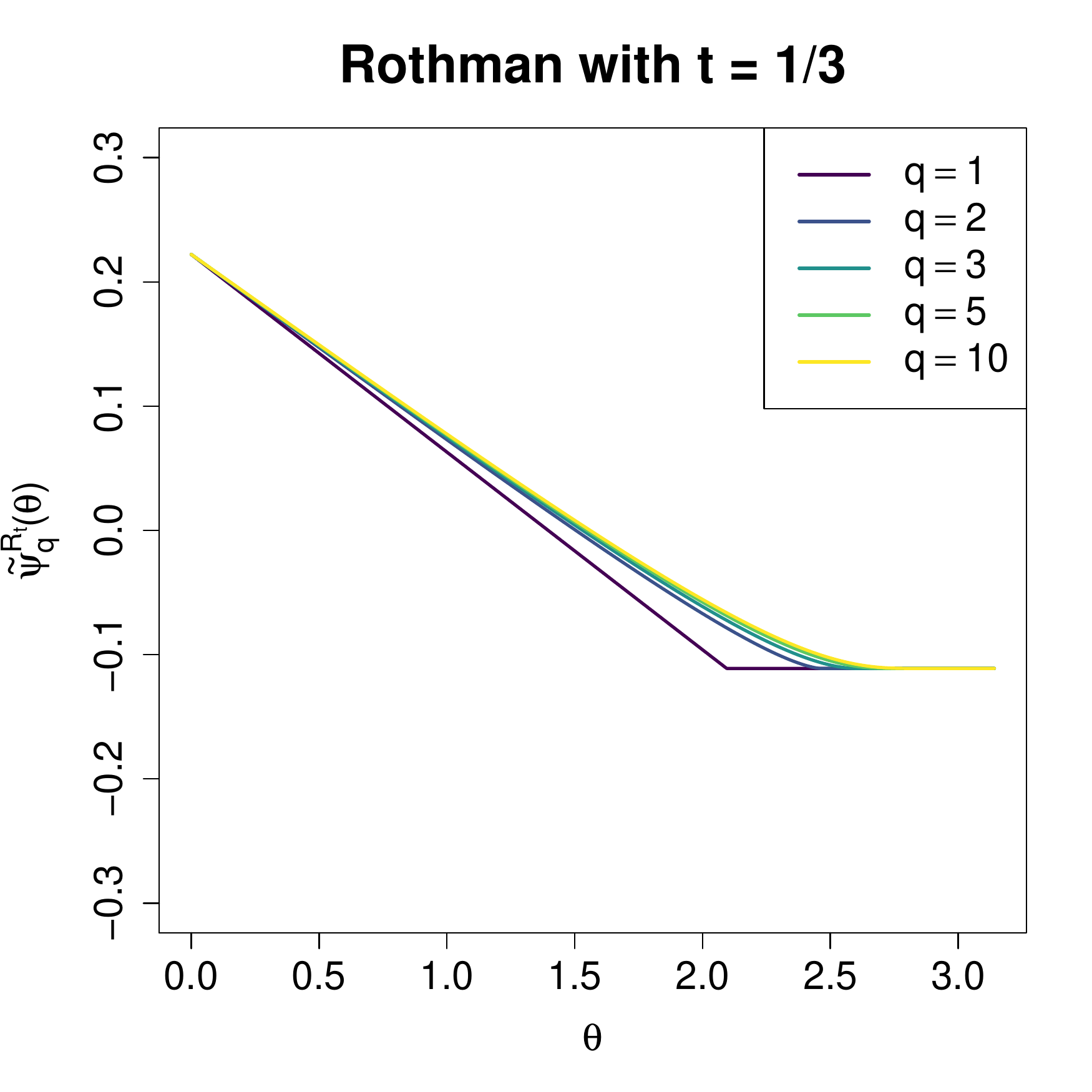}\includegraphics[width=0.33\textwidth]{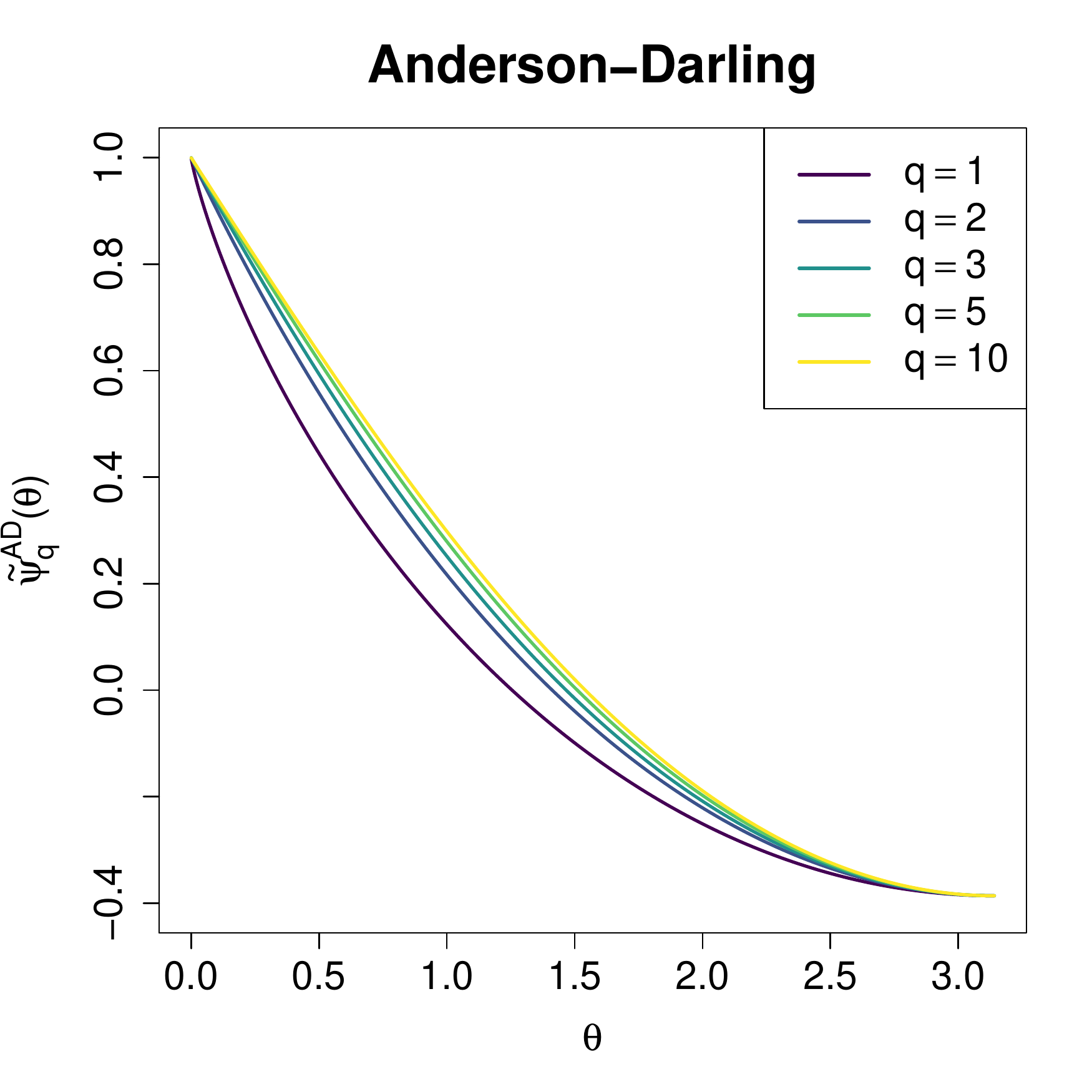}
	\caption{\small From left to right, depiction of $\tilde{\psi}_q^{\mathrm{CvM}}(\theta)=\psi_q^{\mathrm{CvM}}(\theta)-1/3$, $\tilde{\psi}_q^{\mathrm{R}_t}(\theta)=\psi_q^{\mathrm{R}_t}(\theta)+ t_m(1 - t_m)-1/2$, and $\tilde{\psi}_q^{\mathrm{AD}}(\theta)=\psi_q^{\mathrm{AD}}(\theta)+1$, for $q=1,2,3,5,10$ and $t=1/3$. The formulation in \eqref{eq:psi_tilde_theta} is considered to achieve a standardization of the three kernel functions. Recall the difference in vertical scales, indicating the larger variability of $P_{n,q}^{\mathrm{R}_t}$ and $P_{n,q}^\mathrm{AD}$ with respect to $P_{n,q}^\mathrm{CvM}$. \label{fig:psiq}}
\end{figure}

\section{Connection with the Sobolev class}
\label{sec:Sobolev}

\subsection{Sobolev tests of uniformity}
\label{sec:Sobolev_review}

Section \ref{sec:equiv} will show that our class of tests is closely related with the Sobolev class. For that reason, we review next the machinery and results of Sobolev tests later required.\\

Sobolev tests \citep{Beran1968,Gine1975} are based on the eigenfunctions of the Laplacian on $\Om{q}$. 
Intuitively, they inspect the projections of $f\in L^2(\Om{q},\nu_q)$, a pdf on $\Om{q}$ with respect to $\nu_q$, into the sequence of spaces $\{\mathcal{E}_k\}$, where $\mathcal{E}_k$ is the space spanned by the eigenfunctions of the $k$-th non-zero eigenvalue and has dimension  
\begin{align}\label{Eq:d_q,k}
d_{k,q}:=\binom{q+k-2}{q-1}+\binom{q+k-1}{q-1}.
\end{align}

Under $\Hcal_0$, the norm of the projection of $f$ on each $\mathcal{E}_k$ is null. Sobolev statistics incorporate these norms in a sum weighted by a real sequence $\{v_{k,q}\}$ such that $\sum_{k=1}^\infty v_{k,q}^2 d_{k,q}<\allowbreak\infty$. The Sobolev test for $\{v_{k,q}\}$ rejects $\Hcal_0$ for large values of the statistic
\begin{align} 
S_{n,q}(\{v_{k,q}\}):=
\frac{1}{n} \sum_{i,j=1}^n\sum_{k=1}^\infty v_{k,q}^2\langle \mathrm{t}_k(\bX_i),  \mathrm{t}_k(\bX_j) \rangle,\label{eq:sob}
\end{align}
where the addends of \eqref{eq:sob} are given explicitly \citep[Proposition 2.1]{Prentice1978} as
\begin{align} 
\langle\mathrm{t}_k(\bu), \mathrm{t}_k(\bv) \rangle= \begin{cases}
2 T_k(\cos^{-1}(\bu' \bv)), & q=1, \\ 
\big(1+ \frac{2k}{q-1}\big) C_k^{(q-1)/2} (\bu' \bv), & q\geq2,
\end{cases}\quad {\bu},{\bv}\in\Om{q},\label{eq:tkuv}
\end{align}
and where ${\rm t}_k\in\mathcal{E}_k$, $k\geq1$, are certain fixed functions; see \cite{Garcia-Portugues:review} for specifics. In \eqref{eq:tkuv}, $T_k$ represents the $k$-th Chebyshev polynomial of the first kind and $C_k^{(q-1)/2}$ stands for the $k$-th Gegenbauer polynomial of order $(q-1)/2$. \\

Gegenbauer polynomials, thoroughly employed henceforth, form an orthogonal basis, for $q\geq2$, on $L_q^2[-1,1]$, our notation for the space of square-integrable real functions in $[-1,1]$ with respect to 
$z\mapsto (1-z^2)^{q/2-1}$ and $q\geq1$. Therefore, they satisfy, for $k\geq0$,
\begin{align*}
\int_{-1}^1 C_k^{(q-1)/2}(z)C_\ell^{(q-1)/2}(z)(1-z^2)^{q/2-1}\,\mathrm{d}z=\delta_{k\ell}c_{k,q},\; c_{k,q}:=\frac{2^{3-q}\pi\Gamma(q+k-1)}{(q+2k-1)k!\Gamma((q-1)/2)^2},
\end{align*}
and hence any function $g\in L_q^2[-1,1]$, $q\geq 2$, can be uniquely expressed as 
\begin{align*}
g(z)=\sum_{k=0}^\infty b_{k,q}C_k^{(q-1)/2}(z), \quad b_{k,q}=\frac{1}{c_{k,q}}\int_{-1}^1 g(z)C_k^{(q-1)/2}(z)(1-z^2)^{q/2-1}\,\mathrm{d}z.
\end{align*}

Chebyshev polynomials, expressible as $T_k(\cos\theta)=\cos(k \theta)$ for $\theta\in[0,\pi]$, form an orthonormal basis on $L_1^2[-1,1]$ with normalizing constants $c_{k,1}=(1+\delta_{k0})\pi/2$, $k\geq0$. Since from \citet[equation 18.7.25]{NIST:DLMF}
\begin{align}
\lim_{\alpha\to0^+}\frac{1}{\alpha}C_k^\alpha(z)=\frac{2}{k}T_k(z)\;\;\text{for}\;\; k\geq1,   \label{eq:Ck0}
\end{align}
the case $q=1$ in \eqref{eq:tkuv} is a continuous extension of the $q\geq2$ case. For the sake of notation brevity, we henceforth write $\langle\mathrm{t}_k(\bu), \mathrm{t}_k(\bv) \rangle=\big(1+ \frac{2k}{q-1}\big) C_k^{(q-1)/2} (\bu' \bv)$ for $q\geq1$.\\

Alternatively, Sobolev tests are constructed as the locally most powerful rotation-invariant tests for testing $\Hcal_0$ against $f$-specified alternatives $f(\cdot'\bmu)$, $\bmu\in\Om{q}$ \citep{Beran1968,Gine1975}. The construction relies on an application of the Neyman--Pearson lemma and a locality argument assuming $f\approx1$ ($f(\cdot'\bmu)$ is a pdf with respect to $\nu_q$). This gives an equivalent expression for the test statistic \eqref{eq:sob}: 
\begin{align}
S_{n,q}(\{v_{k,q}\})=\frac{1}{n}\int_{\Om{q}}\Big(\sum_{i=1}^nf(\bX_i'\bsigma)-n\Big)^2\,\nu_q(\mathrm{d}\bsigma),\label{eq:fint}
\end{align}
where 
\begin{align}
f(z):=1+\sum_{k=1}^\infty \left(1+\tfrac{2k}{q-1}\right) v_{k,q}C_k^{(q-1)/2}(z),\quad z\in[-1,1]. \label{eq:lmpfp2}
\end{align}

Proposition 2.1 in \cite{Prentice1978} states the existence of $g_f\in L_q^2[-1,1]$ such that 
\begin{align}
g_f(z)=\sum_{k=1}^\infty b_{k,q}C_k^{(q-1)/2}(z), \quad b_{k,q}=(1+2k/(q-1))v_{k,q}^2,\quad k\geq 1,\label{eq:gf}
\end{align}
and such that 
\begin{align}\label{Eq.h_Sobolev}
S_{n,q}(\{v_{k,q}\})= \frac 1 n \sum_{i,j=1}^n g_f \left(\bX_i'\bX_j\right).
\end{align}
When $q=1$, the continuity extension \eqref{eq:Ck0} is assumed in \eqref{eq:lmpfp2} and \eqref{eq:gf},  resulting $f(z)=1+\sum_{k=1}^\infty (2b_{k,1})^{1/2}T_k(z)$ and $g_f(z)=\sum_{k=1}^\infty b_{k,1}T_k(z)$ with $b_{k,1}=2v_{k,1}^2$, $k\geq 1$.

\setcounter{subsection}{4}
\begin{remark}\label{rem:8}
	Equation \eqref{Eq.h_Sobolev} shows that it is rather $\{v_{k,q}^2\}$ who parametrizes $S_{n,q}(\{v_{k,q}\})$. Moreover, $f$ immediately provides $\{v_{k,q}^2\}$, hence $S_{n,q}(\{v_{k,q}\})$. Because of this, we will use the notation $S_{n,q}(f)$ to refer to $S_{n,q}(\{v_{k,q}\})$ when indexing by $f \in \mathcal{F}_q$, where 
	\begin{align*}
	\mathcal{F}_q:= \left\{ f\in L_q^2[-1,1]: f(z)=1+\sum_{k=1}^\infty \lrp{1+\tfrac{2k}{q-1}}v_{k,q}C_k^{(q-1)/2}(z),\,\sum_{k=1}^\infty v_{k,q}^2d_{k,q}<\infty \right\}.
	\end{align*}
	Note that two different functions $f_1,f_2\in\mathcal F_q$ such that $v_{k,q,1}=\pm v_{k,q,2}$, $k\geq1$, determine the same squared coefficients, hence yielding the same statistic.  
\end{remark}

For the sake of reference, we collect in the following theorem the main results on the tests based on \eqref{eq:sob} and \eqref{eq:fint}, as stated in \cite{Gine1975} and \cite{Prentice1978}.
\begin{theorem}[\cite{Gine1975}, \cite{Prentice1978}]\label{theo:Gine}
	Let $\{v_{k,q}\}$ satisfy 
	$\sum_{k=1}^\infty v_{k,q}^2d_{k,q}<\infty$. Let $Y_k\sim \chi^2_{d_{k,q}}$, $k\geq1$, be independent random variables. Then, under $\Hcal_0$, 
	\begin{align*}
	S_{n,q}(\{v_{k,q}\})\inlaw\sum_{k=1}^\infty v_{k,q}^2Y_k.
	\end{align*} 
	In addition, the test that rejects $\Hcal_0$ for large values of $S_{n,q}(\{v_{k,q}\})$ is asymptotically and locally (in $\kappa\to0$) most powerful rotation-invariant (except $O(\kappa^3)$ terms) against the alternative with pdf
	\begin{align}
	f_{\boldsymbol{\mu},\kappa}(\mathbf{x}):=(1-\kappa)\frac{1}{\omega_q}+\kappa \frac{f(\mathbf{x}'\boldsymbol{\mu})}{\omega_q}, \quad 0< \kappa \leq 1,\label{eq:lmpfp}
	\end{align}
	where $\bmu\in\Om{q}$ is unspecified and $f$ is given by \eqref{eq:lmpfp2}.
	Furthermore, if $v_{k,q}\neq0$, for all $k\geq1$, the test is consistent against all non-uniform alternatives with pdf in $L^2(\Om{q},\nu_q)$.
\end{theorem}

\begin{remark}
	By Remark \ref{rem:8}, there might be several alternatives \eqref{eq:lmpfp} against whom the $S_{n,q}(\{v_{k,q}\})$-test enjoys local optimality if differently-signed $\{v_{k,q}\}$ produce bona fide pdfs $f(\cdot'\boldsymbol\mu)$ on $\Om{q}$. This is guaranteed as long as the $f$ obtained in \eqref{eq:lmpfp2} is non-negative. A simple instance of differently-signed sequences is $v_{2k+1,q,1}=-v_{2k+1,q,2}$, whence $f_1(z)=f_2(-z)$, $z\in[-1,1]$, producing a sign flip in the unspecified $\bmu$ in \eqref{eq:lmpfp}. 
\end{remark}

We provide a precise definition of the Sobolev class, required for the next section. 

\begin{definition}[Sobolev class] \label{Def.Sobolev}
	The \emph{Sobolev class} of test statistics is defined as $\mathcal{S} :=    \big\{S_{n,q}(\{v_{k,q}\}) : \{v_{k,q}\}\subset\mathbb{R}, \sum_{k=1}^\infty v_{k,q}^2d_{k,q}<\infty\big\} = \left\{S_{n,q}(f) : f\in\mathcal{F}_q\right\}$. The \emph{$\ell$-finite Sobolev class} is defined as $\mathcal{S}_\ell:=\{S_{n,q}(\{v_{k,q}\})\in\mathcal{S} : \{v_{k,q}\}\text{ has at most $\ell$ non-null terms}\}$. 
\end{definition}

Clearly, $\mathcal{S}_\ell\subset\mathcal{S}$, $\forall\ell\geq1$. The $\ell$-finite Sobolev class has been studied in \cite{Jupp2008} and \cite{JMV19}.

\subsection{Relation between projected-ecdf and Sobolev classes}
\label{sec:equiv}

Despite being rooted on a different motivation, the class of projected-ecdf statistics is  
related with the Sobolev class. In virtue of \eqref{eq:psi_tilde_theta} and \eqref{Eq.h_Sobolev}, both types of statistics have a $U$-statistic form with kernels acting on the sample's shortest angles. Thus, an important issue to address is the existence of a bijection between both classes: \hypertarget{i}{(\textit{i})} for any projected-ecdf statistic with measure $W$, is there an $f^W \in \mathcal{F}_q$ such that the statistic is expressible as a Sobolev one with $g_{f^W}(z)= \tilde{\psi}^W_q(\cos^{-1}(z))$ in \eqref{Eq.h_Sobolev}?; \hypertarget{ii}{(\textit{ii})} for any Sobolev statistic with $f\in\mathcal{F}_q$, is there a measure $W_f$ such that the statistic is expressible as a projected-ecdf one with $\tilde{\psi}^{W_f}_q(\theta)=g_f(\cos(\theta))$ in \eqref{eq:psi_tilde_theta}?\\

To elucidate these queries, we define next several classes of \textit{projected-ecdf} test statistics of varying generality. We exclude the weights with $W(\{0\}) >0$ due to its lack of statistical interest and related technical difficulties. 

\begin{definition}[Projected-ecdf classes] \label{Def.Projected}
	The \emph{projected-ecdf classes} of statistics (\textit{i}) $\mathcal{P}_{+}$,  (\textit{ii}) $\mathcal{P}_{\sigma+}$, and (\textit{iii}) $\mathcal{P}_{\pm}$ are defined as the collection of statistics 
	$P_{n,q}^W$ indexed by $W$, a measure on $[0,1]$, with $W(\{0\})=0$, such that, for each class,  (\textit{i}) $W$ is a probability, (\textit{ii}) $W$ is positive $\sigma$-finite, (\textit{iii}) $W$ is signed, finite, and absolutely continuous with respect to the Lebesgue measure.
\end{definition}

\begin{remark}
	Obviously, $\mathcal{P}_{+}\subset \mathcal{P}_{\sigma+}$. Also,  $P_{n,q}^{\mathrm{CvM}},P_{n,q}^{\mathrm{R}_t}\in \mathcal{P}_{+}$ and $P_{n,q}^{\mathrm{AD}}\in \mathcal{P}_{\sigma+}$. 
\end{remark}

The next theorem answers \hyperlink{i}{(\textit{i})} and \hyperlink{ii}{(\textit{ii})} above in light of the different projected-ecdf classes. Its first statement concludes that ``sensible'' projected-ecdf statistics (associated to probabilities $W$), as well as the Anderson--Darling statistic, are indeed Sobolev statistics. The second shows that finite Sobolev tests are projected-ecdf statistics with absolutely continuous, potentially signed, measures~$W$. 

\begin{theorem}[Projected-ecdf and Sobolev classes relations] \label{theo:Proj_Sobolev}
	It occurs:
	\begin{enumerate}[label=\textit{\roman{*}}., ref=\textit{\roman{*}}]
		\item $\mathcal{P}_{+}\subset \mathcal{S}$ and $P_{n,q}^{\mathrm{AD}}\in\mathcal{S}$. \label{theo:Proj_Sobolev:1}
		\item $\mathcal{S}_\ell\subset \mathcal{P}_{\pm}$, for all $\ell\geq1$. \label{theo:Proj_Sobolev:2}
	\end{enumerate}
\end{theorem}

\begin{remark}
	The contention $\mathcal{P}_{+}\subset \mathcal{S}$ is strict since $P_{n,q}^{\mathrm{AD}}\in\mathcal{S}$ while $P_{n,q}^{\mathrm{AD}}\notin \mathcal{P}_{+}$. 
\end{remark}

\begin{remark}
	The proof of $\mathcal{P}_{+}\subset \mathcal{S}$ does not guarantee the non-negativeness of 
	$\tilde{f}\in\mathcal{F}_q$ such that $P_{n,q}^{W}=S_{n,q}(\tilde{f})$, unless $W$ is a Dirac's delta. This is allowed, as $\mathcal{F}_q$ contains non-positive functions, but obscures the local optimality view of $S_{n,q}(\tilde{f})$. Of course, when $\kappa\to 0$, \eqref{eq:lmpfp} is always well-defined as a pdf, even if $f$ is non-positive. 
\end{remark}

\begin{remark}
	The next counterexample shows that $\mathcal{P}_{\pm} \not\subset\mathcal{S}$. If $q=1$ and $W(x)=\cos(4\pi x)$, then, by Proposition \ref{prop:psi_q:w}, $\psi_1(\theta)=1-\lrp{\cos(\theta)\sin(\theta)}/(2\pi)$ and its Gegenbauer coefficients are $b_{k,1}=((-1)^k-1)/(\pi ^2 (4-k^2))\not\geq0$, hence it can not be written as \eqref{Eq.h_Sobolev}.
\end{remark}

\subsection{Asymptotic null distributions and local optimality}
\label{sec:asymp}

In virtue of Theorem \ref{theo:Proj_Sobolev}, all the projected-ecdf statistics within the class $\mathcal{P}_+$ belong to the Sobolev class. Therefore, the asymptotic distribution and local optimality results stated in Theorem \ref{theo:Gine} readily apply to the class of statistics $\mathcal{P}_+\cup \{P_{n,q}^\mathrm{AD}\}$. This is collected in the next corollary, which follows directly from Theorems \ref{theo:Gine} and \ref{theo:Proj_Sobolev} by noting that the Gegenbauer coefficients of $\psi_q^W$ equal $(1+2k/(q-1))v_{k,q}$ according to \eqref{eq:gf}. 

\begin{corollary}[Asymptotic null distribution and local optimality]\label{coro:asymp}
	Let $W$ be a probability in $[0,1]$ or the Anderson--Darling measure. For $q\geq1$, the Gegenbauer and Chebyshev coefficients of $\psi_q^W$, defined as 
	\begin{align}\label{eq:bkW}
	b_{k,q}^W:=\begin{cases}
	\frac{1}{c_{k,1}}\int_0^\pi\psi_q^W(\theta)T_k(\cos\theta)\,\mathrm{d}\theta,& q= 1,\\
	\frac{1}{c_{k,q}}\int_0^\pi\psi_q^W(\theta)C_k^{(q-1)/2}(\cos\theta)\sin^{q-1}(\theta)\,\mathrm{d}\theta,& q\geq 2
	\end{cases}
	\end{align}
	for $k\geq 0$, are non-negative sequences satisfying $\sum_{k=1}^\infty b_{k,q}^W d_{k,q}<\infty$. Under $\Hcal_0$, 
	\begin{align}
	P_{n,q}^W\inlaw\begin{cases}
	\sum_{k=1}^\infty 2^{-1} b_{k,1}^W Y_k,&q = 1,\\
	\sum_{k=1}^\infty \lrp{1+\frac{2k}{q-1}}^{-1} b_{k,q}^W Y_k,& q\geq 2,
	\end{cases}
	\label{eq:sumchis}
	\end{align}
	where $Y_k\sim \chi^2_{d_{k,q}}$, $k\geq1$, are independent random variables. In addition, the test that rejects $\Hcal_0$ for large values of $P_{n,q}^W$ is asymptotically and locally (in $\kappa\to0$) most powerful rotation-invariant (except $O(\kappa^3)$ terms) against any pdf \eqref{eq:lmpfp} based on $|v_{k,q}|=(b_{k,q}^W)^{1/2}$. 
	Furthermore, if $b_{k,q}^W>0$, for all $k\geq1$, the test is consistent against all non-uniform alternatives with pdf in $L^2(\Om{q},\nu_q)$. 
\end{corollary}

The asymptotic distribution and local optimality of the $(\mathcal{P}_+\cup \{P_{n,q}^\mathrm{AD}\})$-based tests are governed by \eqref{eq:bkW}. The following results are aimed to facilitate these coefficients. 

\begin{theorem}\label{theo:coef_A}
	For $x\in [-1,1]$ and $\theta \in [0,\pi]$, consider $A(\theta,x)$ defined as in \eqref{eq:aij}. Then, for $q\geq1$,
	\begin{align*}
	A(\theta,x)=\sum_{k=0}^\infty a_{k,q}^x C_k^{(q-1)/2}(\cos\theta),
	\end{align*}
	where $a_{0,q}^x:=F_q(x)^2$ and, for $k\geq1$,
	\begin{align*}
	a_{k,q}^x:= \begin{cases}
	\frac{1}{k^2\pi^2}\lrp{1-T_{2k}(x)},&q=1,\\
	\lrp{1+\frac{2k}{q-1}}\lrp{\frac{2^{q-1}\Gamma\lrp{(q+1)/2}^2\Gamma(k)}{\pi\Gamma(k+q)}}^2(1-x^2)^q\lrp{C_{k-1}^{(q+1)/2}(x)}^2,&q\geq2.
	\end{cases}
	\end{align*}
\end{theorem}

\begin{corollary}[Gegenbauer coefficients of $\psi_q^W$]\label{coro:coef_psi} 
	Let $P_{n,q}^W\in\mathcal{P}_{\sigma+}\cup\mathcal{P}_{\pm}$. Then the Gegenbauer coefficients of $\psi_q^W$ are $b^W_{k,q}=\int_{-1}^1a_{k,q}^x\,\mathrm{d}W(F_q(x))$, for $q\geq 1$ and $k\geq 0$. 
\end{corollary}

The well-known Rayleigh \citep{Rayleigh1919} and Bingham \citep{Bingham1974} statistics belong to $\mathcal{S}_1$, the former with $v_{1,q}=\delta_{k1}$ and the latter with $v_{2,q}=\delta_{k2}$. The next result, consequence of Theorem \ref{theo:Proj_Sobolev} and Corollary \ref{coro:coef_psi}, identifies, in the circular case, the statistic in $\mathcal{P}_{\pm}$ equating these and any other statistic in $\mathcal{S}_1$. It also highlights that any $\mathcal{P}_{\pm}$-statistic is decomposable into a weighted difference of two $\mathcal{P}_{+}$-statistics.

\begin{corollary}\label{coro:w1}
	For any circular statistic $S_{n,1}(\{v_{k,1}\})\in\mathcal{S}_\ell$, with $\ell\geq1$, the weight
	\begin{align*}
	w({\{v_{k,1}\}})(x):=\sum_{k=1}^\ell v_{k,1}^2 w_k(x),\quad w_k(x):=-2\pi^2 k^2\cos(2k x \pi),
	\end{align*}
	generates  
	$P_{n,1}^{W({\{v_{k,1}\}})}\in \mathcal{P}_{\pm}$ such that $S_{n,1}(\{v_{k,1}\})=P_{n,1}^{W({\{v_{k,1}\}})}$. 	Also, there exist 
	$P_{n,1}^{W^+({\{v_{k,1}\}})},\allowbreak P_{n,1}^{W^-({\{v_{k,1}\}})}\in \mathcal{P}_{+}$ and $a^+,a^-\geq0$ such that 
	$P_{n,1}^{W({\{v_{k,1}\}})}=a^+P_{n,1}^{W^+({\{v_{k,1}\}})}-a^-P_{n,1}^{W^-({\{v_{k,1}\}})}$.
\end{corollary}

The following results provide relatively explicit control on the Gegenbauer coefficients of $\psi_q^{\mathrm{R}_t}$, $\psi_q^{\mathrm{CvM}}$, and $\psi_q^{\mathrm{AD}}$. The first is a direct consequence of Theorem \ref{theo:coef_A}, whereas the latter involve direct computations on the kernel functions. General closed-form expressions for $b_{k,q}^{\mathrm{AD}}$ are highly challenging to obtain, except for $q=1,2$.

\begin{corollary}[Gegenbauer coefficients for $\psi_q^{\mathrm{R}_t}$]\label{coro:Gegen_Rt}
	Let $t_m$ defined as in Proposition \ref{prop:comp:Roth} and $a_{k,q}^x$ as in Theorem \ref{theo:coef_A}. The Gegenbauer coefficients of $\psi_q^{\mathrm{R}_t}$, $q\geq1$, are
	\begin{align*}	
	b^{\mathrm{R}_t}_{k,q}=\begin{cases}
	\frac{2}{k^2\pi^2} \sin^2\lrp{k\pi t_m},&q=1,\\
	a_{k,q}^{F_q^{-1}(t_m)},
	&q\geq 2,
	\end{cases}
	\quad
	b^{\mathrm{R}_t}_{0,q}=	\frac{1}{2}-t_m(1-t_m), \quad q\geq 1.
	\end{align*}
\end{corollary}

\begin{proposition}[Gegenbauer coefficients for $\psi_q^{\mathrm{CvM}}$]
	\label{prop:Gegen_CvM}
	The Gegenbauer coefficients of $\psi_q^{\mathrm{CvM}}$ for $q=1,2,3$ are 
	\begin{align*}
	b^{\mathrm{CvM}}_{k,q} = \begin{cases}
	\frac{1}{\pi^2k^2}, & q=1,\\ 
	\frac{1}{2(2k+3)(2k-1)}, & q=2,\\
	\frac{35}{72\pi^2}1_{\{k=1\}} + \frac{1}{2\pi^2}\frac{3k^2+6k+4}{k^2(k+1)(k+2)^2}1_{\{k>1\}}, & q=3,
	\end{cases}
	\end{align*}
	and, if $q\geq 2$,
	\begin{align*}
	b^{\mathrm{CvM}}_{k,q} = \frac{(q-1)^2 (2 k+q-1) \Gamma \left(\frac{q-1}{2}\right)^3 \Gamma \left(\frac{3 q}{2}\right)}{8 \pi q^2 \Gamma \left(\frac{q}{2}\right)^3  \Gamma \left(\frac{3 q+1}{2}\right)}{}_4F_3\lrp{1-k,q+k,\tfrac{q+1}{2},\tfrac{3q}{2};q+1,\tfrac{q}{2}+1,\tfrac{3q+1}{2};1}.
	\end{align*}
	Additionally, $b_{0,q}=1/3$ for $q\geq 1$. Therefore, $b_{k,q}^{\mathrm{CvM}}>0$ for $k\geq0$ and $q=1,2,3$.  	
\end{proposition}

\begin{proposition}[Gegenbauer coefficients for $\psi_1^{\mathrm{AD}}$ and $\psi_2^{\mathrm{AD}}$]
	\label{prop:Gegen_AD}
	The Gegenbauer coefficients of $\psi_1^{\mathrm{AD}}$ and $\psi_2^{\mathrm{AD}}$ are
	\begin{align*}
	b_{k,q}^{\mathrm{AD}}=\begin{cases}\frac{1}{\pi k^2}\int_0^\pi\frac{1-\cos(2k\theta)}{(\pi-\theta)\theta}\,\mathrm{d}\theta, & q =1,\\
	\frac{1}{k(k + 1)}, & q =2,
	\end{cases}
	\quad
	b_{0,q}^{\mathrm{AD}}=-1, \quad q \geq 1.
	\end{align*}
	Therefore, $b_{k,q}^{\mathrm{AD}}>0$ for $k\geq1$ and $q=1,2$.
\end{proposition}

The previous results do not guarantee the omnibussness for $q\geq 1$ of the tests based on $P_{n,q}^{\mathrm{CvM}}$ and $P_{n,q}^{\mathrm{AD}}$. The next corollary gives a simple sufficient condition, satisfied by these tests, to guarantee omnibussness. It also shows that non-omnibus tests, related with very specific discrete measures $W$, are somehow a rarity in the projected-ecdf class.

\begin{corollary}[Omnibusness of projected-ecdf tests]\label{coro:omn}
	The statistic $P_{n,q}^W \in\mathcal{P}_{\sigma+}$ generates an omnibus test if and only if $W$ does not concentrate its measure entirely in $F_q(Z_q)$, where $Z_q:=\cup_{k\geq 1}\{x\in[-1,1]:x\in Z_{k,q}\}$ and
	\[
	Z_{k,q}:=\begin{cases}
	\{\cos(m/k\pi):m=0,1,\ldots,k\},&q=1,\\
	\{-1,1\}\cup\big\{x\in(-1,1):C_{k-1}^{(q+1)/2}(x)=0\big\},&q\geq 2.
	\end{cases}
	\]
	In particular, any $W$ that assigns a positive measure to a fixed set of $[0,1]$ with non-zero Lebesgue measure, generates an omnibus projected-ecdf test.
\end{corollary}

\begin{remark}
	Since $F_1(Z_1)
	=\mathbb{Q}\cap[0,1]$, any measure $W$ concentrated on $\mathbb{Q}\cap[0,1]$ generates a non-omnibus projected-ecdf test. This is the case of the $P_{n,1}^{\mathrm{R}_t}$-based test with rational $t\in[0,1]$ (if $t\in[0,1]$ is irrational, the test is omnibus). For $q\geq 2$, the explicit characterization of $F_q(Z_q)$ is more cumbersome, since it depends on the zeros of the Gegenbauer polynomials. Nevertheless, it is easy to see that $\{0,1/2,1\}\subset F_q(Z_q)$ and, as already known, the $P_{n,q}^{\mathrm{R}_{1/2}}$-based test is not omnibus.
\end{remark}

\subsection{Computation of asymptotic tail probabilities}

Computing tail probabilities in the asymptotic distribution $\sum_{k=1}^\infty w_k  Y_k$ in \eqref{eq:sumchis} is not trivial. \cite{Prentice1978} provided approximations, for specific choices of $\{w_{k}\}$, based on inverting the characteristic function \citep{Beran1968}. A more general approach is to compute, for a sufficiently large $K$, the tail probability of $\sum_{k=1}^K w_k Y_k$ by the fast Hall--Buckley--Eagleson (HBE) approximation (three-moment match to a Gamma distribution, see \cite{Buckley1988}) or by \cite{Imhof1961}'s exact method; see \cite{Bodenham2016} for a review of approaches for evaluating the cdf of $\sum_{k=1}^K w_k Y_k$.\\

We consider the following algorithm for computing asymptotic tail probabilities.

\begin{algo}[Asymptotic $p$-value of a test based on 
	$P_{n,q}^W\in\mathcal{P}_{\sigma+},\mathcal{P}_{\pm}$]\label{algo:1}
	\mbox{}
	\begin{enumerate}[label=\textit{\arabic{*}}., ref=\textit{\arabic{*}}]
		\item Compute the sequence $\{b_{k,q}^W\}$ for $k=1,\ldots,K_{\max}$ using the most adequate expression in Corollaries \ref{coro:coef_psi} and \ref{coro:Gegen_Rt} or Propositions \ref{prop:Gegen_CvM} and \ref{prop:Gegen_AD}.\label{algo:1:1}
		\item Reduce $K_{\max}$ to $K_{\delta,x}=\min\{K\geq1:|p^{\mathrm{HBE}}_{K_{\max},x}-p^{\mathrm{HBE}}_{K,x}|\leq\delta\}$ for $\delta\in[0,1]$, where $p^{\mathrm{HBE}}_{K,x}$ represents the HBE-approximated tail probability $\mathbb{P}[\sum_{k=1}^K (v_{k,q}^W)^2 Y_k>x]$, $\{v_{k,q}^W\}$ stem from $\{b_{k,q}^W\}$ using \eqref{eq:gf}, and $x$ equals the observed statistic $P_{n,q}^W$.\label{algo:1:2}
		\item Use \cite{Imhof1961}'s method to compute $\mathbb{P}\big[\sum_{k=1}^{K_{\delta,x}} (v_{k,q}^W)^2 Y_k > x\big]$.\label{algo:1:3}
	\end{enumerate}
\end{algo}

Step \ref{algo:1:2} is useful to reduce the computational burden in \cite{Imhof1961}'s method (run with accuracy $10^{-6}$) for a large number of terms. For example, if $K_{\max}=10^5$ and $\delta=10^{-5}$, then the average $K_{\delta,x}$ is about $4.5\times10^4$ for all $x\in[0,1]$, $q=1,2,3,10$, and statistics $P_{n,q}^\mathrm{CvM}$, $P_{n,q}^\mathrm{AD}$, and $P_{n,q}^{\mathrm{R}_{1/3}}$. In our empirical investigations, we have seen that, for dimensions $1\leq q\leq 10$ and the previous statistics, Step \ref{algo:1:2} can also be effectively omitted. Indeed, $\delta=0$ and $K_{\max}=10^3,10^4,5\times10^4$ give a uniform (in $x\in[0,1]$) tail probability accuracy of two, three, and four digits, respectively. Due to this, we set $K_{\max}=5\times10^4$ and $\delta=0$ in the application of Algorithm \ref{algo:1} in the  next sections. The accuracy deteriorates as $q$ increases, e.g., with $q=50$ a one digit precision is lost. 

\section{Simulation study}
\label{sec:simus}

We compare through simulations the empirical performance of three new projected-ecdf tests: CvM, Anderson--Darling (AD), and Rothman (Rt) for $t=1/3$. We do so by: (\textit{i}) evaluating the accuracy of their asymptotic distributions in Section \ref{sec:null}; (\textit{ii}) comparing their empirical powers with well-known uniformity tests for $\Omega_q$, $q\geq1$, in Section \ref{sec:power}. \\

We consider $M=10^6$ Monte Carlo replicates, dimensions $q=1,2,3,10$, and sample sizes $n=50,100,200$. The implementations of the projected-ecdf test statistics rely on the explicit expressions given in Propositions \ref{coro:psi_q:CvM}, \ref{prop:psi_q:Roth}, and \ref{prop:psi_q:AD}. Gauss--Legendre quadrature with $160$ nodes is used for approximating the integrals in the kernel functions.

\subsection{Asymptotic null distribution accuracy}
\label{sec:null}

Table \ref{tab:null:1} reveals that the exact-$n$ critical values, approximated by $M$ Monte Carlo replicates, quickly converge to the asymptotic critical values, irrespectively of the investigated dimensions, significance levels or projected-ecdf tests. Table \ref{tab:null:2} corroborates that the exact-$n$ rejection frequencies remain within the normal $99\%$ confidence interval for $n=200$ when the asymptotic critical values are employed in the test decision. For $n=50,100$, the rejection frequencies are either inside the confidence interval or quite close to the significance level, on the conservative side.

\begin{table}[!h]
	\centering
	\scriptsize
	\setlength{\tabcolsep}{2.5pt}
	\begin{tabular}{ll|R{1.0cm}R{1.0cm}R{1.0cm}R{1.0cm}|R{1.0cm}R{1.0cm}R{1.0cm}R{1.0cm}|R{1.0cm}R{1.0cm}R{1.0cm}R{1.0cm}}
		\toprule
		\multirow{2}{*}{Test}& \multirow{2}{*}{$q$}& \multicolumn{4}{c|}{$\alpha=0.10$}  & \multicolumn{4}{c|}{$\alpha=0.05$} & \multicolumn{4}{c}{$\alpha=0.01$} \\
		& & $n\!\!\,=\!\!\,50$ & $n\!\!\,=\!\!\,100$ & $n\!\!\,=\!\!\,200$ & $n\!\!\,=\!\!\,\infty$ & $n\!\!\,=\!\!\,50$ & $n\!\!\,=\!\!\,100$ & $n\!\!\,=\!\!\,200$ & $n\!\!\,=\!\!\,\infty$ & $n\!\!\,=\!\!\,50$ & $n\!\!\,=\!\!\,100$ & $n\!\!\,=\!\!\,200$ & $n\!\!\,=\!\!\,\infty$ \\
		\midrule
		CvM & 1 & 0.3025 & 0.3029 & 0.3031 & 0.3035 & 0.3713 & 0.3723 & 0.3731 & 0.3738 & 0.5303 & 0.5336 & 0.5356 & 0.5368\\
		& 2 & 0.2759 & 0.2764 & 0.2769 & 0.2769 & 0.3270 & 0.3277 & 0.3289 & 0.3291 & 0.4412 & 0.4442 & 0.4465 & 0.4469\\
		& 3 & 0.2598 & 0.2605 & 0.2607 & 0.2608 & 0.3010 & 0.3020 & 0.3031 & 0.3029 & 0.3929 & 0.3941 & 0.3960 & 0.3963\\
		& 10 & 0.2202 & 0.2206 & 0.2206 & 0.2208 & 0.2406 & 0.2411 & 0.2412 & 0.2414 & 0.2831 & 0.2844 & 0.2848 & 0.2849\\
		\midrule
		AD & 1 & 1.6824 & 1.6832 & 1.6855 & 1.6875 & 2.0170 & 2.0236 & 2.0272 & 2.0304 & 2.7982 & 2.8122 & 2.8194 & 2.8252\\
		& 2 & 1.5555 & 1.5577 & 1.5613 & 1.5612 & 1.8112 & 1.8156 & 1.8216 & 1.8227 & 2.3856 & 2.3993 & 2.4111 & 2.4122\\
		& 3 & 1.4773 & 1.4805 & 1.4818 & 1.4824 & 1.6869 & 1.6911 & 1.6972 & 1.6961 & 2.1531 & 2.1585 & 2.1690 & 2.1695\\
		& 10 & 1.2780 & 1.2797 & 1.2798 & 1.2810 & 1.3836 & 1.3863 & 1.3863 & 1.3880 & 1.6044 & 1.6100 & 1.6124 & 1.6130\\
		\midrule
		Rt & 1 & 0.4248 & 0.4254 & 0.4259 & 0.4264 & 0.5282 & 0.5297 & 0.5304 & 0.5318 & 0.7666 & 0.7716 & 0.7744 & 0.7764\\
		& 2 & 0.3830 & 0.3837 & 0.3844 & 0.3844 & 0.4585 & 0.4594 & 0.4614 & 0.4617 & 0.6280 & 0.6322 & 0.6355 & 0.6361\\
		& 3 & 0.3584 & 0.3593 & 0.3597 & 0.3598 & 0.4191 & 0.4204 & 0.4220 & 0.4217 & 0.5534 & 0.5556 & 0.5579 & 0.5589\\
		& 10 & 0.2997 & 0.3002 & 0.3002 & 0.3005 & 0.3292 & 0.3299 & 0.3299 & 0.3304 & 0.3907 & 0.3924 & 0.3930 & 0.3933\\
		\bottomrule
	\end{tabular}
	\caption{\small Exact-$n$ and asymptotic critical values for the significance levels $\alpha=0.10,0.05,0.01$ of the CvM, AD, and Rt uniformity tests on $\Om{q}$, for $q=1,2,3,10$. The exact-$n$ critical values are approximated by $M$ Monte Carlo replicates, whereas the asymptotic critical values are computed using Algorithm \ref{algo:1}. \label{tab:null:1}}
\end{table}

\begin{table}[!h]
	\centering
	\scriptsize
	\setlength{\tabcolsep}{2.5pt}
	\begin{tabular}{ll|R{1.1cm}R{1.1cm}R{1.1cm}|R{1.1cm}R{1.1cm}R{1.1cm}|R{1.1cm}R{1.1cm}R{1.1cm}}
		\toprule
		\multirow{2}{*}{Test}& \multirow{2}{*}{$q$}& \multicolumn{3}{c|}{$\alpha=0.10$}  & \multicolumn{3}{c|}{$\alpha=0.05$} & \multicolumn{3}{c}{$\alpha=0.01$} \\
		& & $n\!\!\,=\!\!\,50$ & $n\!\!\,=\!\!\,100$ & $n\!\!\,=\!\!\,200$ & $n\!\!\,=\!\!\,50$ & $n\!\!\,=\!\!\,100$ & $n\!\!\,=\!\!\,200$ & $n\!\!\,=\!\!\,50$ & $n\!\!\,=\!\!\,100$ & $n\!\!\,=\!\!\,200$ \\
		\midrule
		CvM                         & 1                    & 0.0990   & \bf 0.0993 & \bf 0.0996 & 0.0488   & 0.0493     & \bf 0.0497 & 0.0094   & 0.0097     & \bf 0.0099 \\
		& 2                    & 0.0987   & \bf 0.0993 & \bf 0.1000 & 0.0485   & 0.0490     & \bf 0.0499 & 0.0092   & 0.0096     & \bf 0.0099 \\
		& 3                    & 0.0984   & \bf 0.0995 & \bf 0.0999 & 0.0484   & 0.0492     & \bf 0.0502 & 0.0094   & 0.0096     & \bf 0.0099 \\
		& 10                   & 0.0982   & \bf 0.0994 & \bf 0.0996 & 0.0485   & \bf 0.0495 & \bf 0.0496 & 0.0093   & \bf 0.0098 & \bf 0.0100 \\
		\midrule
		AD                          & 1                    & 0.0989   & 0.0991     & \bf 0.0996 & 0.0486   & 0.0493     & \bf 0.0497 & 0.0095   & \bf 0.0097 & \bf 0.0099 \\
		& 2                    & 0.0984   & 0.0991     & \bf 0.1000 & 0.0484   & 0.0490     & \bf 0.0498 & 0.0093   & 0.0097     & \bf 0.0100 \\
		& 3                    & 0.0983   & \bf 0.0994 & \bf 0.0998 & 0.0484   & 0.0492     & \bf 0.0502 & 0.0094   & 0.0096     & \bf 0.0100 \\
		& 10                   & 0.0980   & \bf 0.0992 & \bf 0.0992 & 0.0486   & 0.0494     & 0.0494     & 0.0094   & \bf 0.0098 & \bf 0.0100 \\
		\midrule
		Rt                          & 1                    & 0.0989   & \bf 0.0994 & \bf 0.0996 & 0.0488   & 0.0493     & \bf 0.0495 & 0.0094   & 0.0097     & \bf 0.0099 \\
		& 2                    & 0.0987   & \bf 0.0993 & \bf 0.1000 & 0.0486   & 0.0490     & \bf 0.0499 & 0.0092   & 0.0096     & \bf 0.0099 \\
		& 3                    & 0.0985   & \bf 0.0994 & \bf 0.0999 & 0.0485   & 0.0492     & \bf 0.0502 & 0.0093   & 0.0096     & \bf 0.0099 \\
		& 10                   & 0.0982   & \bf 0.0993 & \bf 0.0993 & 0.0485   & 0.0494     & \bf 0.0495 & 0.0093   & 0.0097     & \bf 0.0099\\
		\bottomrule
	\end{tabular}
	\caption{\small Rejection frequencies using asymptotic critical values for the significance levels $\alpha=0.10,0.05,0.01$ of the CvM, AD, and Rt uniformity tests on $\Om{q}$, for $q=1,2,3,10$. The rejection frequencies are approximated by $M$ Monte Carlo replicates, whereas the asymptotic critical values are computed with Algorithm \ref{algo:1}. Boldfaces denote that the rejection rate is within the normal $99\%$ confidence interval for $\alpha$. \label{tab:null:2}}
\end{table}

\subsection{Empirical power investigation}
\label{sec:power}

\begin{figure}[t]
	\centering
	\includegraphics[width=0.33\textwidth]{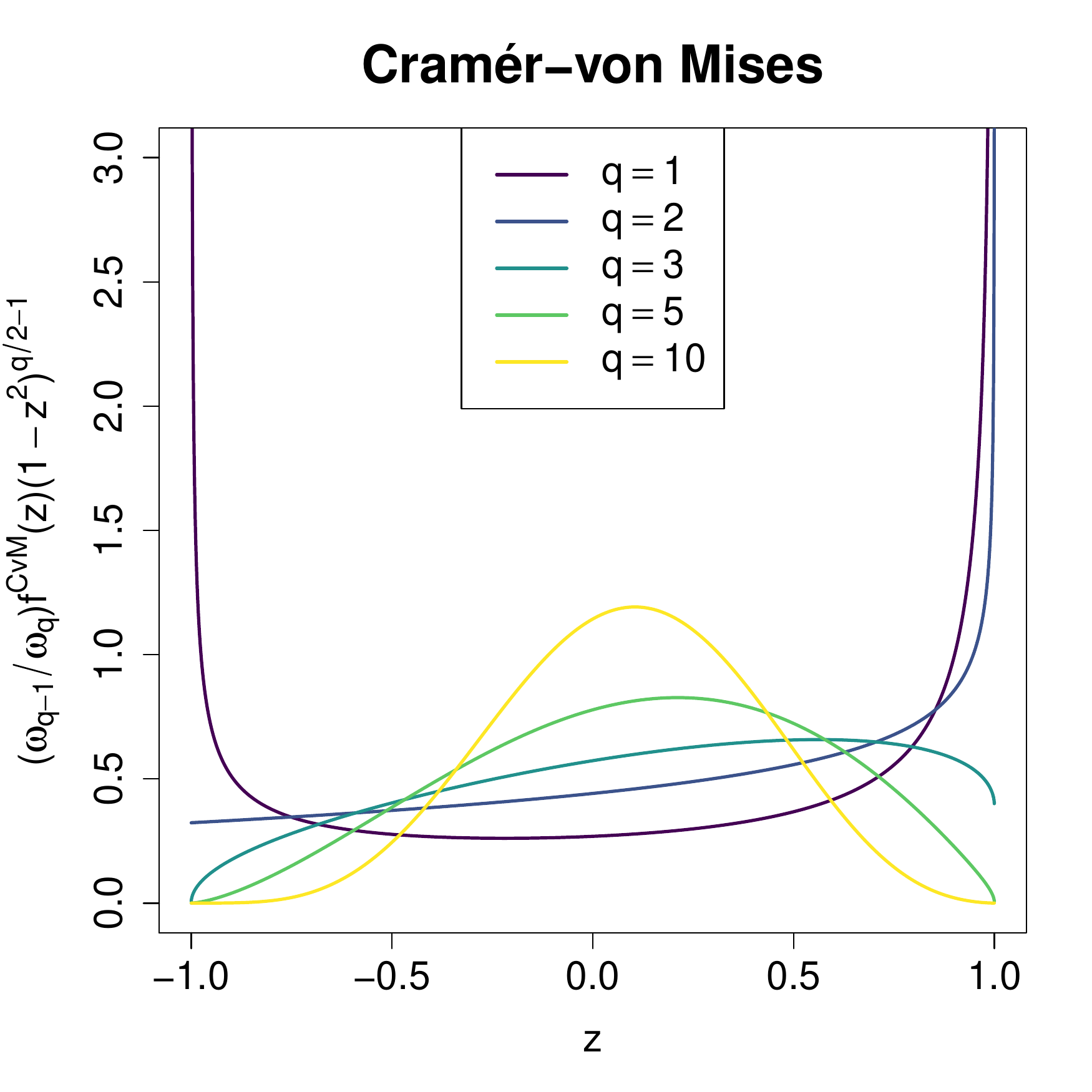}\includegraphics[width=0.33\textwidth]{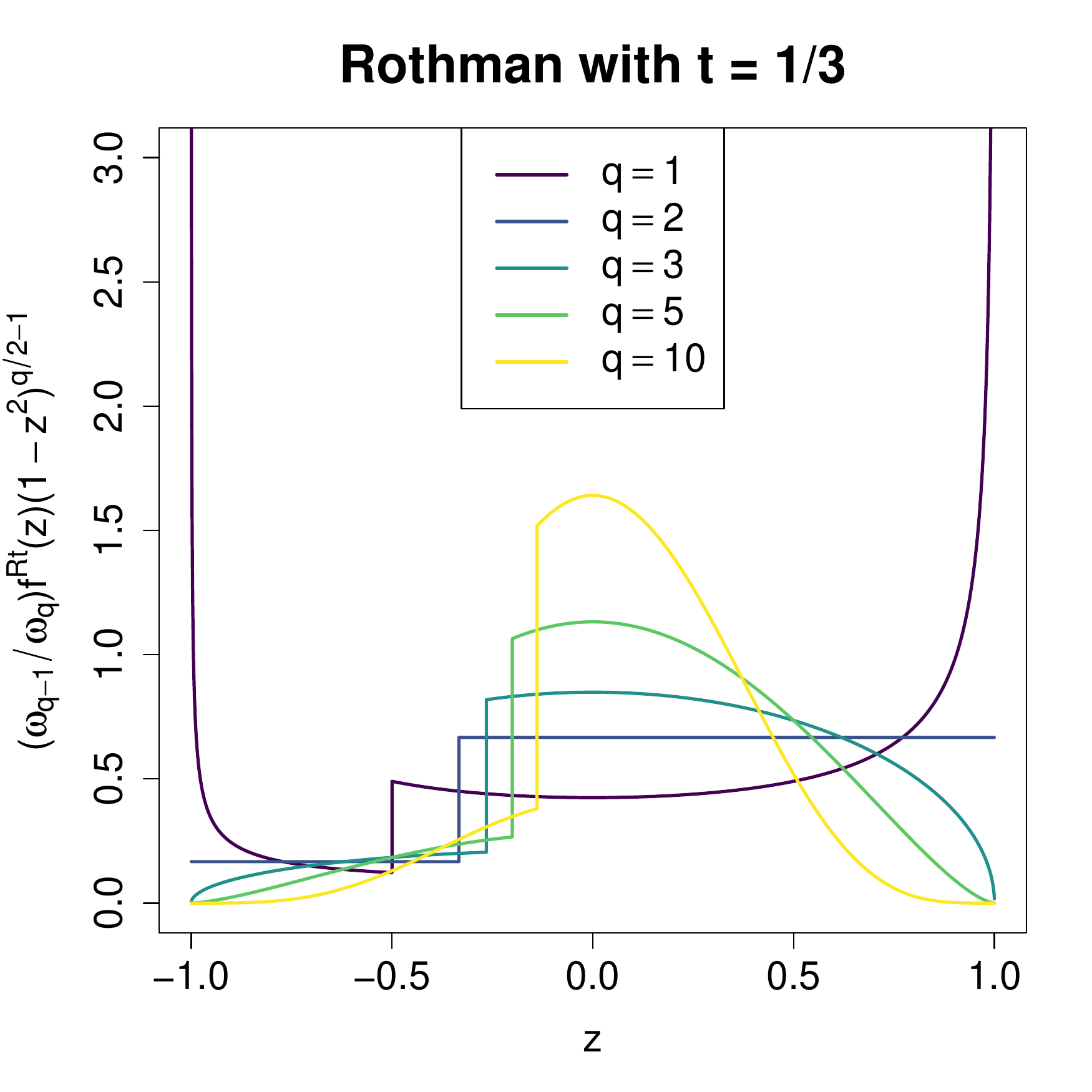}\includegraphics[width=0.33\textwidth]{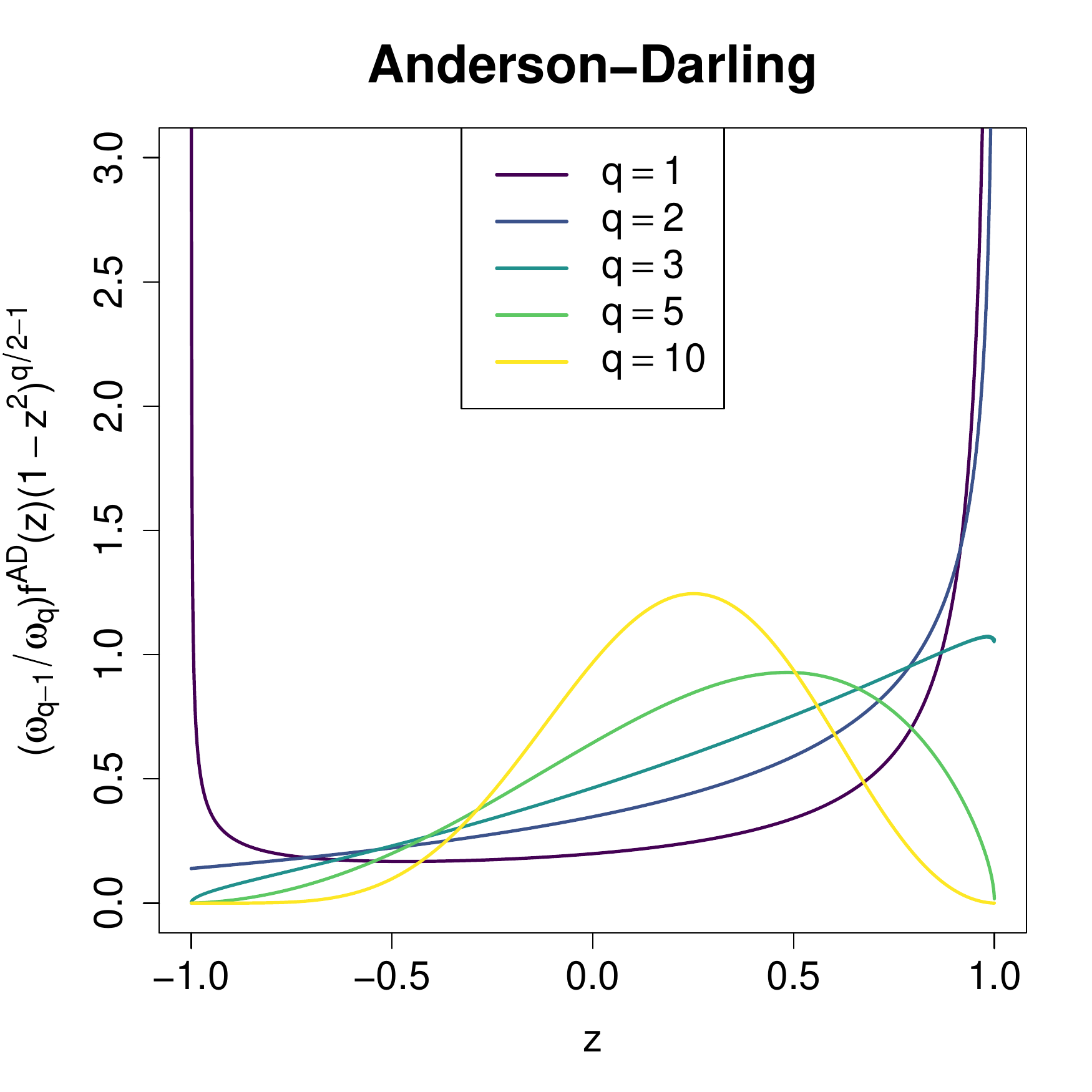}
	\caption{From left to right, depiction of the projected densities $z\mapsto (\omega_{q - 1}/\omega_{q}) f^{W}(z) (1 - z^2)^{q / 2 - 1}$ on $[-1,1]$ generated by $f^{\mathrm{CvM}}$, $f^{\mathrm{R}_{1/3}}$, and $f^{\mathrm{AD}}$, for $q=1,2,3,5,10$. The series in \eqref{eq:dgp} is truncated such that $99.95\%$ of the norm in $L_q^2[-1,1]$ is retained. \label{fig:altsq}}
\end{figure}

We compare the CvM, AD, and Rt projected-ecdf tests with the following classic uniformity tests: Rayleigh, Bingham, Ajne, Gin\'e's $G_n$ \citep{Gine1975,Prentice1978}, Bakshaev, and \cite{Cuesta-Albertos2009}, henceforth abbreviated CCF09. Except the latter (already reviewed in Section \ref{sec:genesis}), all of them belong to the Sobolev class. The main properties of the competing tests are: (\textit{i}) all of them are valid for arbitrary dimensions; (\textit{ii}) only Bakshaev and CCF09 are omnibus tests; (\textit{iii}) the Rayleigh and Ajne tests are not consistent against axial alternatives (symmetric pdfs with respect to $\mathbf{0}$); (\textit{iv}) the Bingham and Gin\'e's $G_n$ tests are designed for axial alternatives, sacrificing power against unimodal alternatives; (\textit{v}) the Rayleigh and Bingham tests are the most powerful rotation invariant tests with respect to von Mises--Fisher and Watson alternatives, respectively; (\textit{vi}) the CCF09 test requires simulating $k$ random directions and then performing a Monte Carlo calibration \emph{conditionally} on those random directions. Following the recommendation in CCF09, we considered $k=50$, then run the simulation study conditionally on a fixed set of $k$ random directions. \\

We employ six different Data Generating Processes (DGPs). The first three are based on the local alternatives for which a $P_{n,q}^W$-based projected-ecdf test is locally asymptotically most powerful rotation-invariant in virtue of Corollary \ref{coro:asymp}: 
\begin{align}
f^W_{\bmu,\kappa}(\bx):=\frac{1-\kappa}{\omega_q}+\frac{\kappa f^W(\bx'\bmu)}{\omega_q},\, f^W(z):=1+\sum_{k=1}^\infty \lrp{1+\tfrac{2k}{q-1}}(b_{k,q}^W)^{1/2}C_k^{(q-1)/2}(z).\label{eq:dgp}
\end{align}
Considering $\bmu=(\zero_{q},1)$, the first three DGPs use \eqref{eq:dgp} with the next coefficients:
\begin{description}
	\item[CvM] $\{b_{k,q}^\mathrm{CvM}\}$ given in Proposition \ref{prop:Gegen_CvM} for $q=1,2,3$, and computed numerically for $q=10$ using Corollary \ref{coro:coef_psi} and Gauss--Legendre quadrature with $5120$ nodes.
	\item[AD] $\{b_{k,q}^\mathrm{AD}\}$ given in Proposition \ref{prop:Gegen_AD} for $q=1$ and computed numerically for $q\geq2$ under the previous conditions.
	\item[Rt] $\{b_{k,q}^{\mathrm{R}_{1/3}}\}$ given in Corollary \ref{coro:Gegen_Rt}.
\end{description}
These DGPs can be seen as ``unimodal'' alternatives: CvM and AD concentrate probability mass about $\bmu$, while Rt does so in a constant cap about $\bmu$. \\

The remaining DGPs include the optimally-detected alternatives for the Rayleigh and Bingham tests, and an alternative without an optimal test among the inspected:
\begin{description}
	\item[vMF] Von Mises--Fisher pdf $\bx\mapsto c^{\mathrm{vMF}}_{q,\eta}\exp(\eta\bx'\bmu)$, with $\eta\geq0$ and $\bmu\in\Om{q}$. We set $\eta=\kappa$ and $\bmu=(\zero_{q},1)$.
	\item[Wat] Watson pdf $\bx\mapsto c^{\mathrm{W}}_{q,\eta}\exp(\eta(\bx'\bmu)^2)$, with $\eta\in\mathbb{R}$ and $\bmu\in\Om{q}$. We set $\eta=2.5\kappa$ and $\bmu=(\zero_{q},1)$.
	\item[SC] Small-circle pdf $\bx\mapsto c^{\mathrm{SC}}_{q,\tau,\eta}\exp(\eta(\bx'\bmu-\tau)^2)$, with $\eta\in\mathbb{R}$, $\tau\in[-1,1]$, and $\bmu\in\Om{q}$. We set $\eta=-1.5\kappa$, $\tau=0.50$, and $\bmu=(\zero_{q},1)$.
\end{description}
The deviation from uniformity is controlled by $\kappa\geq 0$ ($\kappa\leq1$ necessarily for \eqref{eq:dgp}).\\

The simulation of all the alternatives was done through the tangent-normal decomposition implemented in the \texttt{rotasym} package \citep{Garcia-Portugues:rotasym} and the (numerical) inversion method to simulate from the pdfs of the projections along $\bmu$, which for $f^W$ are $z\mapsto (\omega_{q-1}/\omega_q)f^W(z)(1-z^2)^{q/2-1}$ (see Figure \ref{fig:altsq}). The series in $f^W$ is truncated to its first $K_r$ terms explaining $r=99.95\%$ of the $L_q^2[-1,1]$-norm of the series computed with $K_{\max}=5\times10^4$ terms. The simulation from Rt is exact thanks to the closed-form expression $f^{\mathrm{Rt}}(z)=1_{\left\{z\geq F_q^{-1}(t)\right\}} + t$ and its projected quantile function $F_{\mathrm{Rt}}^{-1}(u)=F_q^{-1}((u + t)/(1+t))1_{\{u>t^2\}}+F_q^{-1}(u/t)1_{\{u\leq t^2\}}$. Sampling from CvM for $q=1$ is simplified due to the closed-form expression $f^{\mathrm{CvM}}(z)=1 - \sqrt{2} \log(2(1 - z)) / (2\pi)$.\footnote{Differs from $f(\theta)=\theta^2/(2\pi^2)$ in \citet[page 114]{Mardia2000} for the Watson test, which is not a circular pdf nor generates the Watson statistic from the book's equation (6.3.70).} \\

For the sake of equity, all the tests are calibrated with exact-$n$ critical values approximated by $M$ Monte Carlo replicates at the $\alpha=0.05$ significance level. Table \ref{tab:pow:1} collects the empirical powers for $q=1,2,3,10$, $n=100$, and $\kappa=0.50$. The remaining combinations for $n=50,100,200$ and $\kappa=0.25,0.50,0.75$ are relegated to the Supplementary Material. The following conclusions are extracted from all the tables: 
\begin{enumerate}[label=(\textit{\roman{*}})., ref=(\textit{\roman{*}})]
	
	\item Overall, the CvM test improves over CCF09. It does so with an average (absolute) power gain equal to $0.0315$ for all DGPs but Wat. The ``[$5\%,95\%]$ Interquantile Range'' (IR) of these power gains (taken over all variations of $n$ and $\kappa$, henceforth implicit) is $[0, 0.0930]$. In Wat, the only axial alternative, the IR is $[-0.0703, 0.0085]$. The power gap between CvM and CCF09 stretches with $q$ due to the increasing difficulty of capturing the most $\Hcal_0$-separating directions on $\Om{q}$ by random sampling.\label{conc:cvm}
	
	\item The AD test performs just slightly better ($0.0005$) than CvM on all DGPs except Wat, where AD notably improves CvM. In Wat, the gains for AD have average $0.0553$ and IR $[0.0002, 0.2331]$. AD test an edge against axial alternatives, dominating the CCF09 test for all the DGPs considered (IR: $[0,0.1085]$).\label{conc:ad}
	
	\item The Rt test performs very similarly to CvM in all alternatives except Wat, where it is clearly outperformed by the latter (IR: $[-0.144,0.002]$).\label{conc:rt}
	
	\item The Bakshaev test is slightly ($-0.0026$) outperformed by CvM in all alternatives except vMF, where it behaves similarly to the latter ($0.0002$).\label{conc:bak}
	
	\item Unimodal alternatives are well-detected by projected-ecdf tests: (\textit{a}) the CvM, AD, and Rt tests have very similar performance in their corresponding DGPs; (\textit{b}) in the vMF alternative, the optimal Rayleigh test is barely superior to the projected-ecdf tests; e.g., its average power gain with respect to CvM is just $0.002$.\label{conc:uni}
	
	\item The non-unimodal and non-axial alternative SC is well-detected by the AD test. It outperforms the rest of tests in the majority of situations (especially $n=100,200$).\label{conc:nonuni}
	
	\item The axial alternative Wat is much harder to detect by projected-ecdf tests. Their performance is consistently below the Bingham test, whose average power gain with respect to AD is $0.253$, a sharp contrast with the situation in the vMF alternative.\label{conc:axial}
	
	\item Local asymptotic optimalities have very small effect sizes, and many are actually undetected for the settings considered in the study. Indeed, the AD test is among the most powerful tests in most of the CvM and Rt alternatives. This result can be explained by several factors: (\textit{a}) the very small effect sizes of local optimalities; (\textit{b}) the numerical inaccuracy on sampling exactly \eqref{eq:dgp}; (\textit{c}) the limitation of the explored values for $\kappa$ and $n$; (\textit{d}) the Monte Carlo noise. \label{conc:local}
	
	\item The Rayleigh and Ajne tests perform really similarly. So do the Bingham and Gin\'e's $G_n$ tests. Even in the vMF and Wat alternatives, where Rayleigh and Bingham are respectively the optimal tests, the power difference is minimal.\label{conc:raybing}
	
	\item Overall, the qualitative behaviour of all the tests except Bingham and Gin\'e's $G_n$ is highly similar, a reflection of \ref{conc:uni} and \ref{conc:nonuni}. \label{conc:similar}
	
	\item Though all alternatives are harder to detect when $q$ increases, SC and especially Wat have the larger power dropouts in their optimal tests.\label{conc:misc}
	
\end{enumerate}

Based on the previous conclusions, we regard the AD test as a reference test of uniformity on $\Om{q}$ due to its omnibussness, great performance against unimodal alternatives, and relative robustness against non-unimodal alternatives.

\begin{table}[h!]
	\centering
	\setlength{\tabcolsep}{2.5pt}
	\scriptsize
	\begin{tabular}{ll|R{1.25cm}R{1.25cm}R{1.25cm}R{1.25cm}R{1.25cm}R{1.25cm}R{1.25cm}R{1.25cm}R{1.25cm}}
		\toprule
		\scriptsize DGP  & $q$ & \scriptsize Rayleigh & \scriptsize Bingham & \scriptsize Ajne & Gin\'e & \scriptsize CCF09 & \scriptsize Bakshaev & \scriptsize CvM & \scriptsize AD & \scriptsize Rt \\
		\midrule
		CvM & 1 & 0.2773 & 0.1004 & 0.2793 & 0.1021 & 0.2653 & 0.2879 & 0.2897 & \bf 0.2918 & 0.2879\\
		& 2 & 0.2367 & 0.0794 & 0.2377 & 0.0801 & 0.2083 & 0.2424 & 0.2424 & \bf 0.2434 & 0.2414\\
		& 3 & 0.2064 & 0.0692 & 0.2068 & 0.0696 & 0.1839 & 0.2098 & 0.2095 & \bf 0.2104 & 0.2087\\
		& 10 & 0.1326 & 0.0554 & \bf 0.1327 & 0.0554 & 0.1049 & \bf 0.1328 & \bf 0.1328 & \bf 0.1328 & \bf 0.1328\\
		\midrule
		AD & 1 & 0.9002 & 0.4463 & 0.9071 & 0.4776 & 0.9019 & 0.9234 & 0.9271 & \bf 0.9319 & 0.9225\\
		& 2 & 0.8507 & 0.3110 & 0.8542 & 0.3213 & 0.8037 & 0.8670 & 0.8670 & \bf 0.8710 & 0.8637\\
		& 3 & 0.8092 & 0.2378 & 0.8113 & 0.2426 & 0.7497 & 0.8218 & 0.8209 & \bf 0.8241 & 0.8180\\
		& 10 & 0.6201 & 0.1021 & 0.6204 & 0.1023 & 0.4625 & 0.6241 & 0.6234 & \bf 0.6246 & 0.6224\\
		\midrule
		Rt & 1 & 0.4020 & 0.1282 & 0.4014 & 0.1303 & 0.4124 & 0.4175 & 0.4203 & \bf 0.4225 & 0.4213\\
		& 2 & 0.3362 & 0.0820 & 0.3360 & 0.0833 & 0.3127 & \bf 0.3413 & \bf 0.3413 & \bf 0.3414 & \bf 0.3413\\
		& 3 & 0.2902 & 0.0686 & 0.2903 & 0.0694 & 0.2756 & \bf 0.2924 & \bf 0.2924 & \bf 0.2923 & \bf 0.2922\\
		& 10 & \bf 0.1744 & 0.0543 & \bf 0.1745 & 0.0544 & 0.1410 & 0.1742 & 0.1742 & 0.1740 & \bf 0.1744\\
		\midrule
		vMF & 1 & \bf 0.8867 & 0.0634 & 0.8859 & 0.0634 & 0.8451 & 0.8837 & 0.8816 & 0.8769 & 0.8823\\
		& 2 & \bf 0.6648 & 0.0558 & 0.6638 & 0.0556 & 0.5895 & 0.6606 & 0.6606 & 0.6560 & 0.6628\\
		& 3 & \bf 0.4806 & 0.0530 & 0.4798 & 0.0532 & 0.4247 & 0.4774 & 0.4779 & 0.4749 & 0.4791\\
		& 10 & \bf 0.1265 & 0.0503 & \bf 0.1264 & 0.0503 & 0.1031 & 0.1261 & 0.1262 & 0.1259 & \bf 0.1264\\
		\midrule
		SC & 1 & 0.9843 & 0.3243 & 0.9841 & 0.3212 & 0.9891 & 0.9910 & 0.9918 & \bf 0.9922 & 0.9919\\
		& 2 & 0.8716 & 0.1489 & 0.8711 & 0.1480 & 0.8545 & 0.8858 & 0.8858 & \bf 0.8887 & 0.8836\\
		& 3 & 0.7111 & 0.0933 & 0.7104 & 0.0933 & 0.6767 & 0.7192 & 0.7187 & \bf 0.7202 & 0.7170\\
		& 10 & \bf 0.2092 & 0.0526 & \bf 0.2092 & 0.0525 & 0.1612 & 0.2086 & 0.2088 & 0.2083 & \bf 0.2090\\
		\midrule
		W & 1 & 0.0538 & \bf 0.9785 & 0.0547 & 0.9773 & 0.5358 & 0.3560 & 0.4916 & 0.6396 & 0.4946\\
		& 2 & 0.0536 & \bf 0.8850 & 0.0543 & 0.8823 & 0.2031 & 0.1643 & 0.1643 & 0.2570 & 0.1265\\
		& 3 & 0.0526 & \bf 0.6728 & 0.0529 & 0.6698 & 0.1065 & 0.0980 & 0.0916 & 0.1216 & 0.0769\\
		& 10 & 0.0504 & \bf 0.0832 & 0.0505 & \bf 0.0831 & 0.0515 & 0.0523 & 0.0518 & 0.0526 & 0.0512\\
		\bottomrule
	\end{tabular}
	\caption{\small Empirical powers for the investigated uniformity tests on $\Omega_1$ for $n=100$ and $\kappa=0.50$. Boldfaces indicate the tests whose empirical powers are \textit{not} significantly smaller than the largest empirical power for each row, according to a McNemar's exact one-sided test \citep{Fay2010} performed at $5\%$ significance level. \label{tab:pow:1}}
\end{table}

We conclude by pointing that \ref{conc:local} and \ref{conc:raybing} may seem surprising, yet they had been partially reported in the literature. Related with \ref{conc:local}, \cite{Stephens1969} studied the powers of Ajne and Watson tests under the Rt alternative with $t=1/2$ and different values of $\kappa$, finding that Ajne was only barely more powerful (see his Table 3). With respect to \ref{conc:raybing}, \cite{Figueiredo2003} compared the Bingham and Gin\'e's $G_n$ tests under different dimensions, sample sizes, and concentrations, finding no remarkable differences between them (see their Table 3). \cite{Figueiredo2007} conducted a similar analysis for the Rayleigh and Ajne tests with identical conclusions (see her Tables 2--4). A simulation experiment in the Supplementary Material gives insights about \ref{conc:local}.

\section{Real data applications}
\label{sec:realdata}

We illustrate the practical relevance of the proposed tests with three real data applications in astronomy. The first two build on previous applications in $\Omega_1$ and $\Omega_2$, while the third is a novel case study. 
The end-to-end reproduction of the three applications is possible trough the \texttt{sphunif} package \citep{Garcia-Portugues:sphunif}. The asymptotic $p$-values were computed using Algorithm~\ref{algo:1}.

\subsection{Sunspots}
\label{subsec:sunspots}

Sunspots are darker regions of the Sun generated by local concentrations of the solar magnetic field. They 
appear in a rotationally symmetric fashion emerging due to the wrapping of the field by the Sun's differential rotation \citep{Babcock1961}. As this wrapping advances, sunspots progressively span at lower latitudes until approximately 11 years, when the field reverses its polarity and wrapping is restarted, constituting a \textit{solar cycle}. Non-rotationally symmetric patterns may be triggered by ``preferred zones of occurrence'' where sunspots had originated previously \citep[pages 574 and 581]{Babcock1961}. \\

The significance of non-rotationally symmetric patterns was investigated in \cite{Garcia-Portugues:optimal} using processed data from the Debrecen photoheliographic sunspot catalogue \citep{Baranyi2016,Gyoeri2016}. Their analysis considered tests for rotational symmetry that inspect the circular uniformity of the longitudes of sunspots with respect to an axis $\btheta$. However, due to the non-omnibusness of the tests employed in their analysis, non-rotationally symmetric deviations for which the tests are not consistent may have been undetected. \\

To further investigate the rotational symmetry of sunspots, we applied the CvM, AD, and Rt tests 
to the longitudes about the north pole $\btheta=(0,0,1)'$ of the $5373$ sunspots observed in the cycle 23 (1996--2008), obtaining the asymptotic $p$-values $0.3595$, $0.8393$, and $0.3285$, respectively. We repeated the analysis for the cycle 22 (1986--1996; $4551$ sunspots), obtaining the asymptotic $p$-values $0.0067$, $0.0139$, and $0.0091$. The outcomes of the analysis are coherent with those in \cite{Garcia-Portugues:optimal}, where the $p$-values of a non-omnibus test for rotational symmetry about $\boldsymbol\theta$ are $0.2710$ and $0.0103$ for the cycles 23 and 22, respectively. Therefore, our analysis shows that these outcomes hold when omnibus tests are used and highlights the varying behaviour of different cycles.

\subsection{Long-period comets}
\label{subsec:comet}

Orbits of celestial bodies, such as planets and comets, have attracted scientists' attention for a long time. 
\cite{Bernoulli1735} already discussed whether the clustering of the planets' orbits about the ecliptic, nowadays explained by their origin in the protoplanetary disk, could have happened ``by chance''. The study of comet orbits has been more intricate. Long-period comets (with periods larger than $200$ years) are thought to arise from the roughly spherical Oort cloud, containing icy planetesimals that were ejected from protoplanetary disks by giant planets. These icy planetesimals became heliocentric comets when their orbits were affected by random perturbations of passing stars and the galactic tide (see, e.g., Sections 5 and 7.2 in \citet{Dones2015} and references therein). This conjectured past of the Oort cloud explains the nearly isotropic distribution of long-period comets \citep[evidenced, e.g., in][]{Wiegert1999}, sharply contrasting with the ecliptic-clustered\nopagebreak[4] orbits of short-period comets originating in the flattened Kuiper Belt. \\

As illustrated in \cite{Watson1970} and \cite{Jupp2003}, assessing the uniformity of orbits can be formalized as testing the uniformity on $\Om{2}$ of their directed unit normal vectors. An orbit with \textit{inclination} $i\in[0,\pi]$ and \textit{longitude of the ascending node} $\Omega\in[0,2\pi)$ (see \cite{Jupp2003}) has directed normal vector  $(\sin(i)\sin(\Omega),-\sin(i)\cos(\Omega),\cos(i))'$ to the orbit's plane. The sign of the vector reflects if the orbit is prograde or retrograde. \\

We applied the CvM, AD, and Rt tests to revisit \cite{Watson1970}'s testing of the uniformity of the planets' orbits with updated measurements on $(i,\Omega)$. Unsurprisingly, uniformity is rejected with null Monte Carlo $p$-values. More interesting is the analysis of long-period comets, for which we: (\textit{i}) considered the $208$ long-period elliptic-type single-apparition comets, as of 7th of December 2007, used in \cite{Cuesta-Albertos2009}; (\textit{ii}) performed the same search in \cite{Cuesta-Albertos2009}, restricted to comets with distinct $(i,\Omega)$ up to the second digit, obtaining $438$ comets as of 7th May 2020. The source of both datasets is the JPL Small-Body Database Search Engine ({\small \url{https://ssd.jpl.nasa.gov/sbdb_query.cgi}}). The dynamic nature of the database, with additions of first-ever observed comets and updates on the data for former comets, generated the noticeable differences between (\textit{i}) and (\textit{ii}). \\

In (\textit{i}), the asymptotic $p$-values for the CvM, AD, and Rt tests are, respectively, $0.1011$, $0.0744$, and $0.1207$. Therefore, uniformity is not rejected at significance level $5\%$ and the outcome is in agreement with the analysis in \cite{Cuesta-Albertos2009}. In (\textit{ii}), however, the same tests gave asymptotic $p$-values $0.0041$, $0.0023$, and $0.0052$. Therefore, contrarily to the analysis in (\textit{i}), significant non-uniformity is detected in the orbits of long-period comets with updated records. The observational bias of long-period comets, as described in \cite{Jupp2003}, may explain the leading rejection\nolinebreak[4] cause.

\subsection{Craters on Rhea}
\label{subsec:crater}

Craters are roughly circular depressions resulting from impact or volcanic activity. Impact craters give valuable insights on the planetary subsurface structure, past geologic processes, resurfacing history, and relative surfaces ages \citep{Barlow2015}. Indeed, crater counting is the primary method for determining remotely the relative age of a planetary surface; see \cite{Fassett2016} for a review on crater statistics and their applications.\\

Short-period comets, especially dominant of the cratering process in the outer Solar System, are among the main generators of non-isotropic impact cratering (see \cite{Zahnle2003} and references therein). To evaluate the rareness of uniform crater distributions in the Solar System, we analysed the \textit{named} craters from the Gazetteer of Planetary Nomenclature database ({\small \url{https://planetarynames.wr.usgs.gov/AdvancedSearch}}) of the International Astronomical Union (IUA). As of May 31st 2020, the database contained $5235$ craters for $44$ bodies. Filtering for non-asteroid bodies with at least $30$ craters results in $4818$ observations on $\Omega_2$ containing the planetocentric coordinates of the craters' centers. Table \ref{tab:craters} reveals that, for this dataset, crater uniformity is rejected at significance level $5\%$ in all bodies except Venus and four Saturnian moons. These few non-rejections, however, are suspected to be driven by a uniformity bias in the data: well-separated craters that cover the body are likely more probable to be named than those that cluster (see Figure \ref{fig:Rhea}). Bypassing this source limitation requires from detailed crater databases, available only for certain bodies such as Venus (see, e.g., the analysis in \cite{Garcia-Portugues:cladag}) and Rhea. \\

\newpage
We investigate in detail the crater distribution of Rhea, the second most cratered body in Table \ref{tab:craters} with a uniform-like distribution. Rhea orbits Saturn synchronously, thus it has a \textit{leading hemisphere} that always faces forward into the orbit motion and a \textit{trailing hemisphere} that faces backward (see Figure \ref{fig:Rhea}). Preferred cratering on the leading hemisphere is expected from heliocentric impactors, whereas planetocentric impactors weakly favour the centers of the leading and trailing hemispheres, referred to as \textit{apex} and \textit{antapex}, respectively (see \cite{Hirata2016} and references therein). Both populations of impactors may therefore induce a non-uniform crater distribution. \cite{Hirata2016} found apex-antapex asymmetry for large craters (diameter $D$ larger than $20$ km) and no apparent apex-antapex asymmetry in small craters ($15<D<20$). We assess the significance of these findings, for the stronger hypothesis of uniformity, from his database of $2440$ craters with $D>15$. (The full database contains $3596$ craters, but \cite{Hirata2016}'s analysis only considers those with $D>15$ as the detection of almost all craters above this diameter threshold is guaranteed from the available imagery of Rhea.)

\begin{table}[H]
	\centering
	\footnotesize
	\begin{tabular}{llrrrr}
		\toprule
		Class                            & Name      & Craters &   CvM    &    AD    &    Rt    \\ \midrule
		\multicolumn{1}{l}{\bf Planets}  & Mars      &  $1127$ & $0$ & $0$ & $1\cdot10^{-8}$ \\
		& Venus     &   $881$ & $0.2726$ & $0.2749$ & $0.2806$ \\
		& Mercury   &   $409$ & $0$ & $0$ & $0$ \\ \cmidrule{2-6}
		\multicolumn{1}{r}{\it Dwarf}    & Ceres     &   $115$ & $0.0133$ & $0.0127$ & $0.0150$ \\ \midrule
		\multicolumn{1}{l}{\bf Moons}    & Moon      &  $1578$ & $0$ & $1\cdot10^{-8}$ & $1\cdot10^{-8}$ \\
		& Callisto  &   $141$ & $0$ & $0$ & $3\cdot10^{-8}$ \\
		& Ganymede  &   $129$ & $0.0132$ & $0.0087$ & $0.0184$ \\
		& Europa    &    $41$ & $0.0010$ & $0.0009$ & $0.0010$ \\ \cmidrule{2-6}
		\multicolumn{1}{r}{\it Saturn's} & Rhea      &   $128$ & $0.2793$ & $0.2954$ & $0.2705$ \\
		\multicolumn{1}{r}{\it moons}    & Dione     &    $73$ & $0.5195$ & $0.4989$ & $0.5418$ \\
		& Iapetus   &    $58$ & $0.0034$ & $0.0037$ & $0.0032$ \\
		& Enceladus &    $53$ & $1\cdot10^{-7}$ & $2\cdot10^{-8}$ & $5\cdot10^{-7}$ \\
		& Tethys    &    $50$ & $0.7910$ & $0.8425$ & $0.7199$ \\
		& Mimas     &    $35$ & $0.1701$ & $0.1704$ & $0.1754$ \\
		\bottomrule
	\end{tabular}
	\caption{\small Asymptotic $p$-values of the CvM, AD, and Rt tests when applied to the crater locations of the planets and moons with more than $30$ IUA-named craters.\label{tab:craters}}
\end{table}

\begin{table}[H]
	\centering
	\footnotesize
	\begin{tabular}{lrrrr}
		\toprule
		Diameter  & Craters &                 CvM &        AD &        Rt \\ \midrule
		$15<D<20$ &     $867$ &              $0.1176$ &    $0.0721$ &    $0.1856$ \\
		$D > 20$  &    $1373$ & $2\cdot 10^{-9}$ & $0$ & $3\cdot 10^{-9}$ \\
		$D>15$    &    $2240$ &              $2\cdot 10^{-8}$ & $0$ & $3\cdot10^{-8}$\\
		\bottomrule
	\end{tabular}
	\caption{\small Asymptotic $p$-values of the CvM, AD, and Rt tests when applied to \cite{Hirata2016}'s Rhea crater database. \label{tab:Rhea}}
\end{table}

The tests in Table \ref{tab:Rhea} reveal that uniformity: (\textit{i}) is not rejected for small craters ($15<D<20$) at significance level $5\%$; (\textit{ii}) is emphatically rejected for large craters ($D>20$); (\textit{iii}) is emphatically rejected for all reliable-detected craters ($D>15$). The non-rejection in (\textit{i}) may be attributed to Rhea's ``crater saturation'' \citep{Squyres1997} or to the dominance of planetocentric impactors \citep{Hirata2016}, as the largest craters generated by the debris ejected from large crater impacts is $D\approx 20$ \citep{Alvarellos2005}. In turn, the rejections in (\textit{ii}) and (\textit{iii}) may be explained by the predominantly heliocentric origins of the impactors associated to large craters \citep{Hirata2016}.

\begin{figure}[h]
	\centering
	\includegraphics[width=0.33\textwidth]{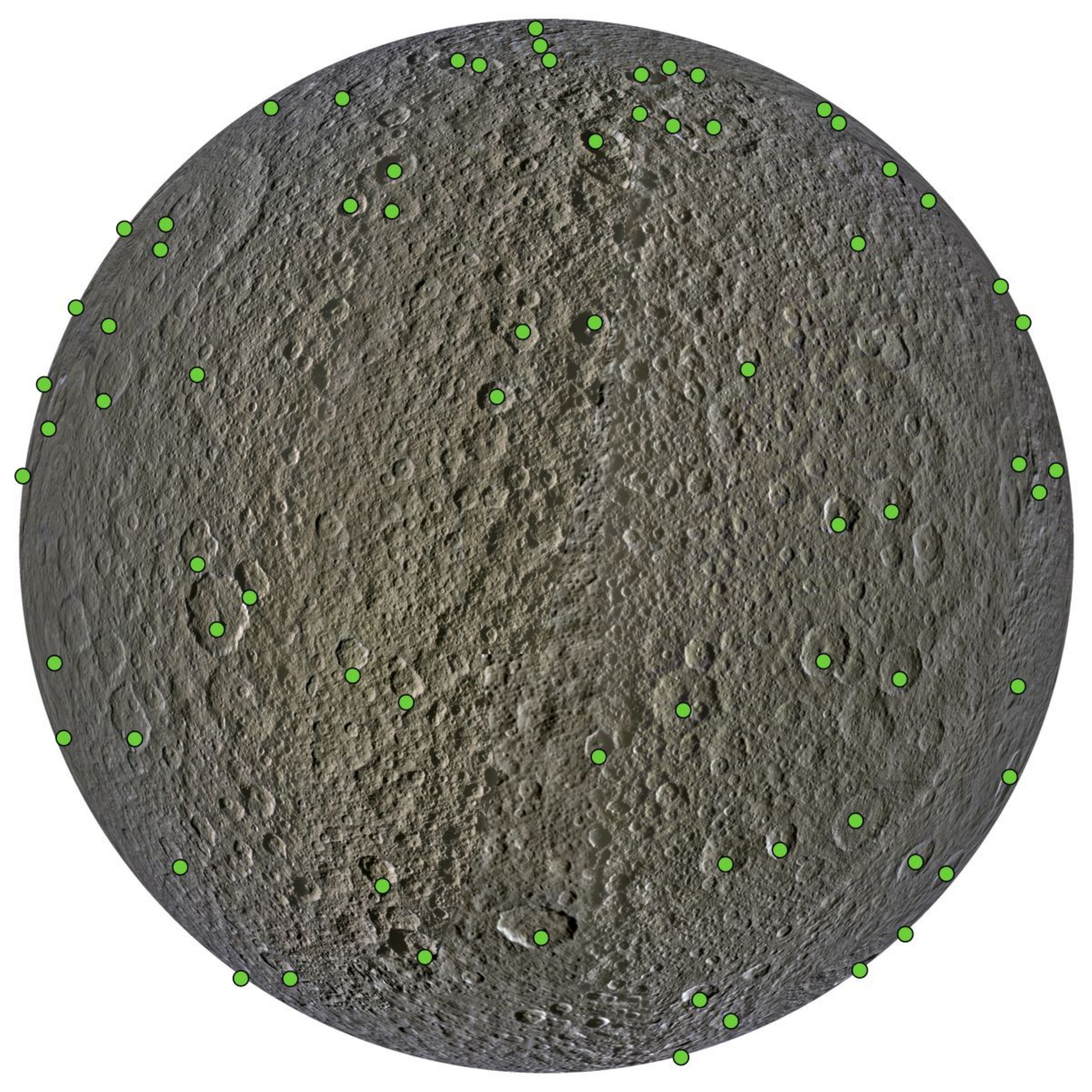}\includegraphics[width=0.33\textwidth]{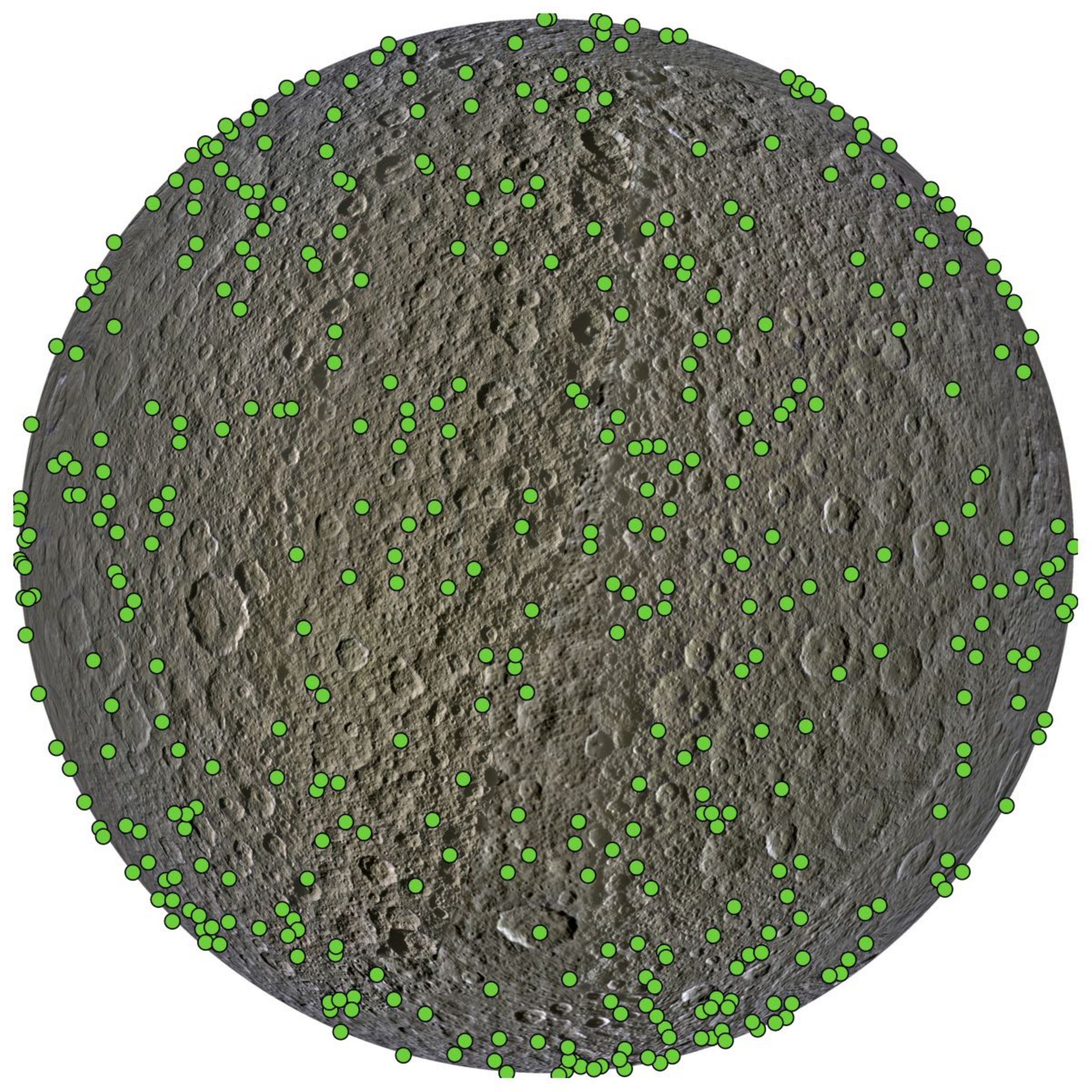}\includegraphics[width=0.33\textwidth]{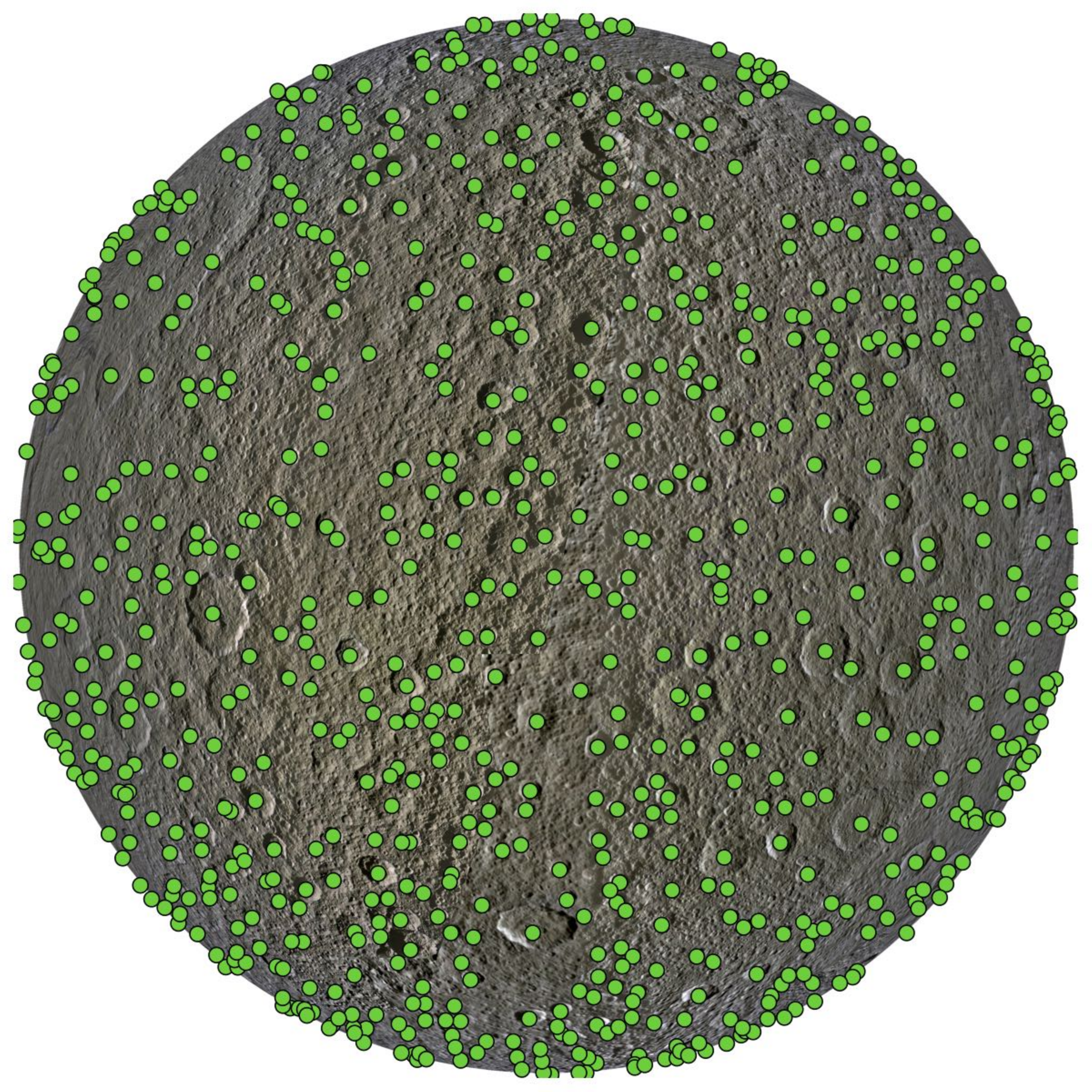}\\
	\includegraphics[width=0.33\textwidth]{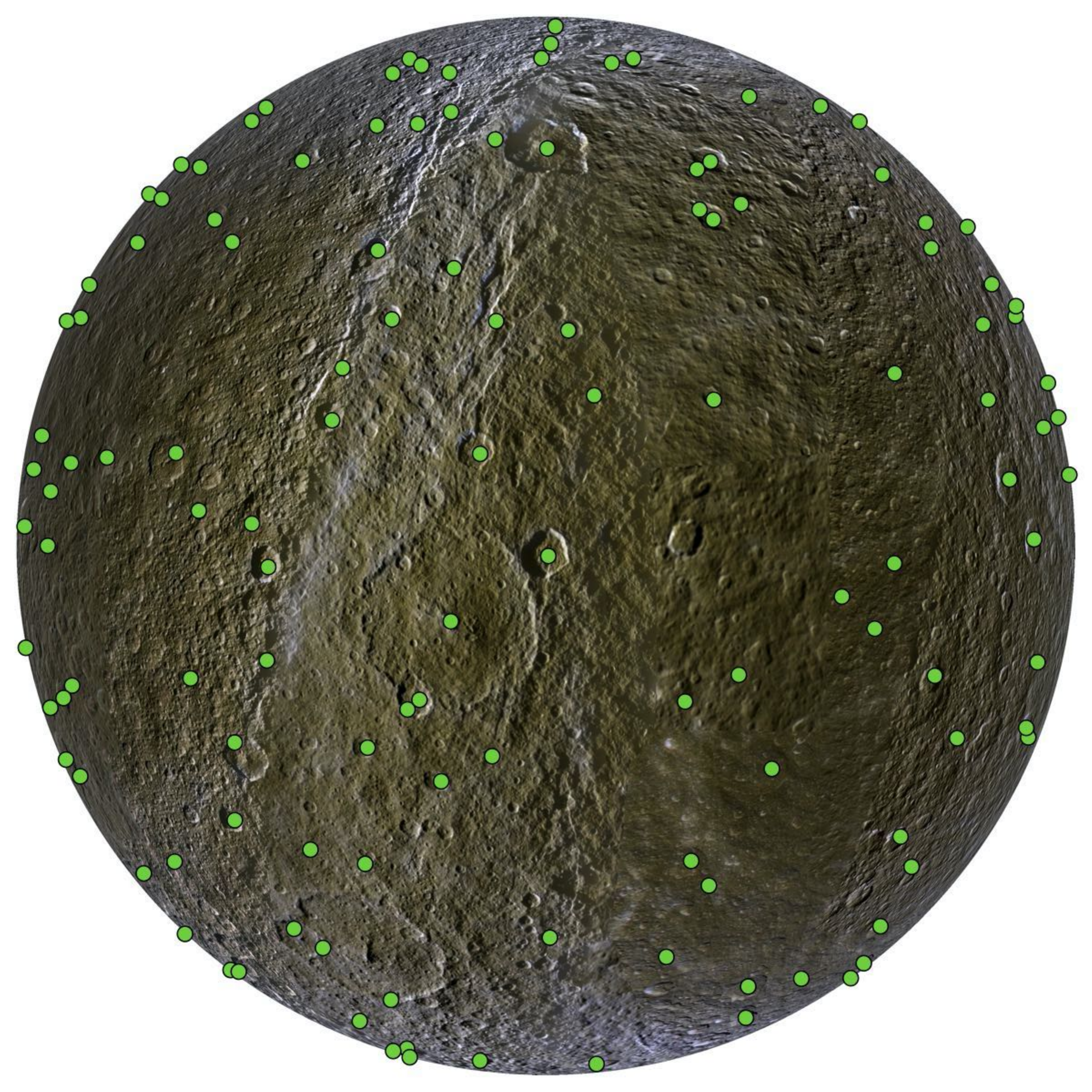}\includegraphics[width=0.33\textwidth]{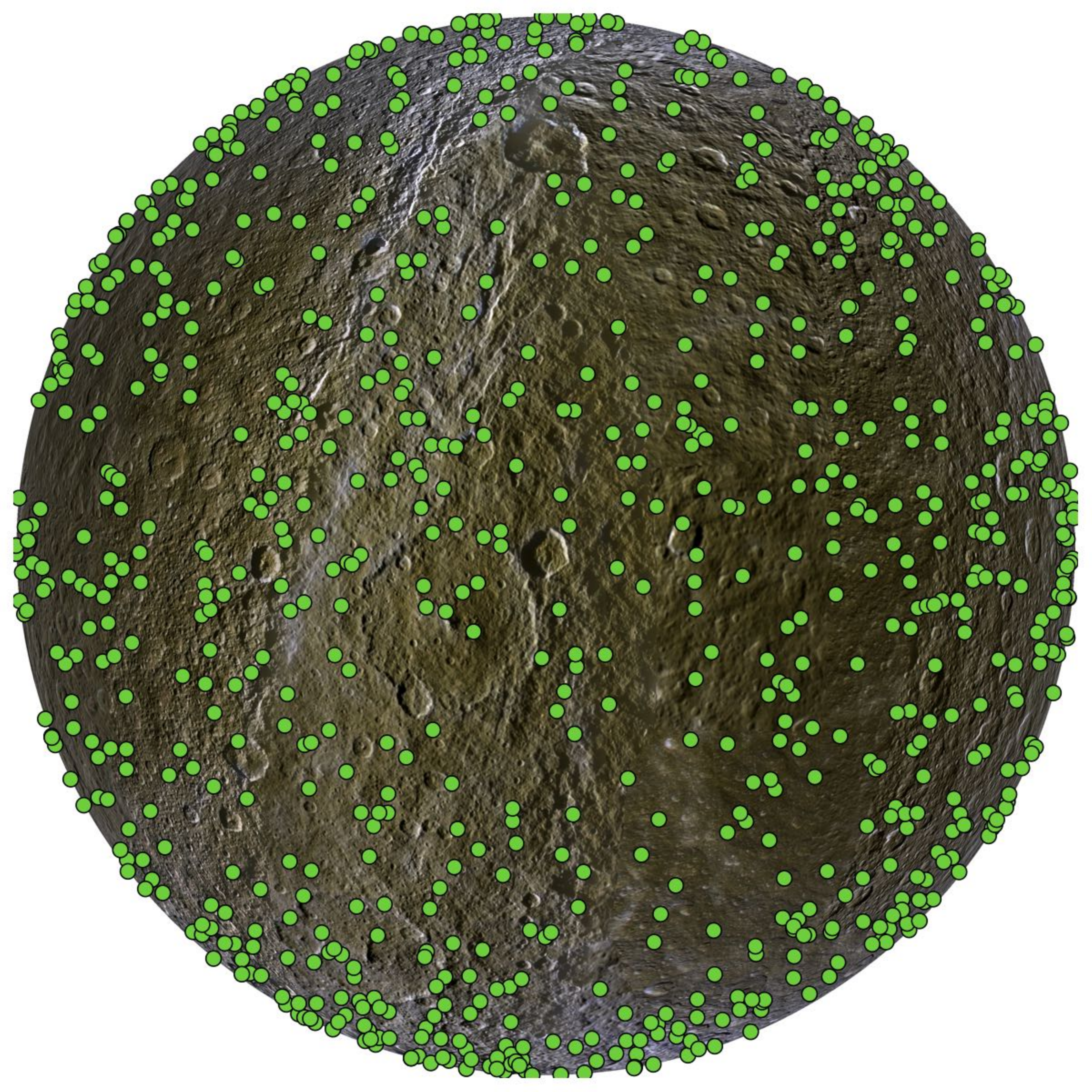}\includegraphics[width=0.33\textwidth]{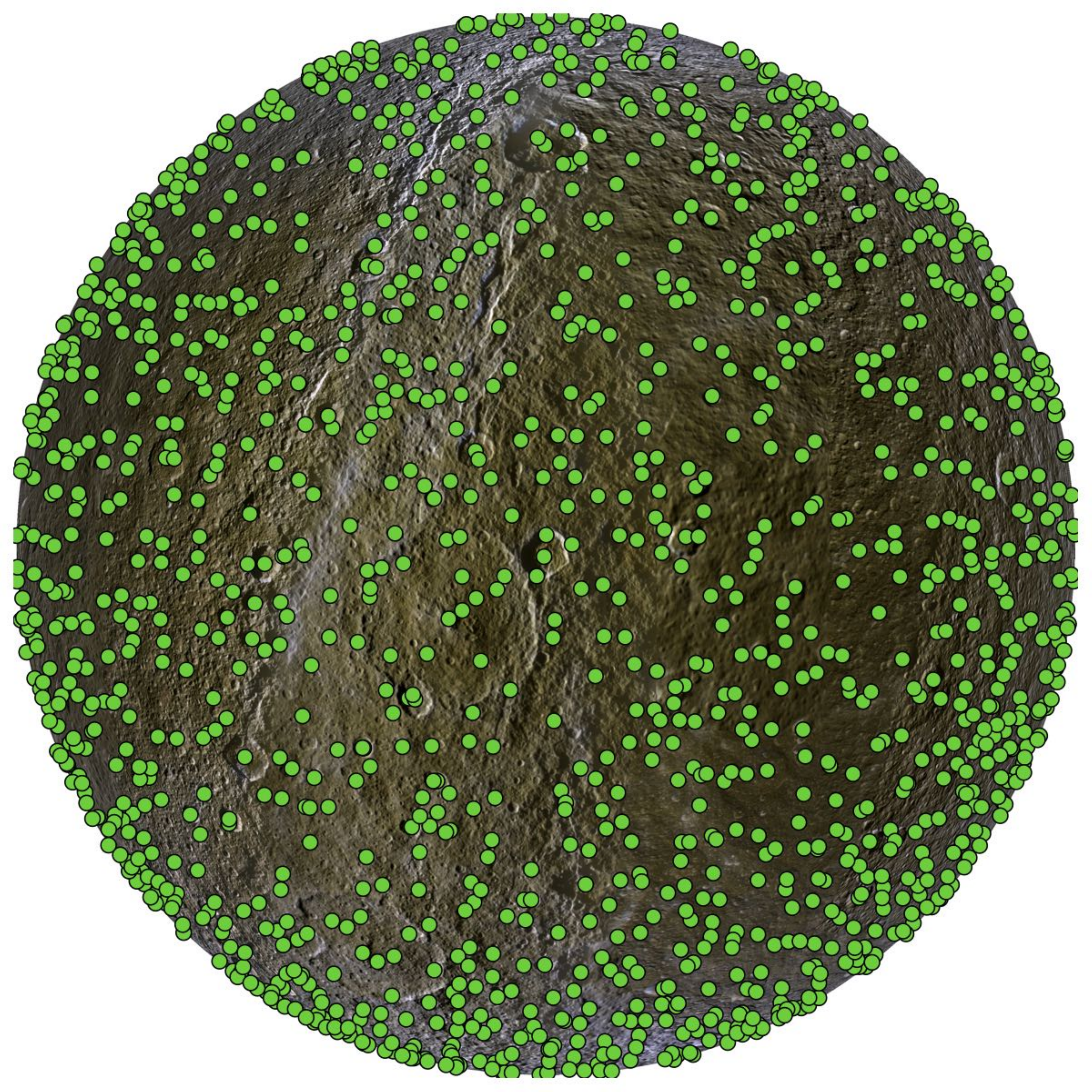}
	\caption{\small Craters on Rhea. The upper and lower rows show the leading and trailing hemispheres, respectively. The north and south poles on each hemisphere correspond to the usual top and bottom positions. From left to right, the columns represent the locations of craters for the IUA-named database, and the \cite{Hirata2016} database for $15<D<20$ and $D>20$, respectively. The locations are superimposed over the PIA 18438 map produced by NASA/JPL-Caltech/Space Science Institute/Lunar and Planetary Institute using data from the Cassini spacecraft. \label{fig:Rhea}}
\end{figure}

\section{Closing remarks}
\label{sec:discussion}

We have introduced the projected-ecdf class of uniformity tests on $\Om{q}$ that bypasses the random exploration of projections in \cite{Cuesta-Albertos2009} and is a close correspondent to the Sobolev class of tests. The novel ecdf optic of the class is pivotal for extending dimension-specific tests and guiding the introduction of the first-ever Anderson--Darling test on $\Om{q}$. Regardless of their generality, the studied projected-ecdf tests have relatively tractable statistics and asymptotic distributions. \\

Empirical evidence showed that the Anderson--Darling test seems to stand out among its competitors due to its competitive performance for unimodal and non-unimodal alternatives. 
The companion \texttt{sphunif} package implements the studied tests and allows the end-to-end replication of the three novel data applications. \\

We conclude mentioning some alternative research directions. A very clear one is to proceed \textit{\`a la} \cite{Escanciano2006} and replace $\nu_q$ by $F_{n,\bga}$ in \eqref{eq:Pn}. This approach avoids the analytically challenging integration on $\Om{q}$, thought at expenses of having an statistic whose computation is $O(n^3)$ and less explicit connection with dimension-specific tests. Another alternative is to replace the CvM norm in \eqref{eq:Pn.W} by the ``$3$-point CvM statistic'' of \cite{Feltz2001}. Finally, goodness-of-fit testing of non-uniform distributions on $\Om{q}$ is possible, yet challenging, by determining the proper substitute for $F_q$ in \eqref{eq:Pn}.

\section*{Supplementary materials}

Supplementary materials provide the proofs of the stated results and contain further simulation results.

\section*{Acknowledgements}

The first author acknowledges financial support from grants PGC2018-097284-B-I00, IJCI-2017-32005 and MTM2016-76969-P, funded by the Spanish Ministry of Economy, Industry and Competitiveness, and FEDER funds. The second and third authors acknowledge financial support from grant MTM2017-86061-C2-2-P from the Spanish Ministry of Economy, Industry and Competitiveness. Comments by Prof. Diego Herranz Muñoz and Josefina F. Ling on the astronomical data applications are kindly acknowledged. The authors gratefully acknowledge the computing resources of the Supercomputing Center of Galicia (CESGA).


\fi

\ifsupplement

\newpage
\title{Supplementary materials for ``On a projection-based class of uniformity tests on the hypersphere''}
\setlength{\droptitle}{-1cm}
\predate{}%
\postdate{}%
\date{}

\author{Eduardo Garc\'ia-Portugu\'es$^{1,2,4}$, Paula Navarro-Esteban$^{3}$, and Juan A. Cuesta-Albertos$^{3}$}

\footnotetext[1]{Department of Statistics, Carlos III University of Madrid (Spain).}
\footnotetext[2]{UC3M-Santander Big Data Institute, Carlos III University of Madrid (Spain).}
\footnotetext[3]{Department of Mathematics, Statistics and Computer Science, University of Cantabria (Spain).}
\footnotetext[4]{Corresponding author. e-mail: \href{mailto:edgarcia@est-econ.uc3m.es}{edgarcia@est-econ.uc3m.es}.}

\maketitle

\begin{abstract}
	These supplementary materials contain four sections. Section \ref{sec:proofs} provides the proofs of the main results of the paper. Section \ref{sec:integral_equations} introduces some required results on integral equations. Section \ref{sec:app_aux_results} gives auxiliary lemmas employed in the proofs. Section \ref{sec_app_powers} contains further simulation results omitted from the main text.
\end{abstract}
\begin{flushleft}
	\small\textbf{Keywords:} Circular data; Directional data; Hypersphere; Sobolev tests; Uniformity. 
\end{flushleft}

\appendix

\section{Proofs}\label{sec:proofs}

\subsection{Proofs of Section \ref{sec:test}}
\label{subsec:proof_sec:test}

\begin{proof}{ of Proposition \ref{prop:sym}}
	A simple change of variables gives
	\begin{eqnarray*}
		\Es{\int_{-1}^{0} \lrp{F_{n,\bga}(x)- F_q(x)}^2\,\mathrm{d}W(F_q(x))}{\bga} 
		=
		\Es{\int_{1}^{0} \lrp{F_{n,\bga}(x^-)- F_q(x)}^2\,\mathrm{d}W(1-F_q(x))}{\bga}
	\end{eqnarray*}
	employing the facts that $F_{n,\bga}(-x)=1-F_{n,-\bga}(x^-)$, $x\in[0,1]$, and that $\bga\sim\nu_q$. However, $F_{n,\bga}(x^-)=F_{n,\bga}(x)$ except for  $x \in \{\bga'\mathbf{X}_1,\ldots,\bga'\mathbf{X}_n\}$. Thus,
	\begin{align*}
	\int_{1}^{0}\lrp{F_{n,\bga}(x^-)- F_q(x)}^2\,\mathrm{d}W(1-F_q(x))
	\neq
	\int_{1}^{0}\lrp{F_{n,\bga}(x)- F_q(x)}^2\,\mathrm{d}W(1-F_q(x))
	\end{align*}
	only if $W(\{F_q(\bga'\mathbf{X}_1),\ldots,F_q(\bga'\mathbf{X}_n)\})>0$. Now, let ${\mathcal D}_W$ denote the points of discontinuity of $x\mapsto W\{[0,x]\}$. This set it at most denumerable, and consequently, 
	\begin{align} \label{eq:term2}
	\Es{\int_{1}^{0}\lrp{F_{n,\bga}(x^-)- F_q(x)}^2\,\mathrm{d}W(1-F_q(x))}{\bga}
	\neq
	\Es{\int_{1}^{0}\lrp{F_{n,\bga}(x)- F_q(x)}^2\,\mathrm{d}W(1-F_q(x))}{\bga}
	\end{align}
	only if 
	$\mathbb{P} \left[
	\bga\in\Om{q}: \{F_q(\bga' \mathbf{X}_1),\ldots,F_q(\bga'\mathbf{X}_n)\} \cap {\mathcal D}_W \neq \emptyset
	\right]>0$. But this probability is bounded by $\sum_{z \in {\mathcal D}_W} 
	\sum_{m=1}^n 
	\mathbb{P} \left\{
	\bga\in\Om{q}: \bga'\mathbf{X}_m =F_q^{-1}(z) 
	\right\}$ and, since each addend represents the probability of $\bga$ belonging to a particular hyperplane and $\bga\sim\nu_q$, the sum equals zero. Inequality \eqref{eq:term2} is therefore impossible.\\
	
	Observing the implicit sign change in $\mathrm{d}W(1-F_q(x))$ and recalling \eqref{eq:Pn.W}, we have that
	\begin{align}\label{eq:proof_defPnqW}
	P_{n,q}^W=2n\Es{\int_{0}^{1}\lrp{F_{n,\bga}(x)-F_q(x)}^2\,\mathrm{d}\tilde{W}(F_q(x))}{\bga}.
	\end{align}
	From \eqref{eq:proof_defPnqW}, undoing the previous change of variables and recalling that $\mathrm{d}\tilde{W}(F_q(x)) = \mathrm{d}\tilde{W}(1-F_q(x))$ by construction, we conclude that $P_{n,q}^W=P_{n,q}^{\tilde{W}}$.
\end{proof}

\begin{proof}{ of Proposition \ref{prop:Aij}}
	The equality $A(\theta,-x)=A(\theta,x)+1-2F_q(x)$ follows from \eqref{eq:aij} and the symmetry of $\nu_q$. 
	Assume $x\geq 0$. If $q=1$, by Lemma \ref{lem:A_q1}, 
	\begin{align}
	A(\theta,x)
	&=1-\frac{2\cos^{-1}(x)}{\pi}+A^*(\theta,x)\notag
	=\begin{cases}
	1-\frac{2\cos^{-1}(x)}{\pi}, & 0\leq \theta< 2\cos^{-1}(x),\\
	1-\frac{2\cos^{-1}(x)}{\pi}+\frac{2\cos^{-1}(x)-\theta}{2\pi}, &  2\cos^{-1}(x)\leq\theta\leq\pi.
	\end{cases}
	\end{align}
	If $q\geq 2$, by \eqref{eq:aij} we have that
	\begin{align}
	A(\theta,0)&=2\int_{-1}^{0}f_q(t)\int_{-1}^1 1_{\lrb{ u\leq t/(1-t^2)^{1/2}\tan\lrp{\theta/2}}}f_{q-1}(u)\,\mathrm{d}u\,\mathrm{d}t\notag\\
	&=2\int_{-\cos\lrp{\theta/2}}^{0}f_q(t)F_{q-1}\lrp{ \frac{t\tan\lrp{\theta/2}}{(1-t^2)^{1/2}}}\,\mathrm{d}t\notag\\
	&=\frac{1}{2}-\frac{\theta}{2\pi},\label{eq:A_0_theta}
	\end{align}
	where \eqref{eq:A_0_theta} is due to Lemma \ref{lemma:aux}. The result follows from \eqref{eq:A_0_theta}, Lemma \ref{lemma:aux}, the equality 
	\begin{align*}
	A(\theta,x)&=A(\theta,0)+2\int_{0}^{x}f_q(t)\int_{-1}^1 1_{\lrb{ u\leq t/(1-t^2)^{1/2}\tan\lrp{\theta/2}}}f_{q-1}(u)\,\mathrm{d}u\,\mathrm{d}t,
	\end{align*}
	and by taking into account that $t/(1-t^2)^{1/2}\tan\lrp{\theta/2}\leq 1$ when $t\in [0,\cos\lrp{\theta/2}]$ and $t/(1-t^2)^{1/2}\tan\lrp{\theta/2}\geq 1$ when $t\in [\cos\lrp{\theta/2},1]$.
\end{proof} 

\begin{proof}{ of Proposition \ref{prop:psi_q:w}}
	From \eqref{eq:pcvm_aux}, we trivially have 
	\begin{align}
	P_{n,q}^W=\frac{2}{n}\sum_{i<j} \psi^W_q(\theta_{ij})+\int_{0}^{1}u(1-nu)\,\mathrm{d}W(u).\label{eq:prop:psi_q:w:1}
	\end{align}
	We separate the first addend of \eqref{eq:prop:psi_q:w:1} for the cases $q=1$ and $q\geq2$. For $q=1$, 
	direct integration of $\int_0^\pi A(\theta,\cos(\alpha))\,\mathrm{d}W(F_1(\cos(\alpha)))$ using Lemma \ref{lem:A_q1} gives
	\begin{align*}
	\psi^w_1(\theta)=& -1+2\int_0^1 u\,\mathrm{d}W(u)+\int_{\frac{\theta}{2\pi}}^{1-\frac{\theta}{2\pi}}\lrp{1-u-\tfrac{\theta}{2\pi}}\,\mathrm{d}W(u)+\int_0^{\frac{\theta}{2\pi}}(1-2u)\,\mathrm{d}W(u)\\
	=&\;1-\int_{\frac{\theta}{2\pi}}^1 W(u)\,\mathrm{d}u-\int_{1-\frac{\theta}{2\pi}}^1 W(u)\,\mathrm{d}u, \end{align*}
	where the second equality follows because $W$ is a cdf. The desired result is deduced after recalling $W(1-t)=1-W(t)$ for $t\in[0,1]$ and $\int_0^1W(u)\,\mathrm{d}u=1/2$.
	
	For $q\ge 2$, simple expressions for $A(\theta_{ij},x)$ are not easy to obtain. We follow thus an alternative approach. First, note that by symmetry
	\begin{align*}
	\omega_q&\lrp{\{\bga\in\Om{q}:\bga'\bX_i\leq x,\bga'\bX_j\leq x\}}\\
	&=\omega_q\lrp{\{\bga\in\Om{q}: \bga'\bX_i\geq \bga'\bX_j, \bga'\bX_i\leq x\}\cup \{\bga\in\Om{q}: \bga'\bX_j\geq \bga'\bX_i, \bga' \bX_j\leq x\}}\\
	&=2\omega_q\lrp{\{\bga\in\Om{q}: \bga'\bX_i\geq \bga'\bX_j, \bga'\bX_i\leq x\}}
	\end{align*}
	and hence
	\begin{align}
	\int_{-1}^{1}A(\theta_{ij},x)\,\mathrm{d}W(F_q(x))=\frac{2}{\omega_q}\int_{-1}^1\int_{\Om{q}}1_{\lrb{\bga'\bX_i\geq \bga'\bX_j, \bga'\bX_i\leq x}}\,\omega_q(\mathrm{d}\bga)\,\mathrm{d}W(F_q(x)). \label{eq:Aij_q2}
	\end{align}
	Consider now the two successive tangent-normal decompositions:
	\begin{align}
	\begin{cases}
	\bga=t\bX_i+(1-t^2)^{1/2}\bB_{\bX_i,q+1}\bxi,\\ \omega_q(\mathrm{d}\bga)=(1-t^2)^{q/2-1}\,\mathrm{d}t\,\omega_{q-1}(\mathrm{d}\bxi),
	\end{cases}
	\quad t\in[-1,1],\,\bxi\in\Om{q-1} \label{eq:tnd1}
	\end{align}
	and 
	\begin{align}
	\begin{cases}
	\bxi=u\bbeta_{ij}+(1-u^2)^{1/2}\bB_{\bbeta_{ij},q}\bzeta,\\
	\omega_{q-1}(\mathrm{d}\bxi)=(1-u^2)^{(q-3)/2}\,\mathrm{d}u\,\omega_{q-2}(\mathrm{d}\bzeta),
	\end{cases}
	\quad u\in[-1,1],\,\bzeta\in\Om{q-2}, \label{eq:tnd2}
	\end{align}
	where $\bB_{\bx,p}$ denotes a semi-orthonormal matrix $p\times (p-1)$ such that $\bB_{\bx,p}\bB_{\bx,p}'=\bI_{p}-\bx\bx'$ and $\bB_{\bx,p}'\bB_{\bx,p}=\bI_{p-1}$ for $\bx\in\Om{p-1}$, and $\bbeta_{ij}:=\bB_{\bX_i,q+1}'\bX_j(1-(\bX_i'\bX_j)^2)^{-1/2}\in\Om{q-1}$. Plugging-in \eqref{eq:tnd1} and \eqref{eq:tnd2} in \eqref{eq:Aij_q2} gives
	\begin{align}
	\int_{-1}^{1}&\,A(\theta_{ij},x)\,\mathrm{d}W(F_q(x))\notag\\
	&=2\frac{\omega_{q-2}}{\omega_{q}}\int_{-1}^{1}\int_{-1}^1\lrb{\int_{-1}^11_{\lrb{t\geq t\cos(\theta_{ij})+u(1-t^2)^{1/2}\sin(\theta_{ij}),t\leq x}}\,\mathrm{d}W(F_q(x))}\notag\\
	&\qquad\times(1-t^2)^{q/2-1}(1-u^2)^{(q-3)/2}\,\mathrm{d}t\,\mathrm{d}u,\notag\\
	&=2\int_{-1}^{1}H_q(t)\lrb{\int_{-1}^11_{\lrb{t\cos(\theta_{ij})+u(1-t^2)^{1/2}\sin(\theta_{ij})\leq t}}\,\mathrm{d}F_{q-1}(u)}\,\mathrm{d}F_q(t)\label{eq:psi_q2}\\
	&=:\psi^W_q(\theta_{ij})\notag
	\end{align}
	since $\int_{t}^1\,\mathrm{d}W(F_q(x))=1-W(F_q(t))=:H_q(t)$. Now, for $\theta\in[0,\pi]$ and $t,u\in[-1,1]$,
	\begin{align*}
	t\cos(\theta)+u(1-t^2)^{1/2}\sin(\theta)\leq t \iff u\leq\frac{t\tan\lrp{\theta/2}}{\lrp{1-t^2}^{1/2}}=:u(\theta,t).
	\end{align*}
	Note that it can occur that $\abs{u(\theta,t)}>1$, so in the integral limits of \eqref{eq:psi_q2} we rather handle
	\begin{align*}
	\tilde{u}(\theta,t):=((u(\theta,t)\vee-1)\wedge1)=\begin{cases}
	1,&\cos\lrp{\tfrac{\theta}{2}}\leq t\leq 1,\\
	u(\theta,t),&-\cos\lrp{\tfrac{\theta}{2}}< t< \cos\lrp{\tfrac{\theta}{2}},\\
	-1,&-1\leq t\leq-\cos\lrp{\tfrac{\theta}{2}}.
	\end{cases}
	\end{align*}
	Therefore, \eqref{eq:psi_q2} becomes
	\begin{align}
	\psi^W_q(\theta)=&\;2\int_{-1}^{1}H_q(t)\bigg\{\int_{-1}^{\tilde{u}(\theta,t)}\,\mathrm{d}F_{q-1}(u)\bigg\}\,\mathrm{d}F_q(t)\notag
	\\
	=&\;2\int_{\cos(\theta/2)}^{1}\bigg\{H_q(t)\int_{-1}^{1}\,\mathrm{d}F_{q-1}(u)\notag\bigg\}\,\mathrm{d}F_q(t)+2\int_{-\cos(\theta/2)}^{\cos(\theta/2)}H_q(t)\bigg\{\int_{-1}^{u(\theta,t)}\,\mathrm{d}F_{q-1}(u)\notag\bigg\}\,\mathrm{d}F_q(t)\notag\\
	=&\;2\int_{\cos(\theta/2)}^{1}H_q(t)\,\mathrm{d}F_q(t)+2\int_{-\cos(\theta/2)}^{\cos(\theta/2)}H_q(t)F_{q-1}\lrp{\frac{t\tan\lrp{\theta/2}}{(1-t^2)^{1/2}}}\mathrm{d}F_q(t).\label{eq.psiq.final}
	\end{align}
	The proposition is proved by applying Lemma \ref{lemma:aux} to the second integral and recalling that $W(1-t)=1-W(t)$ for $t\in[0,1]$ and $\int_0^1W(u)\,\mathrm{d}u=1/2$.
\end{proof}

\begin{proof}{ of Proposition \ref{coro:psi_q:CvM}}
	The second term in \eqref{eq:prop:psi_q:w:1} and the expression for $\psi^\mathrm{CvM}_1$ follow trivially. For $\psi_q^\mathrm{CvM}$, $q\geq 2$, recall that $\psi_q^\mathrm{CvM}$ can be written as
	\begin{align*}
	\psi^\mathrm{CvM}_q(\theta)
	%
	=&-\frac{1}{4}+\frac{\theta}{2\pi}+4\int_0^{\cos(\theta/2)}F_q(t)\lrp{1-F_{q-1}\lrp{\frac{t\tan\lrp{\theta/2}}{(1-t^2)^{1/2}}}}\mathrm{d}F_q(t),
	\end{align*}
	from which the desired expression of $\psi_q^\mathrm{CvM}$ for $q\geq 2$ follows.
	When $q=2$,
	\begin{align*}
	\psi_2^\mathrm{CvM}(\theta)%
	%
	=&-\frac{1}{4}+\frac{\theta}{2\pi}+\frac{\tan^2\lrp{\theta/2}}{\pi}\int_0^1\lrp{\frac{y}{\lrp{y^2+\tan^2\lrp{\theta/2}}^{1/2}}+1}\frac{\cos^{-1}(y)}{\lrp{y^2+\tan^2\lrp{\theta/2}}^{3/2}}\,\mathrm{d}y\\
	=&-\frac{1}{4}+\frac{\theta}{2\pi}+\frac{1}{4\pi}\lrp{\pi-\frac{\pi\tan\lrp{\theta/2}}{\lrp{1+\tan^2\lrp{\theta/2}}^{1/2}}+4\tan^{-1}\lrp{\frac{1}{\tan\lrp{\theta/2}}}}\\
	%
	%
	%
	%
	=&\;\frac{1}{2}-\frac{1}{4}\sin\lrp{\tfrac{\theta}{2}}.
	\end{align*}
	If $q=3$, then
	\begin{align}
	\psi_3^\mathrm{CvM}(\theta)%
	%
	=& -\frac{3}{4}+\frac{\theta}{2\pi}+2F_3^2\lrp{\cos\lrp{\tfrac{\theta}{2}}}-2\int_0^{\cos(\theta/2)}F_3(t)\lrp{1+\frac{t\tan\lrp{\theta/2}}{(1-t^2)^{1/2}}} \, \mathrm{d}F_3(t) \notag\\
	=& -\frac{1}{2}+\frac{\theta}{2\pi}+F_3^2\lrp{\cos\lrp{\tfrac{\theta}{2}}}-\frac{2}{\pi}\cos\lrp{\tfrac{\theta}{2}}\sin\lrp{\tfrac{\theta}{2}}\notag\\
	& -\frac{4}{\pi^2}\tan\lrp{\tfrac{\theta}{2}}\int_0^{\cos(\theta/2)}\lrp{t^2(1-t^2)^{1/2}-t\cos^{-1}(t)}\, \mathrm{d}t, \label{eq:psi_q_proof_2}
	\end{align}
	with
	\begin{align}
	&\int_0^{\cos(\theta/2)}\lrp{t^2(1-t^2)^{1/2}-t\cos^{-1}(t)}\,\mathrm{d}t\notag\\%
	&\qquad=\frac{1}{4}\lrb{\cos^3\lrp{\tfrac{\theta}{2}}\sin\lrp{\tfrac{\theta}{2}}-\theta\cos^2\lrp{\tfrac{\theta}{2}}+\frac{1}{2}\cos\lrp{\tfrac{\theta}{2}}\sin\lrp{\tfrac{\theta}{2}}-\frac{1}{2}\sin^{-1}\lrp{\cos\lrp{\tfrac{\theta}{2}}}}.\label{eq:psi_q_proof_3}
	\end{align}
	Since $F_3\lrp{\cos\lrp{\tfrac{\theta}{2}}}=1+\left.\lrp{\cos\lrp{\tfrac{\theta}{2}}\sin\lrp{\tfrac{\theta}{2}}-\tfrac{\theta}{2}}\middle/\pi\right.$ due to \eqref{eq:Fqrec}, the third term in \eqref{eq:psi_q_proof_2} results
	\begin{align}
	F_3^2\lrp{\cos\lrp{\tfrac{\theta}{2}}}
	=1+\frac{\theta^2}{4\pi^2}-\frac{\theta}{\pi}+\frac{1}{\pi}\lrp{2-\theta}\cos\lrp{\tfrac{\theta}{2}}\sin\lrp{\tfrac{\theta}{2}}+\frac{1}{\pi^2}\cos^2\lrp{\tfrac{\theta}{2}}\sin^2\lrp{\tfrac{\theta}{2}}.\label{eq:psi_q_proof_4}
	\end{align}
	The expression for $\psi_3^\mathrm{CvM}$ arises from combining \eqref{eq:psi_q_proof_2}, \eqref{eq:psi_q_proof_3}, and \eqref{eq:psi_q_proof_4}.
\end{proof}

\begin{proof}{ of Proposition \ref{prop:comp:Roth}}
	The statistic $R_{n,t}$ admits the representation \eqref{eq:sob} for a certain sequence $\{v_{k,1}\}$. Following the arguments in \cite{Watson1967}, \cite{Rothman1972} considered the Fourier expansion of $N(\alpha,t)-nt$ that, adapted to a circular sample $\Theta_1,\ldots,\Theta_n\in[0,2\pi)$, is given by $N(\alpha,t)-nt=\sum_{k=1}^\infty a_k\cos(k\alpha)+b_k\sin(k\alpha)$, where
	\begin{align*}
	a_k=\frac{1}{\pi k}\sum_{i=1}^n[\sin (k\Theta_i)-\sin (k\Theta_i-2\pi k t)],\; b_k=\frac{1}{\pi k}\sum_{i=1}^n[\cos (k\Theta_i-2\pi k t)-\cos (k\Theta_i)].
	\end{align*}
	From the Fourier expansion it readily follows that 
	\begin{align}
	R_{n,t}=\frac{1}{2 n}\sum_{k=1}^\infty [a_k^2+b_k^2]. \label{eq:Rntgood}
	\end{align}
	Expanding the squares and using the cosine addition formula gives
	\begin{align*}
	R_{n,t}&=\frac{1}{2 n}\sum_{i,j=1}^n\sum_{k=1}^\infty\frac{1}{(\pi k)^2}\left\{2\cos(k(\Theta_i-\Theta_j))-\cos(k(\Theta_i-\Theta_j)+2\pi kt)-\cos(k(\Theta_j-\Theta_i)+2\pi kt)\right\}\\
	&=\frac{2}{n}\sum_{i,j=1}^n\sum_{k=1}^\infty \frac{\sin^2(k\pi t)}{(\pi k)^2}\cos(k(\Theta_i-\Theta_j)),
	\end{align*}
	where the last equality follows from the basic trigonometric identities $\cos(x+y)+\cos(x-y)=2\cos(x)\cos(y)$ and $1-\cos(2k\pi t)=2\sin^2(k\pi t)$. As a consequence, $v_{k,1}=\sin(k\pi t)/(\pi k)$ and $R_{n,t}=\frac{1}{n}\sum_{i, j=1}^nh_t(\theta_{ij})$ with 
	\begin{align*}
	h_t(\theta)=2\sum_{k=1}^\infty \frac{\sin^2(k\pi t)}{(\pi k)^2}\cos(k\theta)=\tilde{h}(\theta)-\tfrac{1}{2}\tilde{h}(\theta-2\pi t)-\tfrac{1}{2}\tilde{h}(\theta+2\pi t), \quad \theta\in[0,\pi],
	\end{align*}
	where $\tilde{h}(x):=\sum_{k=1}^\infty \frac{1}{k^2}\cos(kx)=\frac{1}{4}\{(x\mod2\pi-\pi)^2-\tfrac{\pi^2}{3}\}$, for $x\in\mathbb{R}$.
	A case-by-case analysis of the possible values of $\theta\in[0,\pi]$ and $t\in(0,1)$ gives
	\begin{align*}
	h_t(\theta)=\begin{cases}
	\frac{1}{2\pi}(2\pi t(1-t)-\theta),&\text{ if }\theta\in[0,2\pi t_m)\\
	-t_m^2,&\text{ if }\theta\in[2\pi t_m,\pi)\\
	\end{cases}=-\min\lrp{\tfrac{\theta}{2\pi}-t(1-t),t_m^2}
	\end{align*}
	and hence proves \eqref{eq:Rothcomp}. 
\end{proof}

\begin{remark}\label{rem:4}
	Somehow imprecise computational recipes for \eqref{eq:Roth} seem to have proliferated in the literature. We report them here for the benefit of future readers. \cite{Rothman1972} provided in his equation (35) a computational form for $R_{n,t}$, but relying on an undefined notation and without hints to its derivation. Employing his equation (8) produces a factor $2$ difference with respect to the statistic defined in his equation (1), since (8) misses $\frac{1}{2}=\int_0^{1}\cos^2(2k\pi x)\,\mathrm{d}x=\int_0^{1}\sin^2(2k\pi x)\,\mathrm{d}x$ (that does appear in our equation \eqref{eq:Rntgood}). Differently, equation (6.3.63) in \cite{Mardia2000} states that the sequence $\{v_{k,1}\}$ of $R_{n,t}$ (see Section \ref{sec:Sobolev_review}) is $\sin(k\pi t)/(2k\pi t)$, which does not correspond to the statistic as defined in their equation (6.3.50) (equal to our equation \eqref{eq:Roth}). These two misprints introduce new statistics that are proportional to the original definition of $R_{n,t}$ and thus yield the same test decision. However, they may induce spurious test outcomes if the asymptotic distribution of one version is employed with the statistic of another.
\end{remark}

\begin{proof}{ of Proposition \ref{prop:psi_q:Roth}}
	The second addend in \eqref{eq:prop:psi_q:w:1} is 
	\begin{align}
	\int_0^1u(1-nu)\,\mathrm{d}W_t(u)
	=\frac{1}{2}(1-n)+nt(1-t).\label{eq:bias_Rt}
	\end{align}
	Denote $\bar{\theta}:=\theta/(2\pi)$. If $q=1$, the first addend in \eqref{eq:prop:psi_q:w:1} is 
	\begin{align}
	\psi_1^{\mathrm{R}_t}(\theta)=&\;\frac{1}{2}-\bar{\theta}+\int_0^{\bar{\theta}} \lrp{1_{\{t_m\leq u\}}+1_{\{1-t_m\leq u\}}}\,\mathrm{d}u\notag\\
	=&\;\frac{1}{2}-\bar{\theta}+\lrp{\bar{\theta}-t_m}_++\lrp{\bar{\theta}-(1-t_m)}_+\notag\\
	=&\;\frac{1}{2}-\bar{\theta}+\lrp{\bar{\theta}-t_m}_+\label{eq:psi_Rt}.
	\end{align}
	%
	%
	%
	The proof for $q=1$ readily follows from \eqref{eq:bias_Rt} and  \eqref{eq:psi_Rt}. For $q\geq 2$, by Proposition \ref{prop:psi_q:w},
	\begin{align}
	\psi^{\mathrm{R}_t}_q(\theta)=&-\frac{1}{2}+\bar{\theta}+2\int_{0}^{1/2}W_t(u)\,\mathrm{d}u+4\int_0^{\cos(\theta/2)}W_t(F_q(u))\lrp{1-F_{q-1}\lrp{\frac{u\tan\lrp{\theta/2}}{(1-u^2)^{1/2}}}}\,\mathrm{d}F_q(u)\notag\\
	%
	%
	%
	=&-t_m+\frac{1}{2}+2\lrp{F_q\lrp{\cos\lrp{\tfrac{\theta}{2}}}-1+t_m}_+-2F_q\lrp{\cos\lrp{\tfrac{\theta}{2}}}-\bar{\theta}+\frac{3}{2}\notag\\
	&+2\int_0^{\cos(\theta/2)\wedge F_q^{-1}(1-t_m)}F_{q-1}\lrp{\frac{u\tan\lrp{\theta/2}}{(1-u^2)^{1/2}}}\mathrm{d}F_q(u)
	\label{eq:Rt_aux1},
	\end{align} 
	where the last equality follows from Lemma \ref{lemma:aux}. If $\cos(\theta/2)<F_q^{-1}(1-t_m)$, from \eqref{eq:Rt_aux1} and using Lemma \ref{lemma:aux}, then
	\begin{align*}
	\psi^{\mathrm{R}_t}_q(\theta)=-t_m+\frac{1}{2}-2F_q(\cos(\theta/2))-\bar{\theta}+\frac{3}{2}+2F_q(\cos(\theta/2))+\bar{\theta}-\frac{3}{2}=\frac{1}{2}-t_m.
	\end{align*}
	If $\cos(\theta/2)>F_q^{-1}(1-t_m)$, from \eqref{eq:Rt_aux1}, 
	\begin{align*}
	\psi^{\mathrm{R}_t}_q(\theta)=&\;t_m+2+2F_q\lrp{\cos\lrp{\tfrac{\theta}{2}}}-2-2F_q\lrp{\cos\lrp{\tfrac{\theta}{2}}}-\bar{\theta}+2\int_0^{F_q^{-1}(1-t_m)}F_{q-1}\lrp{\frac{u\tan\lrp{\theta/2}}{(1-u^2)^{1/2}}}\mathrm{d}F_q(u)\\
	=&\;t_m-\frac{3}{2}+2F_q\lrp{\cos\lrp{\tfrac{\theta}{2}}}-2\int_{F_q^{-1}(1-t_m)}^{\cos(\theta/2)}F_{q-1}\lrp{\frac{u\tan\lrp{\theta/2}}{(1-u^2)^{1/2}}}\mathrm{d}F_q(u),
	\end{align*}
	where the last equality follows from Lemma \ref{lemma:aux}.\\
	
	Take now $q=2$. Then $F_2^{-1}(1-t_m)=1-2t_m$ and, in addition,
	\begin{align*}
	2\int_0^{ F_2^{-1}(1-t_m)}F_{1}\lrp{\frac{u\tan\lrp{\theta/2}}{(1-u^2)^{1/2}}}\,\mathrm{d}F_2(u)
	=&\;1-2t_m+\frac{\theta-\pi}{2\pi}\\
	&-\cos^{-1}\lrp{\frac{(1-2t_m)\tan(\theta/2)}{2\lrp{t_m(1-t_m)}^{1/2}}}\frac{(1-2t_m)}{\pi}\\
	&+\frac{1}{\pi}\tan^{-1}\lrp{\frac{\lrp{\cos^2\lrp{\theta/2}-(1-2t_m)^2}^{1/2}}{\sin\lrp{\theta/2}}}.
	\end{align*}
	Hence, if $\cos(\theta/2)>F_2^{-1}(1-t_m)$, from \eqref{eq:Rt_aux1}, 
	\begin{align*}
	\psi^{\mathrm{R}_t}_2(\theta)=&\;t_m+\pi+\lrp{1+\frac{1}{2\pi}}\theta+2(1-2t_m)\cos^{-1}\lrp{\frac{(1/2-t_m)\tan\lrp{\theta/2}}{\lrp{t_m(1-t_m)}^{1/2}}}(1-2t_m)\\
	&-2\tan^{-1}\lrp{\frac{\lrp{\cos^2\lrp{\theta/2}-(1-2t_m)^2}^{1/2}}{\sin\lrp{\theta/2}}}.
	\end{align*}
	
	If $q=3$, note that 
	\begin{align*}
	2\int_0^{F_3^{-1}(1-t_m)}F_{2}\lrp{\frac{u\tan\lrp{\theta/2}}{(1-u^2)^{1/2}}}\,\mathrm{d}F_3(u)=&\;\frac{2}{\pi}\int_0^{F_3^{-1}(1-t_m)}\lrp{\tan\lrp{\tfrac{\theta}{2}}u+(1-u^2)^{1/2}}\mathrm{d}u\\
	=&\;\frac{\tan\lrp{\theta/2}}{\pi}\lrp{F_3^{-1}(1-t_m)}^2\\
	&+\frac{1}{\pi}\sin^{-1}\lrp{F_3^{-1}(1-t_m)}\\
	&+\frac{1}{\pi}F_3^{-1}(1-t_m)\lrp{1-\lrp{F_3^{-1}(1-t_m)}^2}^{1/2}.
	\end{align*}
	Then, if $\cos(\theta/2)>F_3^{-1}(1-t_m)$, from \eqref{eq:Rt_aux1}, we have
	\begin{align*}
	\psi^{\mathrm{R}_t}_3(\theta)=&\;t_m-\frac{\theta}{2\pi}+\frac{\tan(\theta/2)}{\pi}\lrp{F_3^{-1}(1-t_m)}^2+\frac{1}{\pi}\sin^{-1}\lrp{F_3^{-1}(1-t_m)}\\
	&+\frac{1}{\pi}F_3^{-1}(1-t_m)\lrp{1-\lrp{F_3^{-1}(1-t_m)}^2}^{1/2}.
	\end{align*}
	The proof ends recalling $F_3^{-1}(1-t_m)=\cos(\theta_{t_m}/2)$ from \eqref{eq:Fqrec} and the definition of $\theta_{t_m}$.
\end{proof}

\begin{proof}{ of Proposition \ref{prop:psi_q:AD}}
	For $q=1$, then by \eqref{eq:pcvm2}:
	\begin{align*}
	P_{n,1}^\mathrm{AD}=&\;\frac{1}{\pi}\int_{0}^{\pi} \bigg\{\frac{1}{n}\sum_{i\neq j}A(\theta_{ij},\cos(\alpha))+F_1(\cos(\alpha))(1-nF_1(\cos(\alpha)))\bigg\}w(F_1(\cos(\alpha)))\,\mathrm{d}\alpha\\
	=&\;\frac{2}{n}\sum_{i<j}\frac{1}{\pi}\int_0^\pi\lrb{\frac{A(\theta_{ij},\cos(\alpha))}{F_1(\cos(\alpha))}+\frac{1}{n-1}(1-nF_1(\cos(\alpha)))}\frac{1}{1-F_1(\cos(\alpha))}\,\mathrm{d}\alpha.
	\end{align*}
	Since $F_1(\cos(\alpha))=1-\alpha/\pi$, then
	\begin{align}
	P_{n,1}^\mathrm{AD}=&\;\frac{2}{n}\sum_{ i<j}\int_0^\pi\lrp{\frac{\pi}{\alpha(\pi-\alpha)}A(\theta_{ij},\cos(\alpha))+\frac{1-n}{(n-1)\alpha}+ \frac{n}{(n-1)\pi}}\,\mathrm{d}\alpha\notag\\
	=&\;n+\frac{2}{n}\sum_{ i<j}\int_0^\pi\lrp{\frac{\pi}{\alpha(\pi-\alpha)}A(\theta_{ij},\cos(\alpha))-\frac{1}{\alpha}}\,\mathrm{d}\alpha \label{eq:PCvM-AD}.
	\end{align}
	Since $\alpha\mapsto 1/(\alpha(\pi-\alpha))$ and $\alpha\mapsto 1/\alpha$ are not integrable on $[0,\pi]$, we first compute the sum in the integrand. By Lemma \ref{lem:A_q1}, we have 
	\begin{align*}
	\frac{\pi}{\alpha(\pi-\alpha)}A(\theta_{ij},\cos(\alpha))-\frac{1}{\alpha}=\begin{cases}
	-\frac{1}{\pi-\alpha},&0\leq\alpha \leq \frac{\theta_{ij}}{2},\\
	-\frac{\theta_{ij}}{2\alpha(\pi-\alpha)},&\frac{\theta_{ij}}{2} < \alpha < \pi-\frac{\theta_{ij}}{2},\\
	-\frac{1}{\alpha},& \pi-\frac{\theta_{ij}}{2}\leq \alpha \leq   \pi.
	\end{cases}
	\end{align*}
	Consequently, from \eqref{eq:PCvM-AD}, 
	\begin{align*}
	P_{n,1}^\mathrm{AD}=&\;n+\frac{2}{n}\sum_{i<j}\bigg\{\log\left(\pi-\tfrac{\theta_{ij}}{2}\right)-\log(\pi)-\frac{\theta_{ij}}{2\pi}\int_{\theta_{ij}/2}^{\pi-\theta_{ij}/2}\lrp{\frac{1}{\alpha}+\frac{1}{\pi-\alpha}}\,\mathrm{d}\alpha\\
	&-\log(\pi)+\log\left(\pi-\tfrac{\theta_{ij}}{2}\right)\bigg\}\\
	=&\;n+\frac{2}{n}\sum_{ i<j}\lrc{-2\log\lrp{\pi}+\frac{1}{\pi}\lrb{\theta_{ij}\log(\theta_{ij})+(2\pi-\theta_{ij})\log(2\pi-\theta_{ij})}},
	\end{align*}
	and thus the expression for $\psi_1^\mathrm{AD}$ follows.\\
	
	For $q\ge 2$, write the statistic as
	\begin{align*}
	P_{n,q}^\mathrm{AD}=\lim_{\varepsilon\to 0}P_{n,q}^{\mathrm{AD},\varepsilon}
	\end{align*}
	where, for $\varepsilon>0$,
	\begin{align}
	P_{n,q}^{\mathrm{AD},\varepsilon}:=&\;\frac{1}{n}\sum_{i\neq j}W_{ij}^{\varepsilon}+\int_{-1+\varepsilon}^{1-\varepsilon}F_q(x)(1-nF_q(x))w(F_q(x))\,\mathrm{d}F_q(x),\label{eq:lim_PAD}\\
	W_{ij}^{\varepsilon}:=&\;\int_{-1+\varepsilon}^{1-\varepsilon}A(\theta_{ij},x)w(F_{q}(x))\,\mathrm{d}F_q(x).\notag
	\end{align}
	For the second term of \eqref{eq:lim_PAD} it follows that 
	\begin{align}
	\int_{-1+\varepsilon}^{1-\varepsilon}\frac{F_{q}(x)(1-nF_{q}(x))}{F_{q}(x)(1-F_{q}(x))}\,\mathrm{d}F_{q}(x)
	=&\;n\lrp{F_{q}(1-\varepsilon)-F_{q}(-1+\varepsilon)}\notag\\
	&+(n-1)\log\lrp{\frac{1-F_{q}(1-\varepsilon)}{1-F_{q}(-1+\varepsilon)}}.\label{eq:bias_AD}
	\end{align}
	By the same arguments used in \eqref{eq:psi_q2} and \eqref{eq.psiq.final}, the first term of \eqref{eq:lim_PAD} is
	\begin{align*}
	W_{ij}^{\varepsilon}=2\int_{-1}^{1}H_q^{\varepsilon}(t)\lrb{\int_{-1}^11_{\lrb{t\cos(\theta_{ij})+u(1-t^2)^{1/2}\sin(\theta_{ij})\leq t}}\,\mathrm{d}F_{q-1}(u)}\,\mathrm{d}F_q(t)=:\psi_q^{\varepsilon}(\theta_{ij}),
	\end{align*}
	where, for $x\in(-1+\varepsilon,1-\varepsilon)$,
	\begin{align*}
	H_q^{\varepsilon}(x) := \int_{x}^{1-\varepsilon}w(F_q(t))\,\mathrm{d}F_q(t)=\log\lrp{\frac{F_{q}(1-\varepsilon)}{1-F_{q}(1-\varepsilon)}}+\log\lrp{\frac{1-F_{q}(x)}{F_{q}(x)}}.
	\end{align*}
	Analogously to \eqref{eq.psiq.final}, 
	\begin{align*}
	\psi_q^{\mathrm{AD},\varepsilon}(\theta)=&\;2\int_{\cos\lrp{\theta/2}}^{1}H_q^{\varepsilon}\lrp{t}\,\mathrm{d}F_q(t)+2\int_{-\cos\lrp{\theta/2}}^{\cos\lrp{\theta/2}}H_q^{\varepsilon}\lrp{t}F_{q-1}\lrp{\frac{t\tan\lrp{\theta/2}}{(1-t^2)^{1/2}}}\,\mathrm{d}F_q(t)\\
	&\!\!\!\!\!\!=:\psi_{q,1}^{\mathrm{AD},\varepsilon}(\theta)+\psi_{q,2}^{\mathrm{AD},\varepsilon}(\theta).
	\end{align*}
	Each term can be computed separately:
	\begin{align}
	\psi_{q,1}^{\mathrm{AD},\varepsilon}(\theta)%
	%
	=&-2F_{q}\lrp{\cos\lrp{\tfrac{\theta}{2}}}\log\lrp{\frac{F_{q}(1-\varepsilon)}{1-F_{q}(1-\varepsilon)}}-2\int_{0}^{\cos\lrp{\theta/2}} \log\lrp{\frac{1-F_{q}(x)}{F_{q}(x)}}\mathrm{d}F_{q}(t)\notag\\
	&-\log(4)-2\log\lrp{1-F_{q}(1-\varepsilon)},\label{eq:psi_q1_1_AD}\\
	\psi_{q,2}^{\mathrm{AD},\varepsilon}(\theta)=&\;\lrp{2F_{q}(\cos\lrp{\tfrac{\theta}{2}})-1}\log\lrp{\frac{F_{q}(1-\varepsilon)}{1-F_{q}(1-\varepsilon)}}\notag\\
	&+2\int_0^{\cos\lrp{\theta/2}}\log\lrp{\frac{1-F_{q}(t)}{F_{q}(t)}}\lrp{2F_{q-1}\lrp{\frac{t\tan\lrp{\theta/2}}{(1-t^2)^{1/2}}}-1}\mathrm{d}F_{q}(t),\label{eq:psi_q1_2_AD}
	\end{align}
	where in the first term of \eqref{eq:psi_q1_2_AD} it is employed that
	\begin{align*}
	\int_{-\cos\lrp{\theta/2}}^{\cos\lrp{\theta/2}}F_{q-1}\lrp{\frac{t\tan\lrp{\theta/2}}{(1-t^2)^{1/2}}}\mathrm{d}F_{q}(t)=F_{q}\lrp{\cos\lrp{\tfrac{\theta}{2}}}-\frac{1}{2}
	\end{align*}
	and in the second that
	\begin{align*}
	\int_{-\cos\lrp{\theta/2}}^{\cos\lrp{\theta/2}}&\log\lrp{\frac{1-F_{q}(t)}{F_{q}(t)}}F_{q-1}\lrp{\frac{t\tan\lrp{\theta/2}}{(1-t^2)^{1/2}}}\mathrm{d}F_{q}(t)\\
	=&\;\int_0^{\cos\lrp{\theta/2}}\log\lrp{\frac{1-F_{q}(t)}{F_{q}(t)}}\lrp{2F_{q-1}\lrp{\frac{t\tan\lrp{\theta/2}}{(1-t^2)^{1/2}}}-1}\mathrm{d}F_{q}(t).
	\end{align*}
	From \eqref{eq:psi:q2}, it results
	\begin{align} \label{eq:psi:q2_epsilon}
	\psi^{\mathrm{AD},\varepsilon}_q(\theta)=\psi^\mathrm{AD}_q(\theta)-\log\lrp{1-F_{q}(1-\varepsilon)}-\log\lrp{F_{q}(1-\varepsilon)}.
	\end{align}
	Consequently, by \eqref{eq:lim_PAD}, \eqref{eq:bias_AD}, \eqref{eq:psi_q1_1_AD}, and \eqref{eq:psi_q1_2_AD}:
	\begin{align*}
	P_{n,q}^\mathrm{AD}=&\;\lim_{\varepsilon\to 0}\bigg[\frac{2}{n}\sum_{i< j}\lrb{\psi_{q}^\mathrm{AD}(\theta_{ij})-\log\lrp{1-F_{q}(1-\varepsilon)}-\log\lrp{F_{q}(1-\varepsilon)}}\\
	&\phantom{lim_{\varepsilon\to 0}}+n\lrp{F_{q}(1-\varepsilon)-F_{q}(-1+\varepsilon)}+(n-1)\log\lrp{\frac{1-F_{q}(1-\varepsilon)}{1-F_{q}(-1+\varepsilon)}}\bigg]\\
	=&\;n+\frac{2}{n}\sum_{i< j}\psi_{q}^\mathrm{AD}(\theta_{ij}).
	\end{align*}
	The particular expression for $q=2$ follows trivially from \eqref{eq:psi:q2}. For $q=3$, from \eqref{eq:psi:q2} and taking into account that $F_3(\cos(\theta/2))=1+(\sin(\theta)-\theta)/(2\pi)$, it follows
	\begin{align*}
	\psi_3^\mathrm{AD}(\theta)
	%
	%
	=&-\log(4)-2\log(\pi)+2\log(2\pi+\sin(\theta)-\theta)+\frac{1}{\pi}(\sin(\theta)-\theta)\log\lrp{\frac{2\pi+\sin(\theta)-\theta}{\theta-\sin(\theta)}}\\
	&-\frac{4\tan\lrp{\theta/2}}{\pi}\int_0^{\cos(\theta/2)}t\log\lrp{\frac{\pi}{\cos^{-1}(t)-t(1-t^2)^{1/2}}-1}\,\mathrm{d}t.
	\end{align*}
\end{proof}	

\subsection{Proofs of Section \ref{sec:equiv}}
\label{subsec:proof_theo:Proj_Sob}

The proof of Theorem \ref{theo:Proj_Sobolev} is split for each statement. We begin with the proof of \ref{theo:Proj_Sobolev:1}.\\

\begin{proof}{ of \ref{theo:Proj_Sobolev:1} in Theorem \ref{theo:Proj_Sobolev}}
	We prove separately $\mathcal{P}_+ \subset \mathcal{S}$ and $P_{n,q}^\mathrm{AD}\in\mathcal{S}$.\\
	
	\textit{Proof of $\mathcal{P}_+ \subset \mathcal{S}$.} The proof goes along two steps. \\
	
	\textit{Step 1.} Assume that $W$ is the Dirac's delta on the point $F_q(x)\in(0,1]$, $\delta_{F_q(x)}$, given by $x \in (-1,1]$. Clearly, $W(F_q(A))=\delta_x(A)$ for $A\subset[-1,1]$, and hence from \eqref{eq:Pn.W} we have
	\begin{align}\label{eq:deltax}
	P_{n,q}^{\delta_{F_q(x)}}= n\Es{\left(F_{n,\bga}(x)- F_q(x)\right)^2}{\bga}
	=\frac{F_q(x)^2}n\Es{\left(\sum_{i=1}^n\frac{1_{\{\bga'\bX_i \leq x\}}}{F_q(x)}-1\right)^2}{\bga}.
	\end{align}
	
	Consider the function $f^x_q(z):=(F_q(x))^{-1} 1_{\{z \leq x\}}, z \in [-1,1]$. For each $x\in[-1,1]$, $f^x_q$ is  bounded, thus $f^x_q \in L_q^2[-1,1]$. Also, its first Gegenbauer coefficient equals one:
	\begin{align*}
	b_{0,q}=\begin{cases}
	\vspace{.2cm}
	\frac{1}{c_{0,1}}\int_{-1}^1 f_1^x(z) (1-z^2)^{-1/2}\,\mathrm{d}z=\frac{\mathrm{B}\lrp{\frac{1}{2},\frac{1}{2}}}{\pi F_1(x)} F_1(x)=1,& q=1,\\
	\frac{1}{c_{0,q}}\int_{-1}^1 f_q^x(z) (1-z^2)^{q/2-1}\,\mathrm{d}z=\frac{2^{q-1}\Gamma\lrp{(q+1)/2}^2}{\pi(q-1)\Gamma(q-1)}\frac{\mathrm{B}\lrp{\frac{1}{2},\frac{q}{2}}}{F_q(x)} F_q(x)=1,& q\geq 2,
	\end{cases}
	\end{align*}
	where we have applied the Legendre duplication formula when $q\geq 2$.
	Based on \eqref{eq:lmpfp2}, denote by $(1+2k/(q-1))v_{k,q}^x$ to the remaining Gegenbauer coefficients of $f_q^x$ for $q\geq1$ (using extension \eqref{eq:Ck0} if $q=1$). By showing $\sum_{k=1}^\infty (v_{k,q}^x)^2 d_{k,q} < \infty$, \eqref{eq:deltax} and \eqref{eq:fint} would give $P_{n,q}^{\delta_{F_q(x)}}=F_q(x)^2S_{n,q}(f^x_q)$ and the proof would be completed for the case $W=\delta_{F_q(x)}$.\\
	
	For $q=1$, $d_{k,1}= 2$ for $k\geq 1$ so, by Lemma \ref{lemma:v_{k,q}_x}, $\sum_{k=1}^\infty 2(v_{k,q}^x)^2 < \infty$. For $q\geq 2$, according to \eqref{Eq:d_q,k}, $d_{k,q}\sim k^{q-1}$, so by Lemma \ref{lemma:v_{k,q}_x} we have that $(v_{k,q}^x)^2 d_{k,q} \sim k^{-2}$ and therefore $\sum_{k=1}^\infty(v_{k,q}^x)^2 d_{k,q}< \infty$ as sought.\\
	
	\textit{Step 2.} Assume now that $W$ is a probability measure defined on the Borel sets on $[0,1]$ with $W(\{0\})=0$. The first step is to make explicit the relation between the functions $\tilde \psi_q^{\delta_{F_q(x)}}$ and $g_{f_q^x}$ (defined in \eqref{eq:psi_tilde_theta} and \eqref{Eq.h_Sobolev}, respectively). Since  $P_{n,q}^{\delta_{F_q(x)}}=F_q(x)^2S_{n,q}(f^x_q)$ holds for every $x\in(-1,1]$ and $n\geq1$, if we apply this relation with $n=1$, from  \eqref{eq:psi_tilde_theta} and \eqref{Eq.h_Sobolev}, we obtain that $\tilde \psi_q^{\delta_{F_q(x)}}(\theta_{ii}) = F_q(x)^2g_{f_q^x}(\bX_i' \bX_i)$ for every $i=1,\ldots, n$. From here, and \eqref{eq:psi_tilde_theta} and  \eqref{Eq.h_Sobolev} with $n=2$, we obtain that $\tilde \psi_q^{\delta_{F_q(x)}}(\theta_{12}) = F_q(x)^2 g_{f_q^x}(\bX_1' \bX_2)$ for every selection of $\bX_1, \bX_2 \in {\Om{q}}$, therefore implying that
	\begin{align} \label{Eq:IgualdadSobProy}
	\tilde \psi_q^{\delta_{F_q(x)}}(\theta) = F_q(x)^2 g_{f_q^x}(\cos \theta), \mbox{ for all } \theta \in  [0,\pi].
	\end{align}
	Additionally, notice that $\tilde \psi_q^{\delta_{F_q(x)}} (\theta)= A(\theta,x) - F_q(x)^2$ due to \eqref{eq:psi_tilde_theta} and since the Gegenbauer coefficients of $g_{f_q^{x}}$ are $0$ and $(1+2k/(q-1))(v_{k,q}^x)^2$ for $k\geq 1$ by \eqref{eq:gf}. \\
	
	Consider now the projected-ecdf statistic $P_{n,q}^{W}$ with kernel $\tilde \psi_q^W$.  We have that
	\begin{align*}
	P_{n,q}^W 
	&=
	\int_{-1}^{1} \lrb{
		n\Es{\lrp{F_{n,\bga}(x)- F_q(x)}^2}{\bga}
	}
	\,\mathrm{d}W(F_q(x))
	\\
	&=
	\int_{-1}^{1} 
	P_{n,q}^{\delta_{F_q(x)}}
	\,\mathrm{d}W(F_q(x))
	\\
	&=
	\frac{1}{n}\sum_{i,j=1}^n \int_{-1}^{1} 
	F_q(x)^2 g_{f_q^x}(\bX_i'\bX_j)
	\,\mathrm{d}W(F_q(x)),
	\end{align*}
	where the last equality follows from (\ref{eq:pcvm_aux}) and \eqref{Eq:IgualdadSobProy}. Consider the function
	\begin{align}
	g^W(z) := \int_{-1}^{1} 
	F_q(x)^2 g_{f_q^x}(z)
	\,\mathrm{d}W(F_q(x)),\quad z \in [-1,1].\label{eq:gW}
	\end{align}
	We have that 
	$$
	|F_q(x)^2 g_{f_q^x}(z)| = \big|\tilde \psi_q^{\delta_{F_q(x)}} (\cos^{-1}(z))\big| = |A(\cos^{-1}(z),x) - F_q(x)^2 | \leq 1,
	$$
	thus $g^W$ is bounded and $g^W \in L_q^2[-1,1]$. Denote by $(1+2k/(q-1))u_{k,q}^W$, $k\geq 1$, to its Gegenbauer coefficients. If we prove that $u_{0,q}^W=0$ and $u_{k,q}^W\geq 0$ for $k\geq 1$, and that $\sum_{k=1}^\infty u_{k,q}^W d_{k,q} < \infty$, from \eqref{Eq.h_Sobolev}, we would have that $P_{n,q}^W = S_{n,q}(\{v_{k,q}^W\})$ for $v_{k,q}^W:=(u_{k,q}^W)^{1/2}$ and $\mathcal{P}_+ \subset \mathcal{S}$ would be proved. Since the map $(x,z)\mapsto F_q(x)^2 g_{f_q^x}(z)$ is bounded, Fubini's theorem gives for $q\ge 1$ and $k\geq 1$ that
	\begin{align*}
	u_{k,q}^W &=
	\tfrac{1}{\lrp{1+2k/(q-1)}c_{k,q}}\int_{-1}^1 g^W(z) C_k^{(q-1)/2}(z) (1-z^2)^{q/2-1}\,\mathrm{d} z
	\\
	&=
	\tfrac{1}{\lrp{1+2k/(q-1)}c_{k,q}}\int_{-1}^1 F_q(x)^2 
	\lrb{{\int_{-1}^1g_{f_q^x}(z)C_k^{(q-1)/2}(z) (1-z^2)^{q/2-1}\,\mathrm{d}z}}{\,\mathrm{d}W(F_q(x))}
	\\
	&=
	\int_{-1}^1
	\left(F_q(x)v_{k,q}^x\right)^2{\,\mathrm{d}W(F_q(x))}\\
	& \geq 0.
	\end{align*}
	This reasoning also shows that $u_{0,q}^W=0$. Moreover, from here and Lemma \ref{lemma:v_{k,q}_x} we have
	\[
	u_{k,q}^W \leq  \int_{-1}^1 \,\mathrm{d}W(F_q(x))\frac{2^{2q-1}\Gamma\lrp{\frac{q+1}{2}}^4}{\pi^2F_q(x)^2} \mathcal{O}\lrp{(k+q/2)^{-(q+1)}}
	\]
	for $q\ge 2$, which implies that $\sum_{k=1}^\infty u_{k,q}^W d_{k,q} < \infty$ for $q\geq1$.\\
	
	\textit{Proof of $P_{n,q}^\mathrm{AD}\in\mathcal{S}$.} 
	For $q\geq 2$, relation \eqref{eq:psi:q2_epsilon} shows that, for every $\varepsilon >0$, there exists a finite and positive measure $W^\varepsilon$ such that
	\[
	\psi^\mathrm{AD}_q(\theta)
	= \psi^{W^\varepsilon}_q(\theta) + h(\varepsilon),
	\]
	where $h$ is a real function depending on $\varepsilon$ but not on $\theta$. Therefore, $\psi^\mathrm{AD}_q$ equals to a constant plus the $L_q^2[-1,1]$ map $z \mapsto \psi^{W^\varepsilon}_q(\cos(z))$. Consequently, $\psi^\mathrm{AD}_q \in L_q^2[-1,1]$ and all its Gegenbauer coefficients for $k\geq1$, denoted by $(1+2k/(q-1))u_{k,q}^\mathrm{AD}$, coincide with those of $\psi^{W^\varepsilon}_q$, which are positive and satisfy $\sum_{k=1}^\infty  u_{k,q}^\mathrm{AD} d_{k,q} < \infty$ because, trivially, the proof when $W$ is a probability can be extended to cover $W^\varepsilon$ for fixed $\varepsilon>0$.\\
	
	With respect to $q=1$, in Proposition \ref{prop:Gegen_AD} we obtain the Gegenbauer coefficients of $\psi_1^\mathrm{AD}$, $\{b_{k,1}^\mathrm{AD}\}$,  which are obviously non-negative. Moreover, 
	Lemma \ref{Lemma_serie_q1} shows that $b_{k,1}^\mathrm{AD}= O(\log (k)/k^2)$, so there exists a function $f \in L_q^2[-1,1]$ whose  Gegenbauer coefficients are $1$ and $(b_{k,1}^\mathrm{AD})^{1/2}$, $k\geq 1$. This shows that  $P_{n,1}^\mathrm{AD}=S_{n,1}(f)$ since $d_{k,1}=2$ for $k\geq 1$.
\end{proof}

\begin{remark} 
	In Step 1 of the proof of  $\mathcal{P}_+ \subset \mathcal{S}$, we set the function  
	$f_q^x(z)=(F_q(x))^{-1}1_{\{z\leq x\}}$ and obtain that 
	$P_{n,q}^{\delta_{F_q(x)}}=F_q(x)^2 S_{n,q}(f_q^x)$. In Step 2, \eqref{eq:gW} with $W=\delta_{F_q(x)}$ gives a  different function, $g^{\delta_{F_q(x)}}$, such that its Gegenbauer coefficients $\big(v_{k,q}^{\delta_{F_q(x)}}\big)^2$, satisfy $P_{n,q}^{\delta_{F_q(x)}}= S_{n,q}\big(\big\{v_{k,q}^{\delta_{F_q(x)}}\big\}\big)$. Taking $f^{\delta_{F_q(x)}}(z)= 1_{\{z\leq x\}} - F_q(x)+1$, it is simple to check that $P_{n,q}^{\delta_{F_q(x)}} = S_{n,q}(f^{\delta_{F_q(x)}})$.
\end{remark}

The proof of \ref{theo:Proj_Sobolev:2} requires some results on integral equations given in Section \ref{sec:integral_equations}.

\begin{proof}{ of \ref{theo:Proj_Sobolev:2} in Theorem \ref{theo:Proj_Sobolev}}

	Under the conditions of Theorem \ref{theo:bijection_2}, if we additionally consider that the first Gegenbauer coefficient of $f$  is one, then there exists an absolutely continuous, finite, potentially signed, Borel measure on $[-1,1]$ such that $S_{n,q}(f) = P_{n,q}^W$. This measure is the one whose density with respect to the Lebesgue measure is $f$, which acts as $g$ in the statement of Theorem \ref{theo:bijection_2}.
\end{proof}

\subsection{Proofs of Section \ref{sec:asymp}}
\label{subsec:proof_sec:asymp}

\begin{proof}{ of Theorem \ref{theo:coef_A}}
	Suppose $q=1$ and $k\geq 1$. By Proposition \ref{prop:Aij}, the Gegenbauer coefficients of $A(\theta,x)$ are
	\begin{align}
	a_{k,1}^x=&\;\frac{1}{c_{k,1}}\int_0^{\pi}A(\theta,x)T_k(\cos(\theta))\sin^{q-1}(\theta)\,\mathrm{d}\theta\nonumber\\
	%
	%
	=&\;\frac{2}{\pi}\left\{\int_0^{2\cos^{-1}(x)}\lrp{1-\frac{\cos^{-1}(x)}{\pi}-\frac{\theta}{2\pi}}\cos(k\theta)\,\mathrm{d}\theta+\int_{2\cos^{-1}(x)}^{\pi}\lrp{1-\frac{2\cos^{-1}(x)}{\pi}}\cos(k\theta)\,\mathrm{d}\theta\right\}\nonumber\\
	=&\;\frac{2}{\pi}\int_0^{2\cos^{-1}(x)}\lrp{\frac{\cos^{-1}(x)}{\pi}-\frac{\theta}{2}} \cos(k\theta)\,\mathrm{d}\theta\label{eq:proof_theo_coefA1},
	\end{align}
	where \eqref{eq:proof_theo_coefA1} follows from the orthogonality of the Gegenbauer polynomials. Therefore, 
	\begin{align*}
	a_{k,1}^x&=\frac{2}{\pi}\lrp{\frac{1}{k\pi}\cos^{-1}(x)\lrc{\sin(k\theta)}_0^{2\cos^{-1}(x)}-\lrc{\frac{1}{2k^2\pi}\cos(k\theta)+\frac{\theta}{2k\pi}\sin(k\theta)}_0^{2\cos^{-1}(x)}}\\
	&=\frac{1}{k^2\pi^2}\lrp{1-\cos(2k\cos^{-1}(x))}.
	\end{align*}
	For $k=0$, Proposition \ref{prop:Aij} gives that
	\begin{align*}
	a_{0,1}^x=\frac{1}{\pi}\int_0^{\pi}A(\theta,x)\,\mathrm{d}\theta
	%
	%
	%
	=2F_1(x)-1+\frac{1}{\pi^2}\lrp{\cos^{-1}(x)}^2
	=F_1(x)^2.
	\end{align*}
	
	Assume $q\geq 2$. From \eqref{Eq:IgualdadSobProy} and the identity $\tilde \psi_q^{\delta_{F_q(x)}} (\theta)= A(\theta,x) - F_q(x)^2$, we have that  
	\begin{align*} 
	A(\theta,x)& = F_q(x)^2 (1 +g_{f_q^x}(\cos \theta))=F_q(x)^2 +\sum_{k=1}^{\infty} \lrp{v_{k,q}^x F_q(x)}^2\lrp{1+\frac{2k}{q-1}}C_k^{(q-1)/2}(\cos\theta),
	\end{align*} 
	and the Gegenbauer coefficients of $A(\theta,x)$ are obtained from substituting the value of $v^x_{k,q}$ given in \eqref{eq:v_{k,q}_x} for $k\geq 1$ and from the equality $C_k^{(q-1)/2}(x)=1$ when $k=0$.
\end{proof}	

\begin{proof}{ of Corollary \ref{coro:coef_psi}}
	The coefficients of $\psi_q^W$ are trivially deduced from \eqref{def:psi_theta} and Theorem \ref{theo:coef_A}.
\end{proof}	

\begin{proof}{ of Corollary \ref{coro:w1}}
	Since $F^{-1}_1(x)=-\cos(\pi x)$, then
	\begin{align*}
	w_k(x)
	=-2\pi^2 k^2\cos(2k x \pi)
	=-2\pi^2 k^2\cos(2k\cos^{-1}(F_1^{-1}(x)))
	=-2k^2\pi^2T_{2k}(F_1^{-1}(x)).
	\end{align*}
	Consider the signed measure $W_k$ associated to $w_k$. From Corollary \ref{coro:coef_psi}, we have that
	\begin{align*}
	b_{j,1}^{W_k}&=\int_{-1}^1 a_{j,1}^x w_k(F_1(x))\,\mathrm{d}F_1(x)\\
	&=-2k^2\pi^2\int_{-1}^1 \frac{1}{j^2\pi^2}\lrp{1-T_{2j}(x)} T_{2k}(x)\,\mathrm{d}F_1(x)\\
	&=\frac{2k^2}{j^2\pi}\int_{-1}^1 T_{2j}(x)T_{2k}(x)(1-x^2)^{-1/2}\,\mathrm{d}x\\
	&=\delta_{jk}.
	\end{align*}
	Therefore, $w(\{v_{k,1}\})(x)=\sum_{k=1}^\ell v_{k,1}^2w_k(x)$ is such that $b_{j,1}^{W({\{v_{k,1}\}})}=v_{k,1}^2$, $k\geq1$, and $S_{n,1}(\{v_{k,1}\})=P_{n,1}^{W(\{v_{k,1}\})}$. 
	Defining $\tilde{w}^\pm({\{v_{k,1}\}})(x):=\max(\pm w(\{v_{k,1}\})(x),0)$, then $w(\{v_{k,1}\})=a^+w^+(\{v_{k,1}\})-a^-w^-(\{v_{k,1}\})$, where $(a^\pm)^{-1}:=\int_{-1}^1 \tilde{w}^\pm({\{v_{k,1}\}})(x)\,\mathrm{d}x\geq 0$ and $w^\pm({\{v_{k,1}\}}):=(a^\pm)^{-1}\tilde{w}(\{v_{k,1}\})$ with associated measure $W^\pm(\{v_{k,1}\})$. Then, 
	$P_{n,1}^{W(\{v_{k,1}\})}=a^+P_{n,1}^{W^+(\{v_{k,1}\})}-a^-P_{n,1}^{W^-(\{v_{k,1}\})}$.
\end{proof}

\begin{proof}{ of Corollary \ref{coro:Gegen_Rt}}
	Assume $q\geq 2$ and $k\geq 1$. Due to the symmetry of $x\mapsto a_{k,q}^x$, by Corollary \ref{coro:coef_psi},
	\begin{align*}
	b^{\mathrm{R}_t}_{k,q}=\int_{-1}^1  a_{k,q}^x\,\mathrm{d}W_t(F_q(x))
	=\frac{1}{2}a_{k,q}^{F_q^{-1}(1-t_m)}+\frac{1}{2} a_{k,q}^{F_q^{-1}(t_m)}=a_{k,q}^{F_q^{-1}(t_m)}.
	%
	\end{align*}
	The case $q=1$ is obtained from $F_1^{-1}(x)=\cos(\pi(1-x))$ and $\sin^2(x)=(1-\cos(2x))/2$.
	For $k=0$, by Proposition \ref{prop:psi_q:Roth}, 
	\begin{align*}
	b^{\mathrm{R}_t}_{0,q}=\int_{-1}^1 F_q(x)^2\,\mathrm{d}W_t(F_q(x))
	=\tfrac{1}{2}\lrb{F_q(F_q^{-1}(1-t_m))^2+F_q(F_q^{-1}(t_m))^2}
	=\tfrac{1}{2}\lrp{(1-t_m)^2+t_m^2}.
	\end{align*}
\end{proof}

\begin{proof}{ of Proposition \ref{prop:Gegen_CvM}}
	For $q=1$, by Corollary \ref{coro:coef_psi}, the Gegenbauer coefficients of $\psi_{1}^{\mathrm{CvM}}$ are, for $k\geq1$,
	\begin{align*}
	b^{\mathrm{CvM}}_{k,1}=&\int_{-1}^1  a_{k,1}^x\,\mathrm{d}F_1(x)\\
	=&\;\frac{1}{k^2\pi^3}\int_{-1}^1  \lrp{1-T_{2k}(x)}(1-x^2)^{-1/2}\,\mathrm{d}x\\
	=&\;\frac{1}{k^2\pi^3}\int_{-1}^1  \lrp{1-\cos\lrp{2k\cos^{-1}(x)}}(1-x^2)^{-1/2}\,\mathrm{d}x\\
	=&\;\frac{1}{\pi^2k^2},
	\end{align*}
	where the last equality follows form the orthogonality of the Gegenbauer polynomials.
	
	For $q\geq 2$, denote $\tau_{k,q}:=\big(2^{q-1}\Gamma\lrp{(q+1)/2}^2\Gamma(k)/\lrp{\pi\Gamma(k+q)}\big)^2$. By Corollary \ref{coro:coef_psi}, the Gegenbauer coefficients of $\psi_{q}^{\mathrm{CvM}}$ are, for $k\geq 1$,
	\begin{align}
	b^{\mathrm{CvM}}_{k,q}=&\;\int_{-1}^1  a_{k,q}^x\,\mathrm{d}F_q(x)\nonumber\\
	=&\;\lrp{1+\tfrac{2k}{q-1}}\tau_{k,q}\frac{1}{\mathrm{B}\lrp{\frac{1}{2},\frac{q}{2}}}\int_{-1}^1(1-x^2)^{3q/2-1}\lrp{C_{k-1}^{(q+1)/2}(x)}^2\,\mathrm{d}x\nonumber\\
	=&\;\lrp{1+\tfrac{2k}{q-1}}\tau_{k,q}\frac{1}{\mathrm{B}\lrp{\frac{1}{2},\frac{q}{2}}}\frac{\mathrm{B}\lrp{\frac{3q}{2},\frac{1}{2}}}{\lrp{(k-1)\mathrm{B}(q+1,k-1)}^2}\nonumber\\
	&\times{}_4F_3\lrp{1-k,q+k,\tfrac{q+1}{2},\tfrac{3q}{2};q+1,\tfrac{q}{2}+1,\tfrac{3q+1}{2};1}\label{eq:proof_coefPCvM}\\
	=&\;\frac{(q-1)^2 (2 k+q-1) \Gamma \left(\frac{q-1}{2}\right)^3 \Gamma \left(\frac{3 q}{2}\right)}{8 \pi q^2 \Gamma \left(\frac{q}{2}\right)^3  \Gamma \left(\frac{3 q+1}{2}\right)}\nonumber\\
	&\times{}_4F_3\lrp{1-k,q+k,\tfrac{q+1}{2},\tfrac{3q}{2};q+1,\tfrac{q}{2}+1,\tfrac{3q+1}{2};1},\label{eq:proof_coefPCvM2}
	\end{align}
	where the last equality follows from the Legendre duplication formula and \eqref{eq:proof_coefPCvM} follows from equations (16) and (18) in \cite{Laursen1981}.  For $q=2$, by equation (21) in \cite{Laursen1981}, we have a special expression of \eqref{eq:proof_coefPCvM}: 
	\begin{align*}
	b^{\mathrm{CvM}}_{k,2}&=\frac{1+2k}{2}\lrp{\frac{\Gamma(k)}{2\Gamma(k+2)}}^2\frac{4\Gamma(k+2)(2+(k-1)(k+2))}{(k-1)!(2k-1)(2k+1)(2k+3)}\\
	&=\frac{1}{2}\frac{\Gamma(k)}{\Gamma(k+2)}\frac{(2+(k-1)(k+2))}{(2k-1)(2k+3)}\\
	&=\frac{1}{2(2k-1)(2k+3)}.
	\end{align*}
	If $q=3$, \eqref{eq:proof_coefPCvM2} becomes
	\begin{align}
	b_{k,3}^{\mathrm{CvM}}:=\;\frac{35\lrp{1+k}}{144\pi^2}{}_4F_3\lrp{1-k,3+k,2,\tfrac{9}{2};4,\tfrac{5}{2},5;1},\nonumber
	\end{align}
	however this expression is not easily tractable. For that reason, we directly work with the definition of the Gegenbauer coefficients, i.e., using $\psi_3^{\mathrm{CvM}}=\psi_1^{\mathrm{CvM}}+\sin(\theta)(\pi-\theta-\sin(\theta))/(4\pi^2(1+\cos(\theta)))$. We employ that the Gegenbauer polynomials of order $1$ coincide with the Chebyshev polynomials of the second type (equation 18.5.2 in \cite{NIST:DLMF}) to obtain the Gegenbauer coefficients of $\psi_3^{\mathrm{CvM}}$ for $k>1$:
	\begin{align*}
	b_{k,3}^{\mathrm{CvM}}=&\;\frac{2}{\pi}\int_0^{\pi}\psi_{3}(\theta)\sin((k+1)\theta)\sin(\theta)\,\mathrm{d}\theta\\
	%
	%
	=&\;\frac{1}{\pi}\left\{\frac{1}{2\pi^2}\int_0^{\pi}(1-\cos(\theta))(\pi-\theta-\sin(\theta))\sin((k+1)\theta)\,\mathrm{d}\theta\right.\\
	\phantom{=}& \phantom{\frac{1}{\pi}()}\left.+\int_0^{\pi}\sin((k+1)\theta)\sin(\theta)\,\mathrm{d}\theta+\frac{1}{2\pi^2}\int_0^{\pi}\theta(\theta-2\pi)\sin((k+1)\theta)\sin(\theta)\,\mathrm{d}\theta\right\}\\
	%
	%
	=&\;\frac{1}{2\pi^2}\frac{3k^2+6k+4}{k^2(k+1)(k+2)^2}.
	\end{align*}
	
	For $k=1$, we have
	\begin{align*}
	b_{1,3}^{\mathrm{CvM}}=&\;\frac{4}{\pi}\int_0^{\pi}\psi_{3}^{\mathrm{CvM}} (\theta)\sin^2(\theta)\cos(\theta)\,\mathrm{d}\theta\\
	=&\;\frac{1}{\pi^3}\left\{\int_0^\pi\theta(\theta-2\pi)\sin^2(\theta)\cos(\theta)\,\mathrm{d}\theta\right.\\
	&\quad\quad\left.+\int_0^\pi(\pi-\theta-\sin(\theta))(1-\cos(\theta))\sin(\theta)\cos(\theta))\,\mathrm{d}\theta\right\}\\
	=&\;\frac{35}{72\pi^2}.
	\end{align*}
	If $k=0$ and $q\geq 1$, by Corollary \ref{coro:coef_psi}, 
	\begin{align*}
	b_{0,q}^{\mathrm{CvM}}=\int_{-1}^1F_q(x)^2f_q(x)\,\mathrm{d}x=\frac{1}{3}\lrc{F_q(x)^3}_{-1}^1=\frac{1}{3}.
	\end{align*}
\end{proof}

\begin{proof}{ of Proposition \ref{prop:Gegen_AD}}
	For $q=1$ and $k\geq 1$, Corollary \ref{coro:coef_psi} gives that
	\begin{align*}
	b_{k,1}^{\mathrm{AD}}=&\;\int_{-1}^1 a_{k,2}^x \frac{f_1(x)}{F_1(x)(1-F_1(x))}\, \mathrm{d}x\\
	=&\;\frac{1}{k^2\pi}\int_0^{\pi}(1-T_{2k}(x))\frac{(1-x^2)^{-1/2}}{(\pi-\cos^{-1}(x))(\cos^{-1}(x))^2}\,\mathrm{d}x\\
	=&\;\frac{1}{\pi k^2}\int_0^\pi\frac{1-\cos(2k\theta)}{(\pi-\theta)\theta}\,\mathrm{d}\theta.
	\end{align*}
	For $q=2$ and $k\geq 1$, Corollary \ref{coro:coef_psi} entails that 
	\begin{align}
	b_{k,2}^{\mathrm{AD}}&=\int_{-1}^1 a_{k,2}^x \frac{f_2(x)}{F_2(x)(1-F_2(x))}\, \mathrm{d}x\nonumber\\
	&=2\lrp{1+2k}\lrp{\frac{2\Gamma\lrp{3/2}^2\Gamma(k)}{\pi\Gamma(k+2)}}^2\int_{-1}^1(1-x^2)\lrp{C_{k-1}^{3/2}(x)}^2\, \mathrm{d}x\nonumber\\
	&=\frac{4}{\pi}\frac{\Gamma(3/2)^2\Gamma(k)}{\Gamma(k+2)}\label{eq:proof_coefAD}\\
	&=\frac{1}{k(k+1)},\nonumber
	\end{align}
	where \eqref{eq:proof_coefAD} follows from equation ET II 281(8) in \cite{Gradshteyn2014} taking $\nu=3/2$. 
	
	For $k=0$ and any $q\geq 1$, by \eqref{eq:psi:q2_epsilon},
	\begin{align*}
	\psi^\mathrm{AD}_q(\theta)&=\lim_{\varepsilon\to 0}\lrp{\psi^{\mathrm{AD},\varepsilon}_q(\theta)+\log\lrp{1-F_{q}(1-\varepsilon)}+\log\lrp{F_{q}(1-\varepsilon)}}\\
	&=\lim_{\varepsilon\to 0}\lrp{\int_{-1+\varepsilon}^{1-\varepsilon}\frac{A(\theta,x)}{F_q(x)(1-F_q(x))}\,\mathrm{d}F_q(x)+\log\lrp{1-F_{q}(1-\varepsilon)}+\log\lrp{F_{q}(1-\varepsilon)}}.
	\end{align*}
	Consequently, by Corollary \ref{coro:coef_psi}, we have that 
	\begin{align}
	b_{0,q}^{\mathrm{AD}}&=\frac{1}{c_{0,q}}\int_{0}^\pi \psi^\mathrm{AD}_q(\theta)\sin^{q-1}(\theta)\, \mathrm{d}\theta\nonumber\\
	&=\frac{1}{c_{0,q}}\int_{0}^\pi\lim_{\varepsilon\to 0}\left(\int_{-1+\varepsilon}^{1-\varepsilon}\frac{A(\theta,x)}{F_q(x)(1-F_q(x))}\,\mathrm{d}F_q(x)+\log\lrp{F_{q}(-1+\varepsilon)}\right.\nonumber\\
	&\phantom{=\frac{1}{c_{0,q}}\int_{0}^\pi\lim_{\varepsilon\to 0}()}+\log\lrp{F_{q}(1-\varepsilon)}\bigg)\sin^{q-1}(\theta)\, \mathrm{d}\theta\nonumber\\
	&=\lim_{\varepsilon\to 0}\left\{\frac{1}{c_{0,q}}\int_{0}^\pi\lrp{\int_{-1+\varepsilon}^{1-\varepsilon}\frac{A(\theta,x)}{F_q(x)(1-F_q(x))}\,\mathrm{d}F_q(x)}\sin^{q-1}(\theta)\, \mathrm{d}\theta\right.\nonumber\\
	&\phantom{=\lim_{\varepsilon\to 0}()}+\log\lrp{F_{q}(-1+\varepsilon)}+\log\lrp{F_{q}(1-\varepsilon)}\bigg\}\label{eq:proof_b0AD_1}\\
	&=\lim_{\varepsilon\to 0}\lrb{\int_{-1+\varepsilon}^{1-\varepsilon}F_q(x)^2 \frac{\mathrm{d}F_q(x)}{F_q(x)(1-F_q(x))}+\log\lrp{F_{q}(-1+\varepsilon)}+\log\lrp{F_{q}(1-\varepsilon)}}\label{eq:proof_b0AD_2}\\
	&=\lim_{\varepsilon\to 0}\lrb{-\lrc{F_q(x)}_{-1+\varepsilon}^{1-\varepsilon}+2\log\lrp{F_q(1-\varepsilon)}}\nonumber\\
	&=-1,\nonumber
	\end{align}
	where \eqref{eq:proof_b0AD_2} follows from Fubini's theorem and \eqref{eq:proof_b0AD_1} is obtained from the dominated convergence theorem. The latter is applicable since $|\psi_q^{\mathrm{AD}}(\theta)|$ is bounded for all $\theta\in[0,\pi]$ due to Proposition \ref{prop:psi_q:AD}:
	\begin{align*}
	|\psi_q^{\mathrm{AD}}(\theta)|&=\abs{-\log(4)+4\int_{0}^{\cos\lrp{\theta/2}} \log\lrp{\frac{F_{q}(t)}{1-F_{q}(t)}} \lrp{1-F_{q-1}\lrp{\frac{t\tan\lrp{\theta/2}}{(1-t^2)^{1/2}}}}\mathrm{d}F_{q}(t)}\\
	&\leq\log(4)+4\int_{0}^{\cos\lrp{\theta/2}}\abs{\log\lrp{\frac{F_{q}(t)}{1-F_{q}(t)}}}\mathrm{d}F_{q}(t)\\
	&=\log(4)+4\int_{1/2}^{F_q(\cos\lrp{\theta/2})}\log\lrp{\frac{u}{1-u}}\mathrm{d}u\\
	&\leq\log(4)+4\int_{1/2}^{1}\log\lrp{\frac{u}{1-u}}\mathrm{d}u\\
	&=6\log(2).
	\end{align*}	
\end{proof}

\begin{proof}{ of Corollary \ref{coro:omn}}
	By Theorem \ref{theo:coef_A}, the $k$-th Gegenbauer coefficient of $\theta\mapsto A(x,\theta)$ is $a_{k,q}^x$. Due to the properties of orthogonal polynomials, $x\mapsto a_{k,q}^x$ has exactly $k - 1$ different real zeros in $(-1,1)$ for $q\geq 2$, hence $k+1$ different zeros in $[-1,1]$. For $q=1$, the $k+1$ different roots for $T_{2k}(x)=1$ with $x\in[-1,1]$ are $x_m=\cos(m/k\pi)$, $m=0,1,\ldots,k$. Then, $Z_{k,q}$ has cardinality $k-1$ and $Z_q\subset[-1,1]$ is a denumerable set. \\
	
	Therefore, for any $x^*\in [-1,1]\backslash Z_q$, $a_{k,q}^{x^*}>0$, for all $k\geq1$. As a consequence, for any $x^*\in Z_q$ there is an associated $W_{x^*}:=\delta_{F_q(x^*)}$ whose kernel $\psi^{W_{x^*}}_q$ has positive coefficients $b^{W_{x^*}}_{k,q}=a_{k,q}^{x^*}$, therefore generating an omnibus test by Theorem \ref{theo:Gine}. In addition, any $\sigma$-finite measure $W$ assigning positive measure to $[0,1]\backslash F_q(Z_q)$ generates a test with coefficients $b^W_{k,q}=\int_{-1}^1a_{k,q}^x\,\mathrm{d}W(F_q(x))>0$, hence omnibus. \\
	
	Finally, since $F_q(Z_q)$ has null Lebesgue measure, trivially an arbitrary set $L\subset[0,1]$ with non-null Lebesgue measure implies $L\cap ([0,1]\backslash F_q(Z_q))\neq \emptyset$, thus a measure $W$ such that $W(L)>0$ gives an omnibus test.
\end{proof}

\section{Required results on integral equations}
\label{sec:integral_equations}

We summarise next relevant results on integral equations that are required in the proof of Theorem \ref{theo:Proj_Sobolev}. They are particularizations for real $L^2[-1,1]$ kernels of results in \citet[Chapters 7 and~8]{Smithies1958}.

\begin{definition}[$L^2$-kernel and its adjoint] \label{Def.Kernel}
	A real measurable function $K$ defined on $[-1,1]\times[-1,1]$ is called an \emph{$L^2$-kernel} if $\int_{-1}^1 \int_{-1}^1 K(s,t)^2\,\mathrm{d}s\,\mathrm{d}t$, $\int_{-1}^1 K(s,t)^2\,\mathrm{d}s$, and $\int_{-1}^1 K(s,t)^2\,\mathrm{d}t$ are finite for every $s,t\in[-1,1]$. The {\em adjoint kernel} of $K$, denoted by $K^*$, is defined as $K^*(s,t) := K(t,s)$, for $s,t\in[-1,1]$.
\end{definition}

\begin{definition}[Singular values and singular functions] \label{Def.Singular}
	Let $K$ be an $L^2$-kernel and $u,v\in L^2[-1,1]$ such that $u,v\neq0$. The real $\mu$ is called  a \emph{singular value} of $K$ and $[u,v]$ are referred to as a pair of \emph{singular functions} of $K$ associated to $\mu$ if
	\[
	u(s)= \mu \int_{-1}^1 K^*(s,t) v(t) \,\mathrm{d}t \text{ and } v(s)= \mu \int_{-1}^1 K(s,t) u(t) \,\mathrm{d}t.
	\]
\end{definition}

The set of non-null singular values of $K$ is either finite or denumerable with no finite limit points. In addition, given a non-null singular value $\mu$, the set of singular functions associated to $\mu$ constitutes a linear subspace of finite dimension $d_\mu$. Then, each singular non-null value $\mu$ admits a finite number of pairs of singular functions, say $\{[u_i,v_i]\}_{i=1}^{d_\mu}$, which are an orthonormal basis of dimension $d_\mu$ of the corresponding singular functions. The union of those bases forms a \emph{full orthonormal system} of singular functions. \\

In what follows we consider ordered systems of singular values $\{\mu_n\}_{n=1}^\infty, 0 < \mu_1 \leq \mu_2 \leq \ldots$, where each singular value is repeated $d_{\mu_n}$ times. With an abuse of notation, the collection of singular values and singular functions $\{([u_n,v_n];\mu_n)\}_{n=1}^\infty$, referred to as \emph{singular system}, is treated as infinite.

\begin{proposition}[Theorem 8.7.1 in \cite{Smithies1958}] \label{Prop.Smithies}
	Let $\{([u_n,v_n];\mu_n)\}_{n=1}^\infty$ be a singular system of the $L^2$-kernel $K$ and let $y\in L^2[-1,1]$. Then, the equation
	\[
	y(s) = \int_{-1}^1 K(s,t) x(t) \,\mathrm{d}t \ \text{ for almost every $s \in [-1,1]$}
	\]
	has a solution $x\in L^2[-1,1]$ if and only if 
	\begin{enumerate}[label=\textit{\alph{*})}, ref=\textit{\alph{*})}]
		\item \label{Condition_a.Smithies}
		$
		\sum_{n=1}^\infty \mu_n^2 \int_{-1}^1 (y(s) u_n(s))^2 \,\mathrm{d}s < \infty
		$;
		\item
		$\int_{-1}^1 y(s) u(s) \,\mathrm{d}s=0$ for every function $u\in L^2[-1,1]$ such that $\int_{-1}^1 K^*(s,t) u(s) \,\mathrm{d}s=0$ for almost every $t\in[-1,1]$. \label{Condition_b.Smithies}
	\end{enumerate}
\end{proposition}

Theorem \ref{theo:bijection_2} applies Proposition \ref{Prop.Smithies} to the kernel $(s,t)\mapsto A(\cos^{-1}(s),t)$ with $A$ defined as in Proposition \ref{prop:Aij}. This map is trivially an $L^2$-kernel due to its boundedness.

\begin{theorem}\label{theo:bijection_2}
	Let be the $L^2$-kernel $K(s,t) = A(\cos^{-1}(s),t)$, $s,t \in [-1,1]$, and let $\{([u_n,v_n];\mu_n)\}_{n=1}^\infty$ be a singular system of $K$. Let $y \in L^2[-1,1]$ satisfying \ref{Condition_a.Smithies} in Proposition \ref{Prop.Smithies}. Then, there exists $g \in L^2[-1,1]$ such that
	\[
	y(s)= \int_{-1}^1 K(s,t) g(t)\,\mathrm{d}t \ \text{ for almost every $s \in [-1,1]$}.
	\]
\end{theorem}

\begin{proof}{ of Theorem \ref{theo:bijection_2}}
	Since condition \ref{Condition_a.Smithies} 
	holds by assumption, only condition \ref{Condition_b.Smithies} remains to be proved. For this, since
	$K^*(s,t)=A(\cos^{-1}(t),s)$, we show that if 
	$\int_{-1}^1A(\cos^{-1}(s),t)u(s)\, \mathrm{d}s=0$ for almost every $t\in [-1,1]$ and $u\in L^2[-1,1]$, then $u(s)=0$ for almost every $s \in [-1,1]$. 
	
	Assume $q\geq 2$. By \eqref{eq:Aij_piecewise} and since $\cos(2\cos^{-1}(s))=2s^2-1$, 
	\begin{align}
	A(\cos^{-1}(s),t)=\begin{cases}
	2F_q(t)-1,&  -1\leq s\leq 2t^2-1,\\
	\frac{1}{2}-\frac{\cos^{-1}(s)}{2\pi}+2\int_{0}^{t}F_{q-1}\lrp{\tfrac{z\tan\lrp{\cos^{-1}(s)/2}}{(1-z^2)^{1/2}}}\mathrm{d}F_q(z),&  2t^2-1\leq s< 1.
	\end{cases}\label{eq:Acos1}
	\end{align}
	Also, by assumption, for any $t\in [-1,1]$, we have that
	\begin{align}\label{eq:proof_cond_b}
	\frac{\partial}{\partial t}\int_{-1}^1A(\cos^{-1}(s),t)u(s)\, \mathrm{d}s=0.
	\end{align}
	Therefore,
	\begin{align}
	0=&\;\frac{\partial}{\partial t} \left(\int_{-1}^{2t^2-1}A(\cos^{-1}(s),t)u(s)\, \mathrm{d}s+\int_{2t^2-1}^1A(\cos^{-1}(s),t)u(s)\, \mathrm{d}s\right)\nonumber\\
	%
	=&\;4tu(2t^2-1)g_q(t) + \int_{-1}^1\frac{\partial}{\partial t}A(\cos^{-1}(s),t)u(s)\, \mathrm{d}s\nonumber\\
	=&\;4tu(2t^2-1)g_q(t),\label{eq:proof_deriv}
	\end{align}
	with 
	\begin{align*}
	g_q(t):= 2F_q(t)-\frac{3}{2}+\frac{\cos^{-1}(2t^2-1)}{2\pi}+2\int_0^tF_{q-1}\lrp{\frac{z\tan\lrp{\cos^{-1}(2t^2-1)/2}}{\lrp{1-z^2}^{1/2}}}\mathrm{d}F_q(z)
	\end{align*}
	where \eqref{eq:proof_deriv} follows from \eqref{eq:proof_cond_b} by exchanging integral and differential. Hence, since $F_q(t)>F_q(0)=1/2$, the result follows as $g_q(t)\geq 1 -3/2+\int_0^t \mathrm{d}F_q(y)> -1/2 + F_q(t)>0$. \\
	
	Showing that $\partial A(\cos^{-1}(s),t)/\partial t $ exists and is bounded is enough to guarantee that
	switching differentiation and integration is possible. 
	From \eqref{eq:Acos1}, 
	\begin{align*}
	\frac{\partial}{\partial t} A(\cos^{-1}(s),t)&=\begin{cases}
	2f_q(t), & -1\leq s\leq 2t^2-1,\\
	2F_{q-1}\lrp{\tfrac{t\tan\lrp{\cos^{-1}(s)/2}}{(1-t^2)^{1/2}}}f_q(t), & 2t^2-1\leq s< 1,
	\end{cases}
	\end{align*}
	which is obviously bounded by the definitions of the functions $f_q$ and $F_q$, and since $q\geq 2$. \\
	
	Assume now $q=1$. Obviously, it suffices to show that $\int_{-1}^1A(\cos^{-1}(s),t)u(s)\, \mathrm{d}s=0$ for almost every $t\in(0,1)$ implies $u\equiv0$ almost everywhere. 
	Then, by assumption, $\int_{-1}^1\tilde{A}(\cos^{-1}(s),t)u(s)\, \mathrm{d}s=0$ for almost every $t\in(0,1)$, with 
	\begin{align*}
	\tilde{A}(\cos^{-1}(s),t):=&\;(1-t^2)A(\cos^{-1}(s),t)\\
	=&\;\begin{cases}
	(1-t^2)\lrp{1-\frac{2}{\pi}\cos^{-1}(t)}, & -1\leq s\leq 2t^2-1,\\
	(1-t^2)\lrp{1-\frac{1}{\pi}\cos^{-1}(t)-\frac{1}{2\pi}\cos^{-1}(s)}, & 2t^2-1\leq s< 1
	\end{cases}
	\end{align*}
	due to Proposition \ref{prop:Aij}. Therefore,
	\begin{align}
	0=&\;\frac{\partial}{\partial t}\int_{-1}^1\tilde{A}(\cos^{-1}(s),t)u(s)\, \mathrm{d}s\nonumber\\
	=&\;\frac{\partial}{\partial t} \left(\int_{-1}^{2t^2-1}\tilde{A}(\cos^{-1}(s),t)u(s)\, \mathrm{d}s+\int_{2t^2-1}^1\tilde{A}(\cos^{-1}(s),t)u(s)\, \mathrm{d}s\right)\label{eq:proof_cond_b_q1}\\
	%
	=&\;4tu(2t^2-1)g_1(t) + \int_{-1}^1\frac{\partial}{\partial t}\tilde{A}(\cos^{-1}(s),t)u(s)\, \mathrm{d}s\nonumber\\
	=&\;4tu(2t^2-1)g_1(t),\nonumber
	\end{align}
	with $g_1(t):=\lrp{1-t^2}^{1/2}/\pi$, because $\cos^{-1}(2t^2-1)=2\cos^{-1}(t)$ since $t\in(0,1)$, and where the last equality follows from \eqref{eq:proof_cond_b_q1} by switching integral and differential. Then, the result follows because $g_1(t)>0$ for $t\in(0,1)$. Since by Proposition \ref{prop:Aij} it is easy to see that
	\begin{align*}
	\frac{\partial}{\partial t} \tilde{A}(\cos^{-1}(s),t)&=\begin{cases}
	-2t\lrp{1-\frac{2\cos^{-1}(t)}{\pi}}+2g_1(t),&  -1\leq s\leq 2t^2-1,\\
	-2t\lrp{1-\frac{\cos^{-1}(t)}{\pi}-\frac{\cos^{-1}(s)}{2\pi}}+g_1(t),&  2t^2-1\leq s< 1
	\end{cases}
	\end{align*}
	is well-defined and bounded, the exchange of integral and differential is possible.
\end{proof}

\section{Auxiliary results}
\label{sec:app_aux_results}

\begin{lemma}\label{lemma:aux}
	Let $q\geq2$. Then:
	\begin{align}
	\int_0^{\cos(\theta/2)}F_{q-1}\lrp{\frac{t\tan\lrp{\theta/2}}{(1-t^2)^{1/2}}}\,\mathrm{d}F_q(t)=F_q\lrp{\cos\lrp{\tfrac{\theta}{2}}}+\frac{\theta}{4\pi}-\frac{3}{4}.\label{eq:lemphi}
	\end{align}
\end{lemma}

\begin{proof}{} 
	Denote $\phi_q(\theta)$ to the left hand side of \eqref{eq:lemphi}. Then:
	\begin{align}
	\phi_q(\theta)&=F_q\lrp{\cos\lrp{\tfrac{\theta}{2}}}-\frac{1}{2}+\int_0^{\cos(\theta/2)}\lrp{F_{q-1}\lrp{\frac{t\tan\lrp{\theta/2}}{(1-t^2)^{1/2}}}-1}\,\mathrm{d}F_q(t)\notag\\
	&=:F_q\lrp{\cos\lrp{\tfrac{\theta}{2}}}-\frac{1}{2}+\phi_q^*(\theta).\label{eq:phi_star1}
	\end{align}
	The derivative of $\phi_q^*(\theta)$ with respect to $\theta$ is
	\begin{align}
	\frac{\partial}{\partial\theta}\phi_q^*(\theta)=&-\frac{1}{2}\sin\lrp{\tfrac{\theta}{2}}\lrp{F_{q-1}(1)-1}f_q\lrp{\cos\lrp{\tfrac{\theta}{2}}}\notag\\
	&+\frac{1}{2}\frac{\sec^2\lrp{\theta/2}}{\mathrm{B}\big(\tfrac{1}{2},\tfrac{q-1}{2}\big)\mathrm{B}\lrp{\tfrac{1}{2},\tfrac{q}{2}}}\int_0^{\cos(\theta/2)}t\lrp{1-\frac{t^2}{\cos^2(\theta/2)}}^{\frac{q-3}{2}}\,\mathrm{d}t\notag\\
	=&\;\frac{1}{4\pi}\label{eq:der_phi_star1}.
	\end{align}
	The proof is concluded from \eqref{eq:phi_star1} and \eqref{eq:der_phi_star1}, and since $\phi_q^*(\pi)=0$.
\end{proof}

\begin{lemma}\label{lem:A_q1}
	Let $\alpha$, $\theta \in [0,\pi]$. If $q=1$, then 
	\begin{align*}
	A(\theta,\cos(\alpha))=1-\frac{2}{\pi}\alpha+A^*(\theta,\cos(\alpha)),
	\end{align*}
	where
	\begin{align}\label{eq:defA_theta_alpha}
	A^*(\theta,\cos(\alpha))&:=\begin{cases}
	0, & 0\leq\alpha \leq \frac{\theta_{ij}}{2},\\
	\frac{2\alpha-\theta_{ij}}{2\pi}, & \frac{\theta_{ij}}{2} < \alpha < \pi-\frac{\theta_{ij}}{2},\\
	\frac{2\alpha}{\pi}-1, & \pi -\frac{\theta_{ij}}{2}\leq \alpha \leq   \pi.
	\end{cases}
	\end{align}	
\end{lemma}

\begin{proof}{} 
	By the definition of $A(\theta,\cos(\alpha))$, we have
	\begin{align*}
	A(\theta,\cos(\alpha))&=1-2A(0,\cos(\alpha))+A^*(\theta,\cos(\alpha))=1-\frac{2\alpha}{\pi}+A^*(\theta,\cos(\alpha))\notag
	\end{align*}
	where 
	\begin{align*} 
	A^*(\theta_{ij},\cos(\alpha))=\frac{1}{2\pi}\int_{\Om{1}}1_{\{\angle(\bga,\bX_i)\leq \alpha,\angle(\bga,\bX_j)\leq \alpha\}}\,\omega_q(\mathrm{d}\bga).
	\end{align*}
	Simple geometric arguments yield \eqref{eq:defA_theta_alpha}.
\end{proof}

\begin{lemma} \label{Lemm:Gamma} 
	If $a>b$ and $\gamma \in \R$, then the order of $\Gamma(\gamma k +a)/\Gamma(\gamma k +b)$ as $k \to \infty$ is $(\gamma k +a)^{a-b}$.
\end{lemma}

\begin{proof}{} 
	Stirling's equivalence gives
	\begin{align*}
	\frac{\Gamma(\gamma k +a)}{\Gamma(\gamma k +b)}
	& \sim
	\frac{e^{-(\gamma k +a)}}{e^{-(\gamma k +b)}}
	\frac{(\gamma k +a)^{\gamma k +a}}{(\gamma k +b)^{\gamma k +b}}
	\frac{(\gamma k +b)^{1/2}}{(\gamma k +a)^{1/2}}
	\\
	& \sim
	e^{b-a} 
	\left(\frac{\gamma k +a}{\gamma k +b}\right)^{\gamma k +b}
	(\gamma k +a)^{a-b}
	\\
	& \sim
	e^{b-a} 
	e^{a-b}
	(\gamma k +a)^{a-b}.
	\end{align*}
\end{proof}

\begin{lemma}\label{lemma:IntGegem_2}
	Let $k\geq 1$ and $q\geq 2$. For $x\in[-1,1]$, 
	\begin{align*}
	\left|C_k^{(q-1)/2}(x)\right|  \leq  
	\frac{\Gamma(k+(q-1)/2)}{\Gamma(q-1)\Gamma(k+1)}.
	\end{align*}
\end{lemma}

\begin{proof}{} 
	The results follows from equation (11) in \cite{Lohofer1991} and equation 18.14.4 in \cite{NIST:DLMF}.
\end{proof}

\begin{lemma}\label{lemma:v_{k,q}_x}
	Let $x\in[-1,1]$ and consider the function $f^x_q(z)= (F_q(x))^{-1} 1_{\{z \leq x\}}, z \in [-1,1]$. For $k\geq 1$, denote by $(1+2k/(q-1))v_{k,q}^x$ and by $2v_{k,q}^x$ to the Gegenbauer ($q\geq 2$) and Chebyshev coefficients ($q=1$) of $f^x_q$, respectively. Then, 
	\begin{align*}
	|v_{k,q}^x|\leq\begin{cases}
	\lrp{k\pi F_1(x)}^{-1}, & q=1,\\
	2^{q-1/2}\Gamma\lrp{\frac{q+1}{2}}^2\lrp{\pi F_q(x)}^{-1} \mathcal{O}\left((k+q/2)^{-(q+1)/2}\right),& q\geq 2.
	\end{cases}
	\end{align*}
\end{lemma}

\begin{proof}{} 
	For $q=1$, we have that 
	\begin{align*}
	|v_{k,1}^x| &= \frac{1}{2c_{k,1}F_1(x)}\left\vert\int_{-1}^x T_k(z) (1-z^2)^{-1/2}\, \mathrm{d}z\right\vert
	=\frac{1}{k\pi F_1(x)}|\sin(k\cos^{-1}(x))|
	\leq \frac{1}{k\pi F_1(x)}.
	\end{align*}	
	For $q\geq 2$, equation 18.17.1 in \cite{NIST:DLMF} gives
	\begin{align}\label{Eq:IntGegem_1}
	\int_0^x C_k^{(q-1)/2}(z)  (1-z^2)^{q/2-1}\,\mathrm{d}z=  \frac{q-1}{k(k+q-1)}
	\left(
	C_{k-1}^{(q+1)/2}(0) - (1-x^2)^{q/2}C_{k-1}^{(q+1)/2}(x)
	\right).
	\end{align}
	The parity of $C_k^{(q-1)/2}$, jointly with the definition of the Gegenbauer coefficients and \eqref{Eq:IntGegem_1}, give that
	\begin{align}
	v_{k,q}^x 
	%
	= &\;
	\frac{1}{c_{k,q}\lrp{1+\frac{2k}{q-1}}F_q(x)}\left(\int_{-1}^0 C_k^{(q-1)/2}(z) (1-z^2)^{q/2-1}\, \mathrm{d}z\right.\nonumber\\
	& \phantom{\frac{1}{c_{k,q}\lrp{1+\frac{2k}{q-1}}F_q(x)}()}+\left.
	\int_{0}^x C_k^{(q-1)/2}(z) (1-z^2)^{q/2-1}\, \mathrm{d}z\right)\nonumber
	\\
	=&\;
	\frac{2^{q-1}\Gamma\lrp{\frac{q+1}{2}}^2}{\pi(q-1)F_q(x)}  \frac{(1 - q)(1-x^2)^{q/2}C_{k-1}^{(q+1)/2}(x)}{k(k+q-1)}\nonumber
	\\
	=&\;
	\frac{-2^{q-1}\Gamma\lrp{\frac{q+1}{2}}^2}{\pi F_q(x)} \frac{(k-1)!}{\Gamma(k+q)}(1-x^2)^{q/2}C_{k-1}^{(q+1)/2}(x)\label{eq:v_{k,q}_x}.
	\end{align}
	Then, from \eqref{eq:v_{k,q}_x}, it follows that
	\begin{align}
	|v_{k,q}^x| &\leq \frac{2^{q-1/2}\Gamma\lrp{\frac{q+1}{2}}^2}{\pi F_q(x)}  \frac{(k-1)!}{\Gamma(k+q)} C_{k-1}^{(q+1)/2}(x)\nonumber\\
	&\leq \frac{2^{q-1/2}\Gamma\lrp{\frac{q+1}{2}}^2}{\pi F_q(x)} \frac{(k-1)!}{\Gamma(k+q)} \sup_{x \in [-1,1]}
	\left|C_{k-1}^{(q+1)/2}(x)\right|\nonumber\\
	&\leq \frac{2^{q-1/2}\Gamma\lrp{\frac{q+1}{2}}^2}{\pi F_q(x)}   \frac{(k-1)!}{\Gamma(k+q)}\frac{\Gamma(k+ (q-1)/2)}{\Gamma(k)} \label{Eq:31416}\\
	&=  \frac{2^{q-1/2}\Gamma\lrp{\frac{q+1}{2}}^2}{\pi F_q(x)}\mathcal{O}\left((k+q/2)^{-(q+1)/2}\right)\label{Eq:1312}
	\end{align}
	where \eqref{Eq:31416} stems from Lemma \ref{lemma:IntGegem_2} and \eqref{Eq:1312} from Lemma \ref{Lemm:Gamma}. 
\end{proof}

\begin{lemma} \label{Lemma_serie_q1}
	It occurs that $b_{k,1}^\mathrm{AD}=\mathcal{O}(\log (k)/k^2)$ and, therefore, that $\sum_{k=1}^\infty b_{k,1}^\mathrm{AD} <\infty$. 	
\end{lemma}

\begin{proof}{} 
	It happens that
	\begin{align}\label{Eq.Lemma_serie_1}
	b_{k,1}^\mathrm{AD}\leq 
	\frac{2}{\pi k^2}
	\left(
	\int_0^{\pi/2}\frac{1-\cos(2k\theta)}{\theta}\,\mathrm{d}\theta
	+
	\int_{\pi/2}^\pi\frac{1-\cos(2k\theta)}{\pi-\theta}\,\mathrm{d}\theta
	\right).
	\end{align}
	We will only consider the first integral in \eqref{Eq.Lemma_serie_1} because the other one is handled similarly. Obviously, if $k\ge 1$, then
	\[
	\int_0^{\pi/2}\frac{1-\cos(2k\theta)}{\theta}\,\mathrm{d}\theta
	=
	\int_0^{k \pi}\frac{1-\cos(u)}{u}\,\mathrm{d}u
	\le 
	I_1 + \sum_{h=1}^{k-1} \frac 1{h \pi}  \int_{h \pi}^{(h+1) \pi}(1-\cos(u)) \,\mathrm{d}u,
	\]	
	where $I_1 = \int_0^{ \pi}\frac{1-\cos(u)}{u}\,\mathrm{d}u< \infty$. Therefore,
	\[
	\int_0^{\pi/2}\frac{1-\cos(2k\theta)}{\theta}\,\mathrm{d}\theta
	\le
	I_1 + \sum_{h=1}^{k-1} \frac {1}{h} =\mathcal{O}(\log (k))
	\]
	and \eqref{Eq.Lemma_serie_1} gives that $b_{k,1}^\mathrm{AD}= \mathcal{O}(\log (k)/k^2)$. Note that $\sum_{k=1}^\infty \frac{\log(k)}{k^2}=-\frac{1}{6}\pi^2(-12\log(A)+\gamma+\log(2)+\log(\pi))<\infty$, where $A$ and $\gamma$ are the Glaisher--Kinkelin and Euler--Mascheroni constants, respectively.
\end{proof}

\begin{remark} 
	Lemma \ref{Lemma_serie_q1} also shows that $\psi^\mathrm{AD}_1 \in L_q^2[-1,1]$.
\end{remark}

\section{Further simulation results}
\label{sec_app_powers}

Tables \ref{tab:pow:2}--\ref{tab:pow:5} contain extended simulation results for the empirical power investigation of Section \ref{sec:power}. The simulation settings are exactly those described in that section.\\

\begin{table}
	\centering
	\setlength{\tabcolsep}{2.5pt}
	\scriptsize
	\begin{tabular}{lll|R{1.25cm}R{1.25cm}R{1.25cm}R{1.25cm}R{1.25cm}R{1.25cm}R{1.25cm}R{1.25cm}R{1.25cm}}
		\toprule
		\scriptsize DGP & $n$ & $\kappa$ & \scriptsize Rayleigh & \scriptsize Bingham & \scriptsize Ajne & \scriptsize Gin\'e & \scriptsize CCF09 & \scriptsize Bakshaev & \scriptsize CvM & \scriptsize AD & \scriptsize Rt \\%
		\midrule
		CvM       &  50 &   0.25   &               0.0746 &              0.0557 &           0.0749 &                     0.0559 &            0.0735 &               0.0755 &          0.0757 &     \bf 0.0760 &         0.0755 \\
		&     &   0.50   &               0.1570 &              0.0743 &           0.1580 &                     0.0749 &            0.1512 &               0.1619 &          0.1627 &     \bf 0.1635 &         0.1617 \\
		&     &   0.75   &               0.3077 &              0.1073 &           0.3104 &                     0.1096 &            0.2949 &               0.3201 &          0.3222 &     \bf 0.3244 &         0.3197 \\
		\cmidrule{2-12}
		& 100 &   0.25   &               0.1006 &              0.0619 &           0.1010 &                     0.0619 &            0.0980 &               0.1025 &      \bf 0.1027 &     \bf 0.1028 &         0.1024 \\
		&     &   0.50   &               0.2773 &              0.1004 &           0.2793 &                     0.1021 &            0.2653 &               0.2879 &          0.2897 &     \bf 0.2918 &         0.2879 \\
		&     &   0.75   &               0.5580 &              0.1709 &           0.5628 &                     0.1760 &            0.5398 &               0.5806 &          0.5845 &     \bf 0.5885 &         0.5801 \\
		\cmidrule{2-12}
		& 200 &   0.25   &               0.1571 &              0.0744 &           0.1582 &                     0.0751 &            0.1509 &               0.1614 &          0.1620 &     \bf 0.1628 &         0.1613 \\
		&     &   0.50   &               0.5063 &              0.1566 &           0.5113 &                     0.1609 &            0.4886 &               0.5284 &          0.5321 &     \bf 0.5365 &         0.5282 \\
		&     &   0.75   &               0.8599 &              0.3066 &           0.8650 &                     0.3194 &            0.8484 &               0.8802 &          0.8833 &     \bf 0.8866 &         0.8799 \\ \midrule
		AD        &  50 &   0.25   &               0.1897 &              0.0933 &           0.1924 &                     0.0960 &            0.1884 &               0.1998 &          0.2020 &     \bf 0.2053 &         0.1995 \\
		&     &   0.50   &               0.6134 &              0.2429 &           0.6235 &                     0.2578 &            0.6117 &               0.6480 &          0.6545 &     \bf 0.6635 &         0.6468 \\
		&     &   0.75   &               0.9377 &              0.4936 &           0.9424 &                     0.5255 &            0.9351 &               0.9521 &          0.9542 &     \bf 0.9568 &         0.9514 \\
		\cmidrule{2-12}
		& 100 &   0.25   &               0.3419 &              0.1409 &           0.3478 &                     0.1471 &            0.3399 &               0.3647 &          0.3696 &     \bf 0.3765 &         0.3642 \\
		&     &   0.50   &               0.9002 &              0.4463 &           0.9071 &                     0.4776 &            0.9019 &               0.9234 &          0.9271 &     \bf 0.9319 &         0.9225 \\
		&     &   0.75   &               0.9989 &              0.8023 &           0.9992 &                     0.8361 &            0.9989 &               0.9994 &          0.9995 &     \bf 0.9996 &         0.9994 \\
		\cmidrule{2-12}
		& 200 &   0.25   &               0.6113 &              0.2438 &           0.6227 &                     0.2586 &            0.6144 &               0.6509 &          0.6588 &     \bf 0.6698 &         0.6501 \\
		&     &   0.50   &               0.9966 &              0.7517 &           0.9973 &                     0.7914 &            0.9972 &               0.9985 &          0.9986 &     \bf 0.9989 &         0.9984 \\
		&     &   0.75   &           \bf 1.0000 &              0.9818 &       \bf 1.0000 &                     0.9894 &        \bf 1.0000 &           \bf 1.0000 &      \bf 1.0000 &     \bf 1.0000 &     \bf 1.0000 \\ \midrule
		Rt        &  50 &   0.25   &               0.0867 &              0.0588 &           0.0867 &                     0.0590 &            0.0845 &           \bf 0.0875 &      \bf 0.0875 &     \bf 0.0876 &     \bf 0.0875 \\
		&     &   0.50   &               0.2136 &              0.0870 &           0.2136 &                     0.0880 &            0.2099 &               0.2179 &          0.2187 &     \bf 0.2190 &     \bf 0.2191 \\
		&     &   0.75   &               0.4514 &              0.1383 &           0.4511 &                     0.1411 &            0.4701 &               0.4707 &          0.4742 &     \bf 0.4780 &         0.4756 \\
		\cmidrule{2-12}
		& 100 &   0.25   &               0.1271 &              0.0678 &           0.1268 &                     0.0682 &            0.1233 &           \bf 0.1286 &      \bf 0.1287 &     \bf 0.1285 &     \bf 0.1287 \\
		&     &   0.50   &               0.4020 &              0.1282 &           0.4014 &                     0.1303 &            0.4124 &               0.4175 &          0.4203 &     \bf 0.4225 &         0.4213 \\
		&     &   0.75   &               0.7776 &              0.2380 &           0.7775 &                     0.2453 &        \bf 0.8305 &               0.8125 &          0.8186 &         0.8250 &         0.8199 \\
		\cmidrule{2-12}
		& 200 &   0.25   &               0.2154 &              0.0871 &           0.2152 &                     0.0881 &            0.2122 &               0.2205 &          0.2210 &         0.2213 &     \bf 0.2221 \\
		&     &   0.50   &               0.7091 &              0.2158 &           0.7091 &                     0.2217 &        \bf 0.7506 &               0.7402 &          0.7455 &     \bf 0.7510 &         0.7470 \\
		&     &   0.75   &               0.9795 &              0.4381 &           0.9798 &                     0.4540 &        \bf 0.9938 &               0.9894 &          0.9905 &         0.9917 &         0.9907 \\ \midrule
		vMF       &  50 &   0.25   &           \bf 0.1816 &              0.0504 &           0.1814 &                     0.0506 &            0.1642 &               0.1804 &          0.1792 &         0.1770 &         0.1795 \\
		&     &   0.50   &           \bf 0.5842 &              0.0562 &           0.5830 &                     0.0562 &            0.5301 &               0.5797 &          0.5767 &         0.5700 &         0.5778 \\
		&     &   0.75   &           \bf 0.9112 &              0.0816 &           0.9105 &                     0.0813 &            0.8747 &               0.9085 &          0.9067 &         0.9026 &         0.9076 \\
		\cmidrule{2-12}
		& 100 &   0.25   &           \bf 0.3302 &              0.0511 &           0.3291 &                     0.0511 &            0.2946 &               0.3272 &          0.3251 &         0.3203 &         0.3257 \\
		&     &   0.50   &           \bf 0.8867 &              0.0634 &           0.8859 &                     0.0634 &            0.8451 &               0.8837 &          0.8816 &         0.8769 &         0.8823 \\
		&     &   0.75   &           \bf 0.9979 &              0.1164 &           0.9978 &                     0.1155 &            0.9951 &               0.9977 &          0.9976 &         0.9973 &         0.9976 \\
		\cmidrule{2-12}
		& 200 &   0.25   &           \bf 0.5991 &              0.0521 &           0.5979 &                     0.0522 &            0.5433 &               0.5943 &          0.5910 &         0.5843 &         0.5927 \\
		&     &   0.50   &           \bf 0.9956 &              0.0782 &       \bf 0.9956 &                     0.0778 &            0.9912 &               0.9954 &          0.9952 &         0.9949 &         0.9953 \\
		&     &   0.75   &           \bf 1.0000 &              0.1925 &       \bf 1.0000 &                     0.1909 &        \bf 1.0000 &           \bf 1.0000 &      \bf 1.0000 &     \bf 1.0000 &     \bf 1.0000 \\ \midrule
		SC        &  50 &   0.25   &               0.3017 &              0.0969 &           0.3014 &                     0.0964 &            0.2891 &               0.3080 &          0.3088 &         0.3085 &     \bf 0.3092 \\
		&     &   0.50   &               0.7906 &              0.1780 &           0.7906 &                     0.1765 &            0.7987 &               0.8170 &          0.8211 &     \bf 0.8237 &         0.8216 \\
		&     &   0.75   &               0.9738 &              0.2405 &           0.9741 &                     0.2380 &            0.9823 &               0.9849 &          0.9861 &     \bf 0.9869 &         0.9862 \\
		\cmidrule{2-12}
		& 100 &   0.25   &               0.5643 &              0.1498 &           0.5634 &                     0.1486 &            0.5584 &               0.5844 &          0.5875 &     \bf 0.5887 &     \bf 0.5883 \\
		&     &   0.50   &               0.9843 &              0.3243 &           0.9841 &                     0.3212 &            0.9891 &               0.9910 &          0.9918 &     \bf 0.9922 &         0.9919 \\
		&     &   0.75   &               0.9999 &              0.4487 &           0.9999 &                     0.4446 &        \bf 1.0000 &           \bf 1.0000 &      \bf 1.0000 &     \bf 1.0000 &     \bf 1.0000 \\
		\cmidrule{2-12}
		& 200 &   0.25   &               0.8767 &              0.2640 &           0.8760 &                     0.2615 &            0.8845 &               0.8982 &          0.9013 &     \bf 0.9033 &         0.9022 \\
		&     &   0.50   &               1.0000 &              0.5913 &           1.0000 &                     0.5868 &        \bf 1.0000 &           \bf 1.0000 &      \bf 1.0000 &     \bf 1.0000 &     \bf 1.0000 \\
		&     &   0.75   &           \bf 1.0000 &              0.7605 &       \bf 1.0000 &                     0.7561 &        \bf 1.0000 &           \bf 1.0000 &      \bf 1.0000 &     \bf 1.0000 &     \bf 1.0000 \\ \midrule
		W        &  50 &   0.25   &               0.0514 &          \bf 0.2580 &           0.0518 &                     0.2559 &            0.0888 &               0.0683 &          0.0746 &         0.0854 &         0.0742 \\
		&     &   0.50   &               0.0540 &          \bf 0.7809 &           0.0551 &                     0.7763 &            0.2482 &               0.1438 &          0.1906 &         0.2662 &         0.1906 \\
		&     &   0.75   &               0.0575 &          \bf 0.9854 &           0.0594 &                     0.9846 &            0.5442 &               0.3535 &          0.4981 &         0.6568 &         0.5024 \\
		\cmidrule{2-12}
		& 100 &   0.25   &               0.0513 &          \bf 0.4808 &           0.0516 &                     0.4767 &            0.1394 &               0.0901 &          0.1075 &         0.1374 &         0.1073 \\
		&     &   0.50   &               0.0538 &          \bf 0.9785 &           0.0547 &                     0.9773 &            0.5358 &               0.3560 &          0.4916 &         0.6396 &         0.4946 \\
		&     &   0.75   &               0.0573 &          \bf 1.0000 &           0.0590 &                 \bf 1.0000 &            0.9283 &               0.8734 &          0.9493 &         0.9826 &         0.9510 \\
		\cmidrule{2-12}
		& 200 &   0.25   &               0.0512 &          \bf 0.7959 &           0.0514 &                     0.7918 &            0.2689 &               0.1576 &          0.2122 &         0.2948 &         0.2131 \\
		&     &   0.50   &               0.0535 &          \bf 0.9999 &           0.0545 &                 \bf 0.9999 &            0.9146 &               0.8528 &          0.9350 &         0.9751 &         0.9370 \\
		&     &   0.75   &               0.0569 &          \bf 1.0000 &           0.0589 &                 \bf 1.0000 &            0.9999 &               0.9999 &          1.0000 &     \bf 1.0000 &         1.0000 \\ \bottomrule
	\end{tabular}
	\caption{\small Empirical powers for the uniformity tests on $\Omega_1$. The description of Table \ref{tab:pow:1} applies. \label{tab:pow:2}}
\end{table}

\begin{table}
	\centering
	\setlength{\tabcolsep}{2.5pt}
	\scriptsize
	\begin{tabular}{lll|R{1.25cm}R{1.25cm}R{1.25cm}R{1.25cm}R{1.25cm}R{1.25cm}R{1.25cm}R{1.25cm}R{1.25cm}}
		\toprule
		\scriptsize DGP & $n$ & $\kappa$ & \scriptsize Rayleigh & \scriptsize Bingham & \scriptsize Ajne & \scriptsize Gin\'e & \scriptsize CCF09 & \scriptsize Bakshaev & \scriptsize CvM & \scriptsize AD & \scriptsize Rt \\%
		\midrule
		CvM       &  50 &   0.25   &               0.0700 &              0.0537 &           0.0700 &                     0.0539 &            0.0672 &           \bf 0.0704 &      \bf 0.0704 &     \bf 0.0705 &     \bf 0.0705 \\
		&     &   0.50   &               0.1360 &              0.0645 &           0.1362 &                     0.0649 &            0.1233 &               0.1383 &          0.1383 &     \bf 0.1390 &         0.1379 \\
		&     &   0.75   &               0.2622 &              0.0840 &           0.2630 &                     0.0851 &            0.2303 &               0.2686 &          0.2686 &     \bf 0.2700 &         0.2673 \\
		\cmidrule{2-12}
		& 100 &   0.25   &               0.0905 &              0.0567 &           0.0905 &                     0.0568 &            0.0844 &           \bf 0.0914 &      \bf 0.0914 &     \bf 0.0916 &         0.0913 \\
		&     &   0.50   &               0.2367 &              0.0794 &           0.2377 &                     0.0801 &            0.2083 &               0.2424 &          0.2424 &     \bf 0.2434 &         0.2414 \\
		&     &   0.75   &               0.4918 &              0.1216 &           0.4941 &                     0.1237 &            0.4330 &               0.5043 &          0.5043 &     \bf 0.5068 &         0.5022 \\
		\cmidrule{2-12}
		& 200 &   0.25   &               0.1359 &              0.0638 &           0.1362 &                     0.0640 &            0.1230 &               0.1377 &          0.1377 &     \bf 0.1380 &         0.1374 \\
		&     &   0.50   &               0.4411 &              0.1118 &           0.4432 &                     0.1137 &            0.3870 &               0.4528 &          0.4528 &     \bf 0.4555 &         0.4500 \\
		&     &   0.75   &               0.8096 &              0.2070 &           0.8124 &                     0.2124 &            0.7499 &               0.8231 &          0.8231 &     \bf 0.8259 &         0.8200 \\
		\midrule
		AD        &  50 &   0.25   &               0.1607 &              0.0759 &           0.1615 &                     0.0766 &            0.1462 &               0.1656 &          0.1656 &     \bf 0.1674 &         0.1645 \\
		&     &   0.50   &               0.5367 &              0.1691 &           0.5408 &                     0.1735 &            0.4824 &               0.5555 &          0.5555 &     \bf 0.5607 &         0.5515 \\
		&     &   0.75   &               0.8969 &              0.3460 &           0.8993 &                     0.3568 &            0.8543 &               0.9075 &          0.9075 &     \bf 0.9096 &         0.9054 \\
		\cmidrule{2-12}
		& 100 &   0.25   &               0.2887 &              0.1038 &           0.2910 &                     0.1056 &            0.2560 &               0.3004 &          0.3004 &     \bf 0.3038 &         0.2979 \\
		&     &   0.50   &               0.8507 &              0.3110 &           0.8542 &                     0.3213 &            0.8037 &               0.8670 &          0.8670 &     \bf 0.8710 &         0.8637 \\
		&     &   0.75   &               0.9969 &              0.6342 &           0.9971 &                     0.6510 &            0.9931 &               0.9977 &          0.9977 &     \bf 0.9978 &         0.9975 \\
		\cmidrule{2-12}
		& 200 &   0.25   &               0.5361 &              0.1669 &           0.5403 &                     0.1715 &            0.4824 &               0.5578 &          0.5578 &     \bf 0.5640 &         0.5525 \\
		&     &   0.50   &               0.9919 &              0.5815 &           0.9924 &                     0.5995 &            0.9853 &               0.9941 &          0.9941 &     \bf 0.9946 &         0.9937 \\
		&     &   0.75   &           \bf 1.0000 &              0.9225 &       \bf 1.0000 &                     0.9320 &        \bf 1.0000 &           \bf 1.0000 &      \bf 1.0000 &     \bf 1.0000 &     \bf 1.0000 \\
		\midrule
		Rt        &  50 &   0.25   &           \bf 0.0785 &              0.0540 &       \bf 0.0785 &                     0.0541 &            0.0749 &           \bf 0.0787 &      \bf 0.0787 &     \bf 0.0786 &     \bf 0.0787 \\
		&     &   0.50   &               0.1778 &              0.0657 &           0.1777 &                     0.0664 &            0.1636 &           \bf 0.1788 &      \bf 0.1788 &     \bf 0.1787 &     \bf 0.1790 \\
		&     &   0.75   &               0.3765 &              0.0865 &           0.3763 &                     0.0883 &            0.3543 &               0.3825 &          0.3825 &     \bf 0.3831 &         0.3821 \\
		\cmidrule{2-12}
		& 100 &   0.25   &               0.1095 &              0.0575 &           0.1093 &                     0.0578 &            0.1013 &           \bf 0.1100 &      \bf 0.1100 &         0.1098 &     \bf 0.1101 \\
		&     &   0.50   &               0.3362 &              0.0820 &           0.3360 &                     0.0833 &            0.3127 &           \bf 0.3413 &      \bf 0.3413 &     \bf 0.3414 &     \bf 0.3413 \\
		&     &   0.75   &               0.7019 &              0.1292 &           0.7019 &                     0.1336 &            0.6965 &               0.7192 &          0.7192 &     \bf 0.7232 &         0.7174 \\
		\cmidrule{2-12}
		& 200 &   0.25   &               0.1798 &              0.0651 &           0.1796 &                     0.0655 &            0.1646 &           \bf 0.1809 &      \bf 0.1809 &     \bf 0.1808 &     \bf 0.1807 \\
		&     &   0.50   &               0.6288 &              0.1186 &           0.6286 &                     0.1221 &            0.6158 &               0.6428 &          0.6428 &     \bf 0.6456 &         0.6411 \\
		&     &   0.75   &               0.9612 &              0.2282 &           0.9612 &                     0.2402 &            0.9696 &               0.9701 &          0.9701 &     \bf 0.9722 &         0.9691 \\
		\midrule
		vMF       &  50 &   0.25   &           \bf 0.1180 &              0.0506 &           0.1176 &                     0.0506 &            0.1065 &               0.1172 &          0.1172 &         0.1167 &         0.1177 \\
		&     &   0.50   &           \bf 0.3622 &              0.0533 &           0.3614 &                     0.0532 &            0.3124 &               0.3595 &          0.3595 &         0.3563 &         0.3610 \\
		&     &   0.75   &           \bf 0.7091 &              0.0645 &           0.7080 &                     0.0642 &            0.6346 &               0.7048 &          0.7048 &         0.7003 &         0.7066 \\
		\cmidrule{2-12}
		& 100 &   0.25   &           \bf 0.1980 &              0.0507 &           0.1977 &                     0.0504 &            0.1725 &               0.1966 &          0.1966 &         0.1950 &         0.1973 \\
		&     &   0.50   &           \bf 0.6648 &              0.0558 &           0.6638 &                     0.0556 &            0.5895 &               0.6606 &          0.6606 &         0.6560 &         0.6628 \\
		&     &   0.75   &           \bf 0.9602 &              0.0809 &           0.9597 &                     0.0805 &            0.9286 &               0.9585 &          0.9585 &         0.9568 &         0.9593 \\
		\cmidrule{2-12}
		& 200 &   0.25   &           \bf 0.3698 &              0.0508 &           0.3690 &                     0.0509 &            0.3174 &               0.3662 &          0.3662 &         0.3629 &         0.3675 \\
		&     &   0.50   &           \bf 0.9396 &              0.0620 &           0.9391 &                     0.0617 &            0.8998 &               0.9374 &          0.9374 &         0.9353 &         0.9383 \\
		&     &   0.75   &           \bf 0.9998 &              0.1140 &           0.9998 &                     0.1132 &            0.9990 &               0.9997 &          0.9997 &         0.9997 &         0.9998 \\
		\midrule
		SC        &  50 &   0.25   &               0.1797 &              0.0672 &           0.1793 &                     0.0669 &            0.1624 &               0.1805 &          0.1805 &         0.1801 &     \bf 0.1808 \\
		&     &   0.50   &               0.5316 &              0.0942 &           0.5313 &                     0.0939 &            0.4932 &               0.5402 &          0.5402 &     \bf 0.5412 &         0.5388 \\
		&     &   0.75   &               0.8378 &              0.1135 &           0.8380 &                     0.1127 &            0.8187 &               0.8521 &          0.8521 &     \bf 0.8548 &         0.8495 \\
		\cmidrule{2-12}
		& 100 &   0.25   &               0.3410 &              0.0857 &           0.3403 &                     0.0855 &            0.3077 &           \bf 0.3458 &      \bf 0.3458 &     \bf 0.3460 &     \bf 0.3456 \\
		&     &   0.50   &               0.8716 &              0.1489 &           0.8711 &                     0.1480 &            0.8545 &               0.8858 &          0.8858 &     \bf 0.8887 &         0.8836 \\
		&     &   0.75   &               0.9942 &              0.1951 &           0.9941 &                     0.1937 &            0.9943 &               0.9964 &          0.9964 &     \bf 0.9968 &         0.9961 \\
		\cmidrule{2-12}
		& 200 &   0.25   &               0.6389 &              0.1293 &           0.6378 &                     0.1285 &            0.6015 &               0.6529 &          0.6529 &     \bf 0.6557 &         0.6504 \\
		&     &   0.50   &               0.9962 &              0.2788 &           0.9961 &                     0.2765 &            0.9962 &               0.9977 &          0.9977 &     \bf 0.9980 &         0.9975 \\
		&     &   0.75   &           \bf 1.0000 &              0.3846 &       \bf 1.0000 &                     0.3813 &        \bf 1.0000 &           \bf 1.0000 &      \bf 1.0000 &     \bf 1.0000 &     \bf 1.0000 \\
		\midrule
		W        &  50 &   0.25   &              0.0514 &          \bf 0.1540 &           0.0514 &                     0.1528 &            0.0597 &               0.0591 &          0.0591 &         0.0632 &         0.0570 \\
		&     &   0.50   &              0.0539 &          \bf 0.5682 &           0.0545 &                     0.5640 &            0.1081 &               0.0947 &          0.0947 &         0.1231 &         0.0824 \\
		&     &   0.75   &              0.0576 &          \bf 0.9315 &           0.0592 &                     0.9294 &            0.2457 &               0.1940 &          0.1940 &         0.3124 &         0.1466 \\
		\cmidrule{2-12}
		& 100 &   0.25   &              0.0513 &          \bf 0.2807 &           0.0514 &                     0.2782 &            0.0705 &               0.0677 &          0.0677 &         0.0774 &         0.0634 \\
		&     &   0.50   &              0.0536 &          \bf 0.8850 &           0.0543 &                     0.8823 &            0.2031 &               0.1643 &          0.1643 &         0.2570 &         0.1265 \\
		&     &   0.75   &              0.0573 &          \bf 0.9991 &           0.0588 &                     0.9990 &            0.5830 &               0.5265 &          0.5265 &         0.7639 &         0.3614 \\
		\cmidrule{2-12}
		& 200 &   0.25   &              0.0509 &          \bf 0.5377 &           0.0510 &                     0.5343 &            0.0988 &               0.0884 &          0.0884 &         0.1138 &         0.0772 \\
		&     &   0.50   &              0.0531 &          \bf 0.9966 &           0.0536 &                     0.9964 &            0.4843 &               0.4183 &          0.4183 &         0.6500 &         0.2816 \\
		&     &   0.75   &              0.0567 &          \bf 1.0000 &           0.0581 &                 \bf 1.0000 &            0.9695 &               0.9741 &          0.9741 &         0.9973 &         0.8987 \\\bottomrule
	\end{tabular}
	\caption{\small Empirical powers for the uniformity tests on $\Omega_2$. The description of Table \ref{tab:pow:1} applies. \label{tab:pow:3}}
\end{table}

\begin{table}
	\centering
	\setlength{\tabcolsep}{2.5pt}
	\scriptsize
	\begin{tabular}{lll|R{1.25cm}R{1.25cm}R{1.25cm}R{1.25cm}R{1.25cm}R{1.25cm}R{1.25cm}R{1.25cm}R{1.25cm}}
		\toprule
		\scriptsize DGP & $n$ & $\kappa$ & \scriptsize Rayleigh & \scriptsize Bingham & \scriptsize Ajne & \scriptsize Gin\'e & \scriptsize CCF09 & \scriptsize Bakshaev & \scriptsize CvM & \scriptsize AD & \scriptsize Rt \\%
		\midrule
		CvM       &  50 &   0.25   &               0.0663 &              0.0522 &       \bf 0.0665 &                     0.0522 &            0.0648 &           \bf 0.0668 &          0.0667 &     \bf 0.0667 &         0.0665 \\
		&     &   0.50   &               0.1214 &              0.0589 &           0.1218 &                     0.0589 &            0.1126 &           \bf 0.1230 &      \bf 0.1230 &     \bf 0.1231 &         0.1224 \\
		&     &   0.75   &               0.2288 &              0.0711 &           0.2295 &                     0.0718 &            0.2039 &               0.2327 &          0.2325 &     \bf 0.2330 &         0.2314 \\
		\cmidrule{2-12}
		& 100 &   0.25   &               0.0835 &              0.0546 &           0.0834 &                     0.0547 &            0.0796 &           \bf 0.0839 &      \bf 0.0839 &     \bf 0.0839 &     \bf 0.0837 \\
		&     &   0.50   &               0.2064 &              0.0692 &           0.2068 &                     0.0696 &            0.1839 &               0.2098 &          0.2095 &     \bf 0.2104 &         0.2087 \\
		&     &   0.75   &               0.4378 &              0.0965 &           0.4387 &                     0.0975 &            0.3835 &               0.4449 &          0.4445 &     \bf 0.4460 &         0.4428 \\
		\cmidrule{2-12}
		& 200 &   0.25   &               0.1198 &              0.0586 &           0.1198 &                     0.0588 &            0.1110 &           \bf 0.1209 &      \bf 0.1209 &     \bf 0.1211 &         0.1205 \\
		&     &   0.50   &               0.3895 &              0.0899 &           0.3903 &                     0.0907 &            0.3414 &               0.3965 &          0.3961 &     \bf 0.3976 &         0.3942 \\
		&     &   0.75   &               0.7583 &              0.1524 &           0.7599 &                     0.1549 &            0.6906 &               0.7676 &          0.7671 &     \bf 0.7688 &         0.7649 \\
		\midrule
		AD        &  50 &   0.25   &               0.1434 &              0.0675 &           0.1442 &                     0.0679 &            0.1317 &               0.1466 &          0.1464 &     \bf 0.1471 &         0.1455 \\
		&     &   0.50   &               0.4837 &              0.1333 &           0.4862 &                     0.1353 &            0.4298 &               0.4962 &          0.4952 &     \bf 0.4984 &         0.4922 \\
		&     &   0.75   &               0.8620 &              0.2639 &           0.8636 &                     0.2688 &            0.8070 &               0.8698 &          0.8694 &     \bf 0.8707 &         0.8676 \\
		\cmidrule{2-12}
		& 100 &   0.25   &               0.2547 &              0.0874 &           0.2557 &                     0.0883 &            0.2268 &               0.2619 &          0.2613 &     \bf 0.2637 &         0.2594 \\
		&     &   0.50   &               0.8092 &              0.2378 &           0.8113 &                     0.2426 &            0.7497 &               0.8218 &          0.8209 &     \bf 0.8241 &         0.8180 \\
		&     &   0.75   &               0.9942 &              0.5102 &           0.9943 &                     0.5193 &            0.9866 &               0.9950 &          0.9950 &     \bf 0.9951 &         0.9948 \\
		\cmidrule{2-12}
		& 200 &   0.25   &               0.4832 &              0.1316 &           0.4852 &                     0.1333 &            0.4290 &               0.4973 &          0.4964 &     \bf 0.5004 &         0.4927 \\
		&     &   0.50   &               0.9860 &              0.4602 &           0.9865 &                     0.4698 &            0.9733 &               0.9884 &          0.9883 &     \bf 0.9888 &         0.9877 \\
		&     &   0.75   &           \bf 1.0000 &              0.8404 &       \bf 1.0000 &                     0.8483 &            1.0000 &           \bf 1.0000 &      \bf 1.0000 &     \bf 1.0000 &     \bf 1.0000 \\
		\midrule
		Rt        &  50 &   0.25   &               0.0732 &              0.0523 &       \bf 0.0734 &                     0.0525 &            0.0713 &           \bf 0.0735 &      \bf 0.0735 &         0.0732 &     \bf 0.0734 \\
		&     &   0.50   &               0.1548 &              0.0586 &       \bf 0.1551 &                     0.0589 &            0.1464 &           \bf 0.1552 &      \bf 0.1553 &         0.1547 &     \bf 0.1552 \\
		&     &   0.75   &               0.3247 &              0.0702 &           0.3252 &                     0.0709 &            0.3094 &           \bf 0.3272 &      \bf 0.3272 &         0.3265 &         0.3269 \\
		\cmidrule{2-12}
		& 100 &   0.25   &           \bf 0.0988 &              0.0544 &       \bf 0.0987 &                     0.0548 &            0.0946 &           \bf 0.0988 &      \bf 0.0989 &     \bf 0.0987 &     \bf 0.0988 \\
		&     &   0.50   &               0.2902 &              0.0686 &           0.2903 &                     0.0694 &            0.2756 &           \bf 0.2924 &      \bf 0.2924 &     \bf 0.2923 &     \bf 0.2922 \\
		&     &   0.75   &               0.6357 &              0.0951 &           0.6360 &                     0.0973 &            0.6285 &               0.6453 &          0.6446 &     \bf 0.6467 &         0.6431 \\
		\cmidrule{2-12}
		& 200 &   0.25   &               0.1558 &              0.0586 &           0.1556 &                     0.0589 &            0.1472 &               0.1562 &      \bf 0.1564 &         0.1560 &         0.1562 \\
		&     &   0.50   &               0.5639 &              0.0893 &           0.5639 &                     0.0911 &            0.5528 &               0.5718 &          0.5714 &     \bf 0.5727 &         0.5699 \\
		&     &   0.75   &               0.9377 &              0.1508 &           0.9380 &                     0.1562 &            0.9450 &               0.9458 &          0.9453 &     \bf 0.9473 &         0.9437 \\
		\midrule
		vMF       &  50 &   0.25   &           \bf 0.0918 &              0.0497 &       \bf 0.0920 &                     0.0498 &            0.0868 &           \bf 0.0918 &      \bf 0.0919 &         0.0913 &     \bf 0.0919 \\
		&     &   0.50   &           \bf 0.2477 &              0.0513 &       \bf 0.2474 &                     0.0513 &            0.2208 &               0.2461 &          0.2464 &         0.2446 &         0.2469 \\
		&     &   0.75   &           \bf 0.5260 &              0.0570 &           0.5255 &                     0.0568 &            0.4667 &               0.5224 &          0.5231 &         0.5193 &         0.5244 \\
		\cmidrule{2-12}
		& 100 &   0.25   &           \bf 0.1406 &              0.0507 &           0.1403 &                     0.0507 &            0.1287 &               0.1399 &          0.1400 &         0.1393 &         0.1403 \\
		&     &   0.50   &           \bf 0.4806 &              0.0530 &           0.4798 &                     0.0532 &            0.4247 &               0.4774 &          0.4779 &         0.4749 &         0.4791 \\
		&     &   0.75   &           \bf 0.8577 &              0.0659 &           0.8570 &                     0.0659 &            0.8012 &               0.8548 &          0.8553 &         0.8525 &         0.8565 \\
		\cmidrule{2-12}
		& 200 &   0.25   &           \bf 0.2513 &              0.0498 &           0.2507 &                     0.0498 &            0.2242 &               0.2492 &          0.2497 &         0.2476 &         0.2503 \\
		&     &   0.50   &           \bf 0.8142 &              0.0558 &           0.8133 &                     0.0559 &            0.7535 &               0.8109 &          0.8115 &         0.8083 &         0.8127 \\
		&     &   0.75   &           \bf 0.9944 &              0.0817 &       \bf 0.9944 &                     0.0815 &            0.9868 &               0.9941 &          0.9942 &         0.9939 &         0.9943 \\
		\midrule
		SC        &  50 &   0.25   &           \bf 0.1305 &              0.0578 &       \bf 0.1305 &                     0.0578 &            0.1223 &           \bf 0.1305 &      \bf 0.1306 &         0.1301 &     \bf 0.1305 \\
		&     &   0.50   &               0.3750 &              0.0695 &           0.3753 &                     0.0694 &            0.3463 &               0.3766 &      \bf 0.3768 &         0.3755 &         0.3764 \\
		&     &   0.75   &               0.6760 &              0.0769 &           0.6765 &                     0.0766 &            0.6424 &           \bf 0.6818 &          0.6816 &         0.6811 &         0.6804 \\
		\cmidrule{2-12}
		& 100 &   0.25   &               0.2337 &              0.0670 &           0.2334 &                     0.0670 &            0.2138 &           \bf 0.2350 &      \bf 0.2350 &         0.2347 &         0.2347 \\
		&     &   0.50   &               0.7111 &              0.0933 &           0.7104 &                     0.0933 &            0.6767 &               0.7192 &          0.7187 &     \bf 0.7202 &         0.7170 \\
		&     &   0.75   &               0.9612 &              0.1109 &           0.9612 &                     0.1106 &            0.9532 &               0.9663 &          0.9660 &     \bf 0.9671 &         0.9648 \\
		\cmidrule{2-12}
		& 200 &   0.25   &               0.4579 &              0.0860 &           0.4569 &                     0.0858 &            0.4238 &               0.4634 &          0.4632 &     \bf 0.4637 &         0.4617 \\
		&     &   0.50   &               0.9685 &              0.1503 &           0.9683 &                     0.1496 &            0.9615 &               0.9732 &          0.9729 &     \bf 0.9740 &         0.9718 \\
		&     &   0.75   &               0.9999 &              0.1942 &           0.9999 &                     0.1929 &            0.9999 &               0.9999 &          0.9999 &     \bf 0.9999 &         0.9999 \\
		\midrule
		W        &  50 &   0.25   &               0.0510 &          \bf 0.1019 &           0.0511 &                     0.1015 &            0.0546 &               0.0547 &          0.0543 &         0.0561 &         0.0532 \\
		&     &   0.50   &               0.0529 &          \bf 0.3584 &           0.0533 &                     0.3555 &            0.0739 &               0.0726 &          0.0701 &         0.0811 &         0.0640 \\
		&     &   0.75   &               0.0560 &          \bf 0.7738 &           0.0570 &                     0.7706 &            0.1322 &               0.1170 &          0.1083 &         0.1508 &         0.0884 \\
		\cmidrule{2-12}
		& 100 &   0.25   &               0.0508 &          \bf 0.1677 &           0.0508 &                     0.1668 &            0.0582 &               0.0584 &          0.0575 &         0.0615 &         0.0553 \\
		&     &   0.50   &               0.0526 &          \bf 0.6728 &           0.0529 &                     0.6698 &            0.1065 &               0.0980 &          0.0916 &         0.1216 &         0.0769 \\
		&     &   0.75   &               0.0559 &          \bf 0.9814 &           0.0567 &                     0.9808 &            0.2860 &               0.2339 &          0.2027 &         0.3514 &         0.1374 \\
		\cmidrule{2-12}
		& 200 &   0.25   &               0.0502 &          \bf 0.3189 &           0.0501 &                     0.3167 &            0.0674 &               0.0665 &          0.0645 &         0.0735 &         0.0595 \\
		&     &   0.50   &               0.0517 &          \bf 0.9511 &           0.0520 &                     0.9497 &            0.2119 &               0.1747 &          0.1547 &         0.2540 &         0.1106 \\
		&     &   0.75   &               0.0548 &          \bf 1.0000 &           0.0557 &                 \bf 1.0000 &            0.7051 &               0.6403 &          0.5543 &         0.8363 &         0.3176\\\bottomrule
	\end{tabular}
	\caption{\small Empirical powers for the uniformity tests on $\Omega_3$. The description of Table \ref{tab:pow:1} applies. \label{tab:pow:4}}
\end{table}

\begin{table}
	\centering
	\setlength{\tabcolsep}{2.5pt}
	\scriptsize
	\begin{tabular}{lll|R{1.25cm}R{1.25cm}R{1.25cm}R{1.25cm}R{1.25cm}R{1.25cm}R{1.25cm}R{1.25cm}R{1.25cm}}
		\toprule
		\scriptsize DGP & $n$ & $\kappa$ & \scriptsize Rayleigh & \scriptsize Bingham & \scriptsize Ajne & \scriptsize Gin\'e & \scriptsize CCF09 & \scriptsize Bakshaev & \scriptsize CvM & \scriptsize AD & \scriptsize Rt \\
		\midrule
		CvM       &  50 &   0.25   &               0.0585 &              0.0506 &       \bf 0.0586 &                     0.0507 &            0.0557 &           \bf 0.0585 &      \bf 0.0585 &     \bf 0.0585 &     \bf 0.0585 \\
		&     &   0.50   &               0.0867 &              0.0524 &       \bf 0.0868 &                     0.0524 &            0.0751 &           \bf 0.0870 &          0.0869 &     \bf 0.0870 &     \bf 0.0869 \\
		&     &   0.75   &               0.1453 &              0.0554 &           0.1456 &                     0.0554 &            0.1124 &           \bf 0.1462 &          0.1460 &     \bf 0.1462 &         0.1458 \\
		\cmidrule{2-12}
		& 100 &   0.25   &           \bf 0.0668 &              0.0517 &       \bf 0.0669 &                     0.0518 &            0.0619 &           \bf 0.0670 &          0.0669 &     \bf 0.0671 &     \bf 0.0669 \\
		&     &   0.50   &               0.1326 &              0.0554 &       \bf 0.1327 &                     0.0554 &            0.1049 &           \bf 0.1328 &      \bf 0.1328 &     \bf 0.1328 &     \bf 0.1328 \\
		&     &   0.75   &               0.2735 &              0.0620 &           0.2736 &                     0.0621 &            0.1978 &           \bf 0.2745 &          0.2743 &     \bf 0.2746 &         0.2742 \\
		\cmidrule{2-12}
		& 200 &   0.25   &               0.0874 &              0.0528 &           0.0874 &                     0.0527 &            0.0757 &           \bf 0.0877 &      \bf 0.0877 &     \bf 0.0877 &         0.0876 \\
		&     &   0.50   &               0.2444 &              0.0604 &           0.2446 &                     0.0605 &            0.1777 &               0.2458 &      \bf 0.2459 &     \bf 0.2460 &         0.2453 \\
		&     &   0.75   &               0.5525 &              0.0751 &           0.5528 &                     0.0752 &            0.4012 &               0.5548 &          0.5546 &     \bf 0.5550 &         0.5541 \\
		\midrule
		AD        &  50 &   0.25   &               0.0999 &              0.0550 &           0.1001 &                     0.0549 &            0.0836 &               0.1005 &          0.1003 &     \bf 0.1008 &         0.1002 \\
		&     &   0.50   &               0.3142 &              0.0735 &           0.3146 &                     0.0734 &            0.2273 &               0.3170 &          0.3164 &     \bf 0.3174 &         0.3157 \\
		&     &   0.75   &               0.6854 &              0.1106 &           0.6858 &                     0.1108 &            0.5219 &               0.6878 &          0.6875 &     \bf 0.6880 &         0.6870 \\
		\cmidrule{2-12}
		& 100 &   0.25   &               0.1636 &              0.0611 &           0.1638 &                     0.0611 &            0.1251 &               0.1649 &          0.1646 &     \bf 0.1650 &         0.1644 \\
		&     &   0.50   &               0.6201 &              0.1021 &           0.6204 &                     0.1023 &            0.4625 &               0.6241 &          0.6234 &     \bf 0.6246 &         0.6224 \\
		&     &   0.75   &               0.9624 &              0.1935 &           0.9624 &                     0.1943 &            0.8790 &           \bf 0.9630 &          0.9629 &     \bf 0.9630 &         0.9628 \\
		\cmidrule{2-12}
		& 200 &   0.25   &               0.3166 &              0.0725 &           0.3168 &                     0.0727 &            0.2272 &               0.3199 &          0.3194 &     \bf 0.3204 &         0.3185 \\
		&     &   0.50   &               0.9341 &              0.1716 &           0.9342 &                     0.1724 &            0.8257 &               0.9361 &          0.9358 &     \bf 0.9363 &         0.9353 \\
		&     &   0.75   &           \bf 0.9999 &              0.3941 &           0.9999 &                     0.3953 &            0.9979 &           \bf 0.9999 &      \bf 0.9999 &     \bf 0.9999 &     \bf 0.9999 \\
		\midrule
		Rt        &  50 &   0.25   &           \bf 0.0619 &              0.0504 &       \bf 0.0620 &                     0.0505 &            0.0583 &           \bf 0.0619 &      \bf 0.0618 &     \bf 0.0620 &         0.0619 \\
		&     &   0.50   &           \bf 0.1026 &              0.0522 &       \bf 0.1027 &                     0.0520 &            0.0884 &           \bf 0.1025 &          0.1024 &     \bf 0.1025 &     \bf 0.1026 \\
		&     &   0.75   &           \bf 0.1915 &              0.0539 &       \bf 0.1916 &                     0.0540 &            0.1533 &           \bf 0.1914 &      \bf 0.1915 &     \bf 0.1913 &     \bf 0.1915 \\
		\cmidrule{2-12}
		& 100 &   0.25   &           \bf 0.0745 &              0.0515 &       \bf 0.0744 &                     0.0517 &            0.0680 &           \bf 0.0745 &      \bf 0.0745 &     \bf 0.0745 &     \bf 0.0745 \\
		&     &   0.50   &           \bf 0.1744 &              0.0543 &       \bf 0.1745 &                     0.0544 &            0.1410 &               0.1742 &          0.1742 &         0.1740 &     \bf 0.1744 \\
		&     &   0.75   &               0.4020 &              0.0594 &           0.4021 &                     0.0596 &            0.3161 &               0.4023 &      \bf 0.4025 &         0.4021 &     \bf 0.4026 \\
		\cmidrule{2-12}
		& 200 &   0.25   &               0.1053 &              0.0524 &       \bf 0.1054 &                     0.0524 &            0.0904 &               0.1053 &      \bf 0.1055 &         0.1053 &     \bf 0.1054 \\
		&     &   0.50   &               0.3553 &              0.0584 &           0.3554 &                     0.0587 &            0.2797 &               0.3557 &      \bf 0.3560 &         0.3555 &     \bf 0.3559 \\
		&     &   0.75   &               0.7788 &              0.0698 &           0.7791 &                     0.0699 &            0.6686 &           \bf 0.7810 &      \bf 0.7809 &     \bf 0.7810 &         0.7803 \\
		\midrule
		vMF       &  50 &   0.25   &           \bf 0.0579 &              0.0498 &       \bf 0.0580 &                     0.0497 &            0.0555 &           \bf 0.0580 &      \bf 0.0579 &     \bf 0.0580 &     \bf 0.0579 \\
		&     &   0.50   &           \bf 0.0840 &              0.0498 &       \bf 0.0842 &                     0.0497 &            0.0739 &               0.0838 &          0.0839 &         0.0838 &     \bf 0.0841 \\
		&     &   0.75   &           \bf 0.1372 &              0.0501 &       \bf 0.1373 &                     0.0501 &            0.1099 &               0.1370 &          0.1370 &         0.1370 &         0.1371 \\
		\cmidrule{2-12}
		& 100 &   0.25   &           \bf 0.0657 &              0.0505 &       \bf 0.0658 &                     0.0505 &            0.0610 &           \bf 0.0657 &      \bf 0.0657 &     \bf 0.0656 &     \bf 0.0657 \\
		&     &   0.50   &           \bf 0.1265 &              0.0503 &       \bf 0.1264 &                     0.0503 &            0.1031 &               0.1261 &          0.1262 &         0.1259 &     \bf 0.1264 \\
		&     &   0.75   &           \bf 0.2585 &              0.0509 &           0.2583 &                     0.0510 &            0.1927 &               0.2573 &          0.2576 &         0.2569 &         0.2580 \\
		\cmidrule{2-12}
		& 200 &   0.25   &           \bf 0.0847 &              0.0500 &       \bf 0.0846 &                     0.0500 &            0.0747 &           \bf 0.0846 &      \bf 0.0847 &     \bf 0.0846 &     \bf 0.0847 \\
		&     &   0.50   &           \bf 0.2315 &              0.0506 &       \bf 0.2314 &                     0.0507 &            0.1749 &               0.2309 &      \bf 0.2313 &         0.2307 &     \bf 0.2314 \\
		&     &   0.75   &           \bf 0.5305 &              0.0519 &           0.5302 &                     0.0519 &            0.3967 &               0.5287 &          0.5296 &         0.5282 &         0.5301 \\
		\midrule
		SC        &  50 &   0.25   &               0.0655 &              0.0508 &       \bf 0.0657 &                     0.0508 &            0.0609 &           \bf 0.0657 &      \bf 0.0656 &     \bf 0.0657 &     \bf 0.0656 \\
		&     &   0.50   &               0.1161 &              0.0516 &       \bf 0.1163 &                     0.0515 &            0.0968 &           \bf 0.1160 &      \bf 0.1161 &         0.1160 &     \bf 0.1162 \\
		&     &   0.75   &               0.2096 &              0.0519 &       \bf 0.2098 &                     0.0519 &            0.1624 &               0.2088 &          0.2090 &         0.2085 &         0.2094 \\
		\cmidrule{2-12}
		& 100 &   0.25   &           \bf 0.0846 &              0.0511 &       \bf 0.0846 &                     0.0511 &            0.0744 &           \bf 0.0845 &      \bf 0.0845 &     \bf 0.0845 &     \bf 0.0846 \\
		&     &   0.50   &           \bf 0.2092 &              0.0526 &       \bf 0.2092 &                     0.0525 &            0.1612 &               0.2086 &          0.2088 &         0.2083 &     \bf 0.2090 \\
		&     &   0.75   &           \bf 0.4415 &              0.0533 &       \bf 0.4413 &                     0.0532 &            0.3315 &               0.4400 &          0.4405 &         0.4394 &         0.4411 \\
		\cmidrule{2-12}
		& 200 &   0.25   &           \bf 0.1283 &              0.0527 &       \bf 0.1283 &                     0.0526 &            0.1050 &               0.1282 &      \bf 0.1283 &     \bf 0.1282 &     \bf 0.1284 \\
		&     &   0.50   &           \bf 0.4360 &              0.0557 &           0.4359 &                     0.0558 &            0.3280 &               0.4354 &      \bf 0.4360 &         0.4351 &     \bf 0.4361 \\
		&     &   0.75   &               0.8191 &              0.0571 &           0.8190 &                     0.0571 &            0.6824 &               0.8191 &      \bf 0.8196 &         0.8188 &     \bf 0.8196 \\
		\midrule
		W        &  50 &   0.25   &               0.0499 &          \bf 0.0533 &           0.0500 &                 \bf 0.0533 &            0.0499 &               0.0502 &          0.0500 &         0.0503 &         0.0500 \\
		&     &   0.50   &               0.0501 &          \bf 0.0655 &           0.0502 &                     0.0653 &            0.0504 &               0.0511 &          0.0508 &         0.0514 &         0.0505 \\
		&     &   0.75   &               0.0506 &          \bf 0.0964 &           0.0507 &                     0.0962 &            0.0517 &               0.0531 &          0.0523 &         0.0536 &         0.0517 \\
		\cmidrule{2-12}
		& 100 &   0.25   &               0.0502 &          \bf 0.0561 &           0.0502 &                 \bf 0.0560 &            0.0507 &               0.0506 &          0.0505 &         0.0507 &         0.0504 \\
		&     &   0.50   &               0.0504 &          \bf 0.0832 &           0.0505 &                 \bf 0.0831 &            0.0515 &               0.0523 &          0.0518 &         0.0526 &         0.0512 \\
		&     &   0.75   &               0.0510 &          \bf 0.1581 &           0.0511 &                     0.1578 &            0.0536 &               0.0559 &          0.0545 &         0.0569 &         0.0531 \\
		\cmidrule{2-12}
		& 200 &   0.25   &               0.0504 &          \bf 0.0633 &           0.0505 &                 \bf 0.0632 &            0.0508 &               0.0513 &          0.0511 &         0.0514 &         0.0509 \\
		&     &   0.50   &               0.0507 &          \bf 0.1263 &           0.0507 &                     0.1259 &            0.0526 &               0.0544 &          0.0535 &         0.0552 &         0.0524 \\
		&     &   0.75   &               0.0512 &          \bf 0.3150 &           0.0512 &                     0.3144 &            0.0572 &               0.0615 &          0.0587 &         0.0637 &         0.0555\\\bottomrule
	\end{tabular}
	\caption{\small Empirical powers for the uniformity tests on $\Omega_{10}$. The description of Table \ref{tab:pow:1} applies. \label{tab:pow:5}}
\end{table}

We perform next an independent simulation study to elucidate the reasons of the somehow surprising conclusion \ref{conc:local} in Section \ref{sec:power}. The tests of the simulation study in such section are benchmarked with respect to the Invariant Likelihood Ratio Test (ILRT) for testing uniformity against the alternative \eqref{eq:dgp}. If $f_0$ denotes the uniform pdf on $\Om{q}$, the ILRT for testing uniformity against \eqref{eq:dgp} for a specified $0<\kappa<1$, that is, for testing
\begin{align}
\Hcal_0: f=f_0\quad \text{vs.}\quad \Hcal_{1,\kappa}: f\in\{f_{\bmu,\kappa}:\bmu\in\Om{q}\}
\label{eq:LRT}
\end{align}
is the test that rejects for large values of the ILRT statistic:
\begin{align*}
\mathrm{L}_{\kappa}:=\int_{\Om{q}} \prod_{i=1}^n f_{\bmu,\bga}(\bX_i) \,\omega_q(\mathrm{d}\bga).
\end{align*}

We focus on the simplest DGP \eqref{eq:dgp} among CvM, AD, and Rt that admits a tractable ILRT. This DGP is Rt with $t=1/2$ for $q=1$, therefore coinciding with \cite{Ajne1968}'s ``semicircle deviation''. In this setting, each sample observation can be parametrized as $\bX_i=(\cos\Theta_i,\sin\Theta_i)'$ for $\Theta_i\in[0,2\pi)$ and $f^{\mathrm{Rt}}(z)=1_{\left\{z\geq 0\right\}} + 1/2$. Thus the ILRT statistic becomes
\begin{align*}
\mathrm{L}_{\kappa}
=\int_0^{2\pi} \prod_{i=1}^n g_\kappa(\Theta_i-\theta) \,\mathrm{d}\theta
=\sum_{j=1}^{2n}\int_{I_j} \prod_{i=1}^n g_\kappa(\Theta_i-\theta)\,\mathrm{d}\theta
=\sum_{j=1}^{2n} \prod_{i=1}^n g_\kappa(\theta_{j}) \ell_j,
\end{align*}
where $g_\kappa(\varphi):=\frac{1}{2\pi}\lrb{1+\kappa \left(1_{\{\cos(\varphi)\geq 0\}} - \tfrac{1}{2}\right)}$, $\{I_j\}_{j=1}^{2n}$ are certain intervals defined below, and $\ell_j$ is the length of $I_j$ and $\theta_j$ its midpoint. The intervals $\{I_j\}_{j=1}^{2n}$ are constructed by first augmenting the sample $\{\Theta_i\}_{i=1}^n$ to $\{\tilde{\Theta}_i\}_{i=1}^{2n}$, where $\tilde{\Theta}_{i}=(\Theta_i-\pi/2)\mod2\pi$ and $\tilde{\Theta}_{i+n}=(\Theta_i+\pi/2)\mod2\pi$ for $i=1,\ldots,n$, and then setting $I_j:=[\tilde{\Theta}_{(j)},\tilde{\Theta}_{(j+1)})$, $j=1,\ldots,2n$, where $\tilde{\Theta}_{(2n+1)}:=\tilde{\Theta}_{(1)}+2\pi$.\\

We consider $M=10^8$ Monte Carlo replicates to reduce the Monte Carlo noise and capture smaller power effects. We employ the tests considered in Section \ref{sec:power} (Ajne is omitted since it coincides with Rt for $t=1/2$) plus the ILRT for \eqref{eq:LRT}. We use sample size $n=50$ and the local deviations $\kappa=0.05k$, $k=0,\ldots,20$. As in Section \ref{sec:power}, the statistics are calibrated under the null hypothesis by Monte Carlo. The obtained empirical powers are collected in Figures \ref{fig:onehundred:1} and \ref{fig:onehundred:2} and  give the following conclusions:

\begin{enumerate}[label=(\textit{\alph{*}})., ref=(\textit{\alph{*}})]
	
	\item The optimality of the ILRT is verified and evidenced to be smaller than $10^{-3}$ for the investigated $\kappa$'s (Figure \ref{fig:onehundred:1}). Therefore, the power gap between the optimal test for \eqref{eq:LRT} and other tests is fairly small, as is also reflected in the virtual equivalence of the powers shown in Figure \ref{fig:onehundred:2}. The Monte Carlo noise explains that the empirical power of Rt is larger than the power of the ILRT.
	\item The Rt test is locally equivalent to the ILRT for $\kappa\approx0$, both being indistinguishable (at the $95\%$ confidence) within the Monte Carlo noise until $\kappa$ approaches $0.10$. The Rt test clearly outperforms the remaining tests except the ILRT. 
	\item An apparently high number of Monte Carlo replicates such as $10^6$ is still insufficient to fully capture optimalities in the investigated DGP. We conjecture this is a prevalent issue with all the alternatives \eqref{eq:dgp} investigated in Section \ref{sec:power}.
	\item Unsurprisingly, the Bingham and Gin\'e tests are blind against this alternative and have the nominal significance level as power. A difference in power is evidenced for the Rayleigh and Ajne test (here acting as the Rt test), yet again it is fairly small.
	
\end{enumerate}

We conclude mentioning that this kind DGP was already considered in \cite{Stephens1969}. In particular, his Table 3 compares the powers of Ajne, Watson, and \cite{Kuiper1960} tests for the circle at significance level $10\%$ using $5000$ Monte Carlo replicates. However, his study does not show that the Ajne test is significantly (with a $95\%$ confidence) more powerful than the competing tests for this alternative.

\begin{figure}[H]
	\centering
	\includegraphics[width=0.85\textwidth]{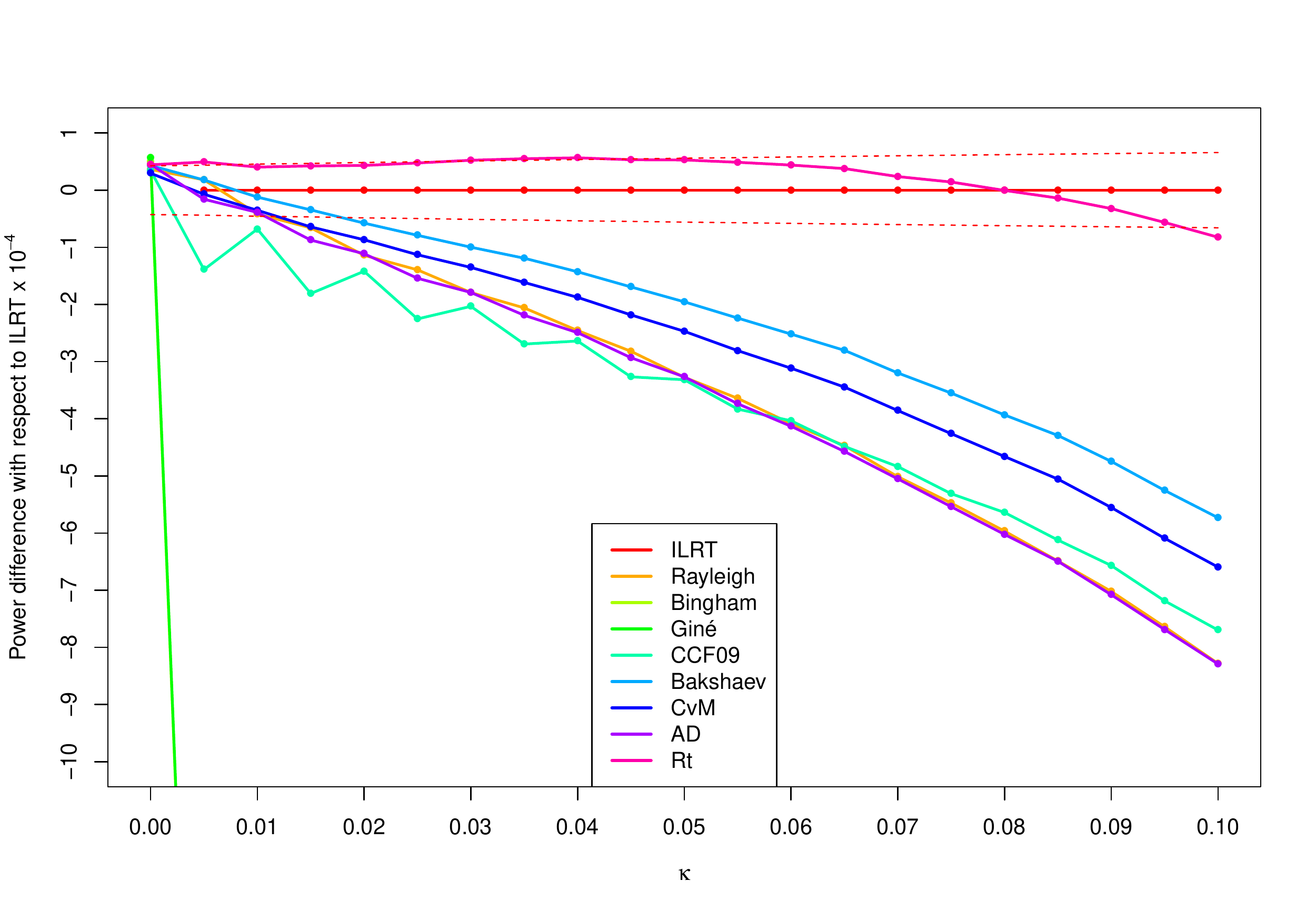}
	\vspace*{-0.5cm}
	\caption{\small Difference of empirical powers with respect to the ILRT for different deviations $\kappa$. The dashed lines represent the $99\%$ confidence interval about the ILRT power. For $\kappa=0$, the testing problem \eqref{eq:LRT} is undefined, and its power is replaced by the significance level, $5\%$. The vertical axis is on the scale $10^{-4}$. \label{fig:onehundred:1}}
\end{figure}

\begin{figure}
	\centering
	\includegraphics[width=0.85\textwidth]{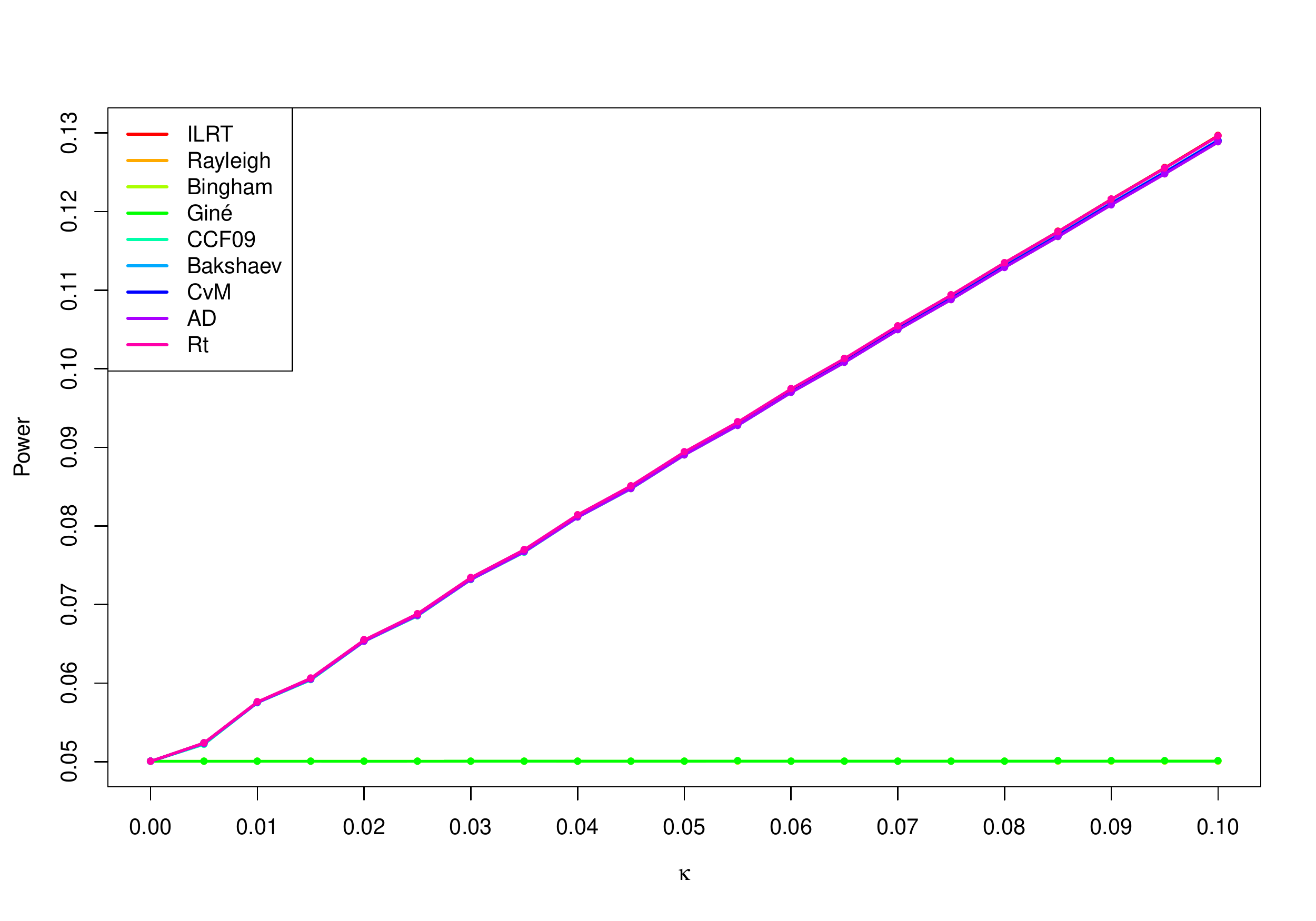}
	\vspace*{-0.5cm}
	\caption{Empirical powers for different deviations $\kappa$.  \label{fig:onehundred:2}}
\end{figure}


\fi

\end{document}